\documentclass[12pt,eqsecnum,preprint]{aastex}

\shorttitle{Multi-Wavelength Flares on HR~1099}
\shortauthors{Osten et al.}

\begin{document}
 
\title{A Multi-Wavelength Perspective of Flares on HR~1099: Four Years of Coordinated Campaigns 
\label{sec:hr1099}}

\author{Rachel A. Osten\altaffilmark{1}}
\affil{National Radio Astronomy Observatory, 
Charlottesville, VA 22903}

\author{Alexander
Brown, Thomas R. Ayres }
\affil{Center for Astrophysics and Space Astronomy, University of
Colorado, Boulder, CO 80309-0389 }

\author{Stephen A. Drake}
\affil{HEASARC, NASA$/$GSFC, Greenbelt, MD 20771} 

\author{ Elena Franciosini, Roberto Pallavicini}
\affil{INAF$/$Osservatorio Astronomico di Palermo \\
%Piazza del Parlamento 1 \\
%90134 Palermo \\
Italy}

\author{Gianpiero Tagliaferri}
\affil{INAF/Osservatorio Astronomico di Brera \\
%via Bianchi 46 \\
%23807 Merate \\
Italy}

\author{Ron T. Stewart}
\affil{Australia Telescope National Facility, Retired \\
%CSIRO \\
%P.O. Box 76 \\
%Epping \\
%NSW2121 \\
Australia}

\author{Stephen L. Skinner}
\affil{Center for Astrophysics and Space Astronomy, University of Colorado, Boulder, CO 80309-0389}

\author{Jeffrey L. Linsky}
\affil{JILA, University of Colorado \& NIST, Boulder, CO 80309-0440 }

\altaffiltext{1}{Jansky Fellow, National Radio Astronomy Observatory}

\begin{abstract}
We report on four years of multiple wavelength observations of the RS~CVn system V711~Tau (HR~1099)
from 1993, 1994, 1996 and 1998.  This combination of radio, ultraviolet (UV), extreme ultraviolet (EUV), 
and X-ray observations allows us
to view, in the most comprehensive manner currently possible, the coronal and upper atmospheric variability
of this active binary system.  We report on the changing activity state of
the system as recorded in the EUV and radio across the four years of observations, and study the 
high energy variability using an assemblage of X-ray telescopes.  
We find: \\
(1) evidence for coherent emission at low radio frequencies ($\leq$ 3 GHz) which appears to be both
highly time variable and persistent for several hours.  
Such phenomena are relatively common, occurring $\approx$ 30\% of the time
HR~1099 was observed at L-band.
The measured polarizations of these bursts are left
circularly polarized, in contrast with behavior at higher frequencies which has the opposite
helicity.  The polarizations are consistent with a variable source that is 100\% left circularly
polarized, along with a steady level of flux and polarization which is 0 or slightly right circularly
polarized.  There appears to be a low degree of correlation between bursts at 20 cm and higher frequency
gyrosynchrotron flares, and also between 20 cm bursts and large EUV/soft X-ray (SXR) outbursts.
\\
(2) Higher frequency (5--8 GHz) flares show an inverse relationship between flux and polarization levels 
as the flare evolves; this behavior is consistent with flare emission which is initially unpolarized.
Large variations in spectral index are observed, suggesting changes in optical depths of the flaring plasma
as the burst progresses.  
%{\bf (REMOVE: There is generally poor correlation between high- and  low-frequency radio flares, 
%cementing previous inferences of a different emission mechanism at work in active binary systems at low frequencies.)}
Quiescent polarization spectra show an increase of polarization with frequency, a pattern typically seen
in active binary systems but still not understood.
\\
(3) EUV observations reveal several large flares, in addition to numerous smaller enhancements.  The
total range of variability as gleaned from light curve variations is only a factor of 7, however.
Observations in different years provide evidence of a change in the quiescent, not obviously flaring, 
luminosity, by a factor of up to 2.  From an analysis of time-resolved spectral variations, we are able 
to infer evidence for the creation of high-temperature plasma during flare intervals compared with quiescent
intervals.  Interpretation of EUV spectral variations is hindered by the lack of ability to diagnose
continuum levels and activity-related abundance changes, which are known from higher energy observations.
Electron densities determined by line ratios of density-sensitive emission lines are high
(10$^{12}$--10$^{13}$ cm$^{-3}$) and there is no evidence for large density enhancements during flare intervals,
compared with quiescent intervals.  \\
(4) X-ray observations reveal several flares, and provide evidence of energy-dependent flare
evolution:  harder X-ray energies show faster temporal evolution than at softer energies.  Time-resolved X-ray
spectral analysis shows the presence of hot plasma,  T$_{e}\sim$ 30 MK, during flares compared to quiescent intervals, 
as well as evidence for changing abundances during flares.  The abundance of
iron (which is subsolar) shows an enhancement of a factor of three at the peak of a moderate flare seen by {\it ASCA}
relative to the pre-flare level; abundances
decrease during the flare decay. 
No hard ($> 15$ keV) emission is detected by either {\it RXTE} or {\it
BeppoSAX}.\\
(5) The luminosity ratios L$_{EUV}$/L$_{R}$ in quiescence determined from several time intervals
during the four campaigns are consistent with previously determined ratios from a sample of active stars and solar 
flares.  The range of L$_{EUV}$/L$_{R}$ from three EUV/radio (3.6 cm) flares is the same as the values
obtained during quiescence, which points to a common mechanism for producing both flaring
and not flaring emission.\\
(6) Seventeen flares were observed in the EUV and$/$or SXR during the four campaigns; 
of the eight flares that had radio coverage,
three show 3.6 cm radio flares, which are generally consistent with the Neupert effect.  Five EUV$/$SXR flares had
partial UV coverage; all show UV responses, particularly in the C~IV transition.  
The UV flare enhancements can occur at the same time as the 3.6 cm radio flares, in two cases where radio, UV, and EUV/SXR
flare coverage overlapped.\\
(7) For SXR 
%high energy 
flares, we find that the contrast between flare emission and quiescent emission increases as expected
towards higher energies, making flare detections easier at harder X-ray energies.  This is due to the creation
of high temperature plasma during flares, which shows up predominantly in high energy continuum emission.
%{\bf (REMOVE There are no detections of hard X-ray emission ($>$ 20 keV) during quiescence or flares.)}
We find a discrepancy between the implied flaring rate based on EUV observations, and higher energy observations;
the lower
%softer 
energies tend to miss many of the flares, due to the lack of sufficient contrast with quiescent emission. \\

\end{abstract}


\section{Introduction}
As a bright and variable system in almost all wavelength regions, HR~1099 (V711~Tau; HD~22468) has been
the subject of many past observations from radio to X-ray wavelengths.
HR~1099 is a short-period (P$_{\rm orb} =$ 2.83774 d) binary composed of a G5 dwarf and a K1 subgiant.
The system lies at a Hipparcos distance of 29 pc \citep{hipparcos}.
The orbital and rotational periods of the stars are tidally synchronized.
The mass ratio is 1.3, yet the K subgiant outsizes the G subgiant by a factor 
of three in radius.  The orbital inclination is $\approx$ 33$^{\circ}$ 
\citep[][and references therein]{cabs}
and there are no eclipses.  
Because of its spectroscopic characteristics and highly active chromospheric, transition region,
and coronal emissions, HR~1099 is a member of the RS~CVn class of binary systems.
Many of the phenomena seen on RS~CVn systems have solar counterparts, and invite the comparison
between ``hyper-active'' stars such as these and less active stars like the Sun.

A schematic of the generally accepted model for solar flares proceeds as follows:  
A flare begins with a source of free energy, thought to originate from magnetic reconnection high in the 
solar corona.  Some of this energy is used to 
accelerate electrons to moderately relativistic speeds. 
Electron beams can be generated, and propagate outward or 
into the atmosphere; the electron beams emit plasma radiation (type III bursts), whose frequencies
trace the ambient plasma density the beams encounter as they propagate either outward or
into the atmosphere \citep{aschwandenbenz1997}.  Downward-directed accelerated electrons
can become magnetically trapped in the coronal loop or arcade, and emit gyrosynchrotron emission
at microwave frequencies.
As the electrons impact the denser regions of the lower
atmosphere, they deposit their energy, emitting hard X-rays via nonthermal bremsstrahlung 
emission and white light continuum emission at the footpoints of the loops.  
The energy deposited by the electrons in the lower atmosphere heats plasma to coronal temperatures
on a timescale short compared to the hydrodynamic expansion time,
ablating material at 10$^{6}$--10$^{7}$ K up the coronal loops, where it
emits soft X-ray radiation.  As the coronal density increases coronal material
can stop the energetic electrons higher up in the atmosphere thereby heating the corona
directly.  
%As the amount of heat deposited in the lower atmosphere decreases, the emission
%from those regions dies away and the corona emits soft X-rays.  
Once the nonthermal energy input has
ceased, the material in the loop condenses into the chromosphere and the soft X-ray emission
returns to its preflare state.  This scenario implies a set of related and
correlated variations that should be observable in different spectral regions.

Despite the relatively advanced state of understanding of multi-wavelength solar flare emissions,
there is a paucity of observations on the stellar side.  Part of this may result from
the difficulties inherent in organizing such an observing campaign
for stellar flares; but the potential 
benefits to understanding stellar flares must surely outweigh the enormous effort needed to
coordinate such observations.  \citet{gudelbenz} showed that there existed
an almost linear relationship between stellar quiescent radio and soft X-ray emission,
suggesting that coronal heating and particle acceleration are closely linked in a way that may be
similar to solar flares.  In one of the earliest studies of multi-wavelength emission, \citet{weileretal}
 examined optical, UV, and radio observations of two RS~CVn systems 
(one of them HR~1099) and said ``there are suggestive coincidences between peak radio flux density
and optical-UV emission activity from both systems.''  Other studies of multi-wavelength stellar
flares have revealed a zoo of phenomena, leading to split opinions as to whether such
correlations even exist.  In one of the earliest simultaneous studies, \citet{kundu1988}
examined radio and X-ray observations of four flare stars.  While they noted some X-ray bursts
coincided in time or were preceded by 10--15 minutes by 20 cm radio flares, they still claimed
the degree of correlation was low.  \citet{foxetal1994} observed the decay of a radio
flare on the RS~CVn system EI~Eri but did not see an accompanying X-ray or EUV flare.
Yet \citet{stern1992} found UV and microwave flaring occuring during
an X-ray outburst on the RS~CVn system $\sigma^{2}$~CrB.  \citet{gagne}
investigated the M dwarf binary system EQ~Peg with radio, optical, EUV and X-ray wavelengths
and found two populations of X-band (3.6 cm) flares:  highly polarized flares with no counterparts
at shorter wavelengths, and moderately polarized flares which do have shorter wavelength
counterparts.  Recently, \citet{ostenetal2000} established a correlation
of radio flares with X-ray flares on $\sigma^{2}$~CrB, with X-ray flares peaking up to 1.4 hours
before the radio peak.  And most recently, \citet{ayres2001} conducted
coordinated UV, EUV, X-ray and radio observations on HR~1099
and detected a large UV flare not seen in higher
energy emissions.

The flare mechanism is a poorly understood phenomenon, even when considered in a single wavelength region
(or on a very well-studied star like the Sun).  
Expectations that stellar flares mimic the behavior of solar flares can introduce biases
into the interpretation of multi-wavelength studies.  Also, it is necessary to
recognize the importance of time delays between different wavelength regions 
when studying flares in a multi-wavelength context and the necessity
of long exposure times to catch and observe flares in their entirety.  

This paper describes the results of four campaigns which observed HR~1099 in multiple wavelength 
regions, in 1993, 1994, 1996, and 1998.  A summary of these campaigns is given in
Table~\ref{table1}.  Section~\ref{sec:prevobs} summarizes previous
observations of HR~1099; section~\ref{sec:ch5datred} describes the observations and initital data reduction;
Section~\ref{sec:anal} examines 
the wavelength regions individually across the time span of these
four campaigns; \S~\ref{sec:multi} investigates comparisons between the different wavelength regions
investigated; and \S~\ref{sec:ch5conc} concludes.

\section{Previous Studies \label{sec:prevobs}}
Much is known about the starspot distribution on HR~1099: 
The subgiant primary of HR 1099 is one of the brightest stars to show evidence for 
spots in its photosphere, and has been the subject of numerous photometric and
Doppler imaging campaigns since 1975.  \citet{vogtetal1999}  presented a study
of 11 years of Doppler images of the stellar surface that display the long
term stability of a cool polar spot and lower latitude spots that vary on less than
one year timescales.  The polar spot, together with a low degree of shearing and
nearly solid body rotation of the other spots detected, led \citeauthor{vogtetal1999} to suggest
the presence of a global multi-kilogauss dipolar magnetic field on the star.  \citet{donatietal1990}
used the technique of Zeeman Doppler imaging to detect magnetic
features, and deduced magnetic fields whose average strength was 985 G covering 8\%
of the total stellar surface.  A longer term study of magnetic features by \citet{donati1999}
found that a significant fraction of the magnetic flux comes from regions at
photospheric temperatures; i.e., there was no spatial correlation between brightness
and magnetic inhomogeneities.  They determined variations in the orbital period with
a period of 18 years, and deduced a magnetic activity cycle period of $\geq$ 12
years from a lack of polarity reversal during their 6 years of observations.  

Other attempts to determine cycle periods from photometry have indicated similarly
long timescales:  \citet{henryetal1995}  from 18 years of photometry determined two
periodicities in the mean V magnitudes from their spot solutions of 5.5 and 16
years.  \citet{olahetal2000} used 22 years of monitoring and
found a cycle period of 16.5 years along with a secondary period of 3.5 years.  This
second period is in agreement with \citet{vogtetal1999}'s finding of a weak periodicity in
the area of the polar spot of about 3 years.  While there is a consensus that
long-term cyclic behavior is at work on HR 1099, the lack of consensus on a
timescale for this periodicity hampers any attempt to place our observations in the
context of an activity trend.

\citet{linskyetal1989} reported on {\it International Ultraviolet Explorer (IUE)}
observations of a flare on HR~1099 in 1981, in
which the Mg II $k$ line profile displayed a broad component of 66 km s$^{-1}$ and was
redshifted by 90$\pm$30 km s$^{-1}$.  The electron density at 10$^{4}$ K was a factor
of 15 larger during the peak of the flare than in the quiescence preceding the
flare.  
\citet{woodetal1996} made an association
between broad wings in ultraviolet (UV) emission lines as seen by the
{\it Goddard High-Resolution Spectrograph (GHRS)} and micro-flaring in the
chromosphere and transition region of HR~1099.
\citet{busaetal1999} 
used {\it IUE} observations from 1992 to investigate velocity variations in
the Mg II $h$ line which they attributed to emission from a localized region in the
stellar chromosphere of the primary.  In coordinated {\it IUE} and {\it GHRS} observations in
1993, \citet{dempseyetal1996} saw several flare enhancements in UV line emission; one
such enhancement lasted more than 24 hours as seen with {\it IUE}.  

\citet{barstow1985} discussed a pointed EXOSAT observation of HR~1099
in
1981, in which a small flare with factor of two enhancement occurred.   The light
curve also displayed smaller amplitude variability on timescales of 500--1000
seconds, which \citet{vandenoordbarstow1988} explained as evidence of flare-like
heating in the corona of HR~1099.  From spectral fits to the data, \citet{pasquinietal1989}
found evidence for coronal temperatures of 3 and 
25 MK.  \citet{drakeetal1994} saw modulation in the light curve of {\it Extreme Ultraviolet
Explorer (EUVE)} observations
at the level of 40\%, which they attributed to rotational modulation of a starspot.
Contemporaneous optical photometry showed a light minimum which leads the extreme ultraviolet (EUV)
maximum by 90 degrees in phase.  

\citet{feldmanetal1978} presented radio data featuring several large flares on HR~1099
 in 1978 with flux densities reaching up to 1 Jy at 10.5 GHz.  \citet{lestradeetal1984}
performed 8.4 GHz Very Long Baseline Interferometry (VLBI) observations and found radio emission from a source
comparable
in size to that between the surfaces of the two stars in the binary system
(0.9 mas)
during a large radio outburst of 400 mJy.  \citet{jonesetal1996} modeled 
Very Large Array (VLA) and Australia Telescope Compact Array (ATCA)
quiescent observations and determined from gyrosynchrotron model fits that a highly
organized magnetic field was necessary to explain the data.  \citet{trigilioetal1993}
reported on VLBI 5 GHz observations during a large flare whose source size
increased to a size comparable to the entire binary system.   \citet{umanaetal1995}
report the results of a radio monitoring campaign at 5 GHz from 1990--1993 which detected several
large flares with flux densities of up to 800 mJy.  
 \citet{trigilioetal2000} performed a 5 GHz European VLBI Network (EVN) observation of HR 1099
simultaneously with our observing campaign in 1998. They found an overall
size of the radio source, during quiescence, comparable to the angular
diameter of the active K star and thus much smaller than the active period
radio source previously observed by \citet{trigilioetal1993}.
More recently, \citet{richardsetal}
presented results from five years of radio monitoring at 2.3 and 8 GHz with the 
Green Bank Interferometer, and found periodicities of flaring cycles at $\sim$ 120 days,
and a weaker period of $\sim$ 81 days, with 8.3 GHz flare fluxes reaching a peak of 1.44 Jy.
In a related note, \citet{schmittetal2003} now have succeeded in spatially resolving an X-ray limb flare on the magnetically
active star in Algol, so we may soon see the first image of a flare on HR~1099.

%Because of thermal temperature structuring of the atmospheres of cool stars, observations 
%in multiple wavelength regions effectively allows us to observe the characteristics 
%and dynamics of plasma
%at different vertical heights in the atmosphere.  On the Sun, spatially resolved
%multiple wavelength observations add a third dimension; on unresolved stars we get a sense
%of the interrelation between different atmospheric layers.  The solar community has
%a vigorous program of examining relationships between emissions across the electromagnetic 
%spectrum.  This is especially useful in flare studies, where correlated dynamics
%gives a clue to the underlying physical processes at work during a flare.  Multi-wavelength
%correlations of flare emission on the Sun have been studied since 1961 (Kundu 1961 JGR 66, 4308) 
%when Kundu showed a remarkable correlation between centimeter wavelength emission and
%hard X-ray emission in flares.  Studies have continued, and it is now known that centimeter
%wavelength, soft X-ray, and hard X-ray emission in solar flares are both temporally and
%spatially  well-correlated (Kundu et al. 1994 ApJS 590, 599).  Other studies have
%concentrated on the relationship between UV, EUV, and hard X-ray emission (Cheng, Tandberg-Hanssen,
%\& Orwig 1984 ApJ 278, 853; McClymont \& Canfield 1986 ApJ 305, 936).  While the
%observations still generate a fair amount of speculation, there is enough agreement to
%generate a ``standard model'' of a solar flare (Dennis \& Schwartz 1989 Solar Physics 121, 75).
%

Finally, quiescent and flaring X-ray emission from HR 1099 has been
recently observed with {\it Chandra} and {\it XMM-Newton}
\citep{brinkmanetal2001,audard2001a,drake2001,ayres2001,nessetal2002,audard2003}
with emphasis on high-resolution grating spectroscopy (see
\S~\ref{sec:hr1099ascasec} for a discussion).

%%%%%%%%%%%%%%%%%%%%%%%%%%%%%%%%%%%%%%%%%%%%%%%%%%%%%%

\section{Observations and Data Reduction \label{sec:ch5datred}}
Since the nature of this study is to investigate both short- and long-term variability
in this active binary system, the number of individual observations that comprise the
whole study is large and spans many years.  Such a collaborative effort involves
numerous people, and several papers have been written analyzing individual datasets
within this large study.  
The current investigation does present a reanalysis of several datasets which have been considered by
others.  
Papers and conference proceedings describing the radio datasets are:
\citet{jonesetal1996,brownetal1994,brownetal1996,brownetal1997,brownetal1998}.
For the radio datasets, the 1993 and 1994 data 
\citep[previously published in][]{jonesetal1996,brownetal1994} are reanalyzed along with the 1996 (presented only at conferences) and
1998 datasets for a uniform and consistent analysis.
The UV data has been discussed in \citet{dempseyetal1996,brownetal1994,brownetal1996}, and is presented
here for multi-wavelength comparisons.
The {\it EUVE} data (lightcurves and spectra) have been discussed in a number of papers,
including: \citet{ostenbrown1999,griffithsjordan1998,sf2002}.  
The context of the {\it EUVE} analysis in the present paper is to examine the light curve and spectral
variability on a flare-to-flare basis, which was not the focus of previous investigations.
Preliminary analysis of
the various X-ray datasets has been provided in 
\citet{brownetal1996,brownskinner1996,brownetal1994,brownetal1997,brownetal1998,pallavic2001}.
The current analysis of the X-ray datasets is a detailed examination of both spectral and light curve changes.
We discuss the observations in order of decreasing wavelength.  
A summary of all the observations appears in Table~\ref{table1}.

\subsection{{\it VLA} Observations}
HR~1099 was observed with the NRAO\footnote{The National Radio Astronomy Observatory is a facility of the
National Science Foundation operated under cooperative agreement by Associated Universities, Inc.}
VLA\footnote{For more information on the VLA, see http://www.vla.nrao.edu} radio array for all four of the campaigns. The observations were performed in two subarrays, in 
order to obtain simultaneous multi-frequency observations.  Typically one subarray observed at 3.6 cm,
while the other switched between 6 and 20 cm.  Sparse coverage at 2 cm was included to extend
frequency coverage for spectral index determinations, but only general information on activity levels
at 2 cm can be gleaned.
The flux calibrator for all frequencies for all VLA observations was 0137+331, with
a flux density of 1.81 Jy at 2 cm, 3.29 Jy at 3.6 cm, 5.52 Jy at 6 cm, and 15.97 Jy at 20 cm.
The phase calibrator for all frequencies for all the VLA observations was 0323+055, with
measured flux densities at 2 cm of 0.22 Jy, 3.6 cm of 0.43 Jy, 6 cm of 0.83 Jy, and 20 cm of 2.79 Jy.
The 1993 VLA data were obtained at 3.6, 6, and 20 cm (8.41, 4.89, 1.46 GHz), while the remaining three years 
of observations were at 2, 3.6, 6, and 20 cm ( 14.96, 8.41, 4.89, 1.46 GHz).  
In 1993, the observations were taken in DnC array
on September 13.6--18.4;
with 45.8 hours of time on source at 3.6 and 6 cm.  Observations at 20 cm were compromised
due to radio frequency interference.
The 1994 observations were obtained on August 25.3--28.6, in the B array.
Time spent on source at 2, 3.6, 6, and 20 cm was 1.3, 30.1, 13.1, and 13.4 hours, respectively.
During the 1996 campaign on September 2.3--7.7, the VLA was in D array; time spent on the source at
2, 3.6, 6, and 20 cm was 0.75, 48.9, 13.2, and 22.6 hours, respectively. 
The 1998 VLA observations were collected on September 7.3--11.7, in the B array.  Time spent on the source
at 2, 3.6, 6, and 20 cm was 1.7, 41, 16.9, and 18.5 hours, respectively. 
Data were calibrated and reduced using AIPS; light curve generation was performed
using the DFTPL task.  

\subsection{{\it ATCA} Observations}
Observations using the ATCA\footnote{For more information on the ATCA, see http://www.atnf.csiro.au} were obtained during all four
years of the campaigns, in 1993, 1994, 1996, and 1998.  The data comprise four wavelengths: 3.6, 6,
13, and 20 cm (8.64, 4.8, 2.378, and 1.38 GHz, respectively).  
In 1993, 1994, and 1996 the phase calibrator at all four frequencies was
0336-019; while in 1998 the phase calibrator for all four frequencies was 0323+055.  The primary flux 
calibrator was 1934-638, with a flux density of 2.84 Jy at 3.6 cm, 5.83 Jy at 6 cm, 11.152 Jy at 13 cm, 
and 14.95 Jy at 20 cm.
In 1993, 44 hours of data were collected on September 14.5--17.9.
The 1994 observations were obtained on August 23.7--27.0, and 38 hours of data were collected.
In 1996, ATCA observations spanned September 2.6--7.0 for a total of 63 hours on source.
In 1998, 93 hours of data were collected on September 8.6--12.9 on the source.  
The description of the 1993 and 1994 datasets are given in \citet{jonesetal1996}; the 1996 ATCA dataset
was reduced similarly.  The 1998 data was reduced and calibrated using MIRIAD, and light curve
generation was performed using the $uvfit$ task.
The nominal flux density of 1934-638 was revised in 1994 August, and all observations prior
to this time used a slightly different flux scale, which was changed to provide
better agreement with flux scales used at the VLA.  The changes are $<$ 10\%, and we
apply a multiplicative scaling to the ATCA light curves in 1993 to account for this.

\subsection{{\it IUE} Observations}
In 1993 and 1994, {\it IUE} \citep{boggess1978b,boggess1978a} observed HR~1099 with multiple exposures.
Spectra from 1150--1980 \AA\ and 1900--3100 \AA\ 
were collected (on 1993 September 16.4--19.2, and on
1994 August 23.5--28.7).   
The reduction is described more fully in \citet{brownetal1994} and \citet{dempseyetal1996}.

\subsection{{\it GHRS} Observations}
Observations with the HST GHRS instrument \citep{brandt1994,heap1995} occurred on 1993 September 15.0--19.7.  
They are comprised mostly of G160M and Echelle-B observations of the Ly $\alpha$ (1216 \AA), Mg~II $h$ and $k$
lines (2803 and 2795 \AA, respectively), C~IV (1548 and 1550 \AA), and C~III] 1909 \AA\,
and Si~III] 1892 \AA\ lines.
These data have been reduced and analyzed by \citet{dempseyetal1996};
and further description of the observations can be found in that paper.
We use the time variation of the extracted fluxes in \S~\ref{sec:multi}.

\subsection{{\it EUVE} Observations}
{\it EUVE} observations were obtained on 1993 September 16.4--21.6, 1994 
August 24.0--28.0, 1996 September 1.5--11.1, and 1998 September 3.0--11.7.  Total
exposure times for the four observations were 135, 102, 250, and 226 ks, respectively.  For more
information on the {\it EUVE} satellite, see Malina \& Bowyer (1991).  The Lexan$/$boron filter
on the Deep Survey$/$Spectrometer (DS$/$S) is sensitive over the range 80--150 \AA.  
Initial reduction and processing were performed using the standard {\it EUVE} General Observer
(EGO) software in IRAF\footnote{IRAF is distributed by the National Optical Astronomy
Observatories, which is operated by the Association of Universities for Research in Astronomy,
Inc., under contract to the National Science Foundation (USA).}.  Corrections were made for
detector ``primbsching'' and dead time.
The primbsch algorithm distributes telemetry equally
to operating detectors to maximize scientific gain.  Losses from primbsching
can occur from observations of bright sources when there is a high background
rate in any detector (Miller-Bagwell \& Abbott 1995\footnote{Also online at
http:$//$archive.stsci.edu$/$euve$/$data$\_$product.html.}).  
The 1998 observation 
suffered from extremely high
background due to increased solar radiation as the Sun approached a maximum in
its cycle.   Good time intervals were generated based on filtering times of high
background and large deadtime and primbsch corrections, and the exposure time
for the spectrometer
data were corrected for these effects by dividing by the correction factors.
%Earlier in the {\it EUVE} mission, these correction factors were of the order
%of a few percent or less and were not applied to the light curve data.  For more recent
%observations, however, the corrections are becoming appreciable due to the
%increased solar background radiation and appear in the light curve giving rise to an
%apparent one day periodicity.   Following the light curve generation 
%procedure described in \citet{ostenbrown1999} and \citet{ostenetal2000}, 
%we applied the primbsch and deadtime
%corrections to the 1998 {\it EUVE} light curve by multiplying the number of 
%photons in each light curve bin by the average correction factor appropriate
%for the good time intervals in that bin.  We investigated the magnitude of the
%background, deadtime, and primbsch corrections during the other three
%observations, and found that they are not significant.  The exposure time for
%the spectrometer data for all the observations 
%is corrected, however we did not
%apply the deadtime and primbsch corrections to the light curve data from 1993, 1994, and 1996
%as the
%correction was less than one percent.  
%The resultant light curves are shown in Fig.~\ref{euvelc}.

Spectral extraction of the short wavelength (SW; 80--180 \AA) and medium
wavelength (MW; 180--380 \AA) spectrometer data followed the procedures used in
\citet{ostenetal2000} and references therein.
There was no preset distinction between quiescent and flaring times based on
flux levels.  Rather, the attempt was made to investigate flares on an individual
basis.  Flares can have a range in peak luminosity, as is known from solar and other
stellar studies 
\citep{kucera1987,pallavic1990},
and the ``quiescent'' level can also vary, so a cut-off based on count rate or luminosity
would miss the early and late stages of a flare, or miss a flare with a small
enhancement above quiescence entirely.  Light curves were examined and the beginning
and ending of flares were identified based on a smooth increase to peak and smooth decrease from
peak.  This allows us to examine low-level flares as well as larger events, attempt to
characterize them, and compare and contrast derived properties.  This is more subjective
that using a threshold, but allows for a more nuanced approach to studying coronal variability.
We divided the observations into segments according
to activity state.  
We then summed time segments to provide spectra with decent statistics
while still being able to discriminate between times of varying activity.  %Figure~\ref{euvelc}
%shows the temporal segments used in the spectral extraction.
%Initially I did not include any contribution from continuum emission in
%the spectral fits; as seen later, the predicted continuum fluxes for V478~Lyr and BY~Dra are
 %below the detection limit.  
%A summary of exposure times for the different
%segments and spectrometers is given in Table~\ref{table2}.  

%%%%%%%%%%%%%%%%%%%%%%%%%

In addition to performing the time-resolved spectroscopy, we used the segments from
across the four years to build up spectra that represent a composite of the different activity 
states of the system.  We also used DS$/$S pointed observations of the system in 1992 October
and 1999 September to boost the signal to noise.
We summed all the quiescent spectra 
to obtain a high-quality spectrum of quiescence.  In the same way we summed all the flare
spectra to obtain a high-quality composite spectrum of flaring.  In this way we can determine
gross differences between quiescence and flaring, and compare these to differences that may arise
from flare to flare.  %The exposure times for the composite quiescent and flare spectra are listed in
%Table~\ref{table2}.

\subsection{{\it ASCA} Observations}
{\it Advanced Satellite for Cosmology and Astrophysics (ASCA)} observations of
HR~1099 in 1994 August covered 0.9 days,
with an exposure time of 37.7 ks.  
For more information on
the {\it ASCA} satellite, refer to \citet{tanakaetal1994}.  The SIS
detectors are sensitive over the range 0.6--10 keV, while
the GIS detectors are sensitive to 
%higher 
energies 0.8--12 keV. 
The observations with the SIS
detectors were performed in 1-CCD mode, while the GIS detectors were operated
in PH mode.

The raw data were processed using the HEASOFT 5.0
software package, creating final output light
curves  and spectra.  Standard screening was applied to the data.
Source spectra were extracted using a
circular extraction region around the star with radius 3.9' for SIS0,
3.3' for SIS1, 16.8' for GIS2, and 17'
for GIS3, excluding off-chip areas on the SIS
detectors.  Blank-sky background spectra were obtained from the {\it ASCA}
Guest Observer Facility and were matched to the values of the standard
screening parameters adopted here.  Light curves for all detectors showed the
same features.
%The background in the SIS0 light curves, measured on the same CCD as
%the source, was extremely low, averaging xx counts s$^{-1}$.
%The SIS0 light curve is
%shown in Fig.~\ref{1994ascalc}.   
A small flare occurred during the observation, so we
extracted spectra in time slices corresponding to different levels of activity.

\subsection{{\it RXTE} Observations}
Observations with the {\it Rossi X-ray Timing Explorer (RXTE)} \citep[see][]{bradtetal1990}
were obtained on 1996 September 2.8--6.2
and 1998 September 7.6--11.2.  The exposure times in 1996 and 1998 were
38 and 72 ks.
The Proportional Counter Array (PCA) uses five identical 
Xe$/$methane multi-anode proportional counters to collect photons
over the energy range 2--100 keV with peak effective area $\approx$
7,000 cm$^{2}$ around
10 keV.  Each Proportional Counter Unit (PCU) has six Xenon layers
and one propane layer.  Due to the soft nature of HR~1099
compared to more typical {\it RXTE} X-ray sources, the majority of
photons were 
collected in the outer, most sensitive Xe layer.
The energy resolution is $\sim$18\% at 6 keV for the Xenon
layer.  We examined data from the PCA only; the High Energy X-ray Timing
Experiment (HEXTE), sensitive from 20 -- 250 keV, did not detect the
source.  Processing of the data was performed at the {\it RXTE}
Guest Observer Facility (GOF), using standard FTOOLS (version 4.1)
software.  Only data collected during times of the satellite orbit when
elevation of the target above the Earth's limb was $>$ 10$^{\circ}$ were
used; data taken during South Atlantic Anomaly (SAA) passes (and for 30 minutes after 
passage) were discarded, due to the large increase in background
during such time intervals.  Screening also was performed based on 
contamination from magnetospheric$/$solar flare electrons, which
also cause an increase in the background at low energies.  
The 1996 observations were performed with all five PCUs on; during the 1998
observation, PCUs number 3 and 4 turned off at different times.  
Science data taken 
with different numbers of PCUs must be analyzed separately.  However, for the 1998
observation the amount of time operated in 3 PCU-mode was small compared to the time
operated in 5 PCU-mode, so we restricted our analysis to the 5 PCU-mode observations.
We also extracted light curves in the 2--5 keV and
5--12 keV energy ranges.
Based on a two temperature MEKAL fit in XSPEC \citep{arnaud1996}
to the extracted 5 PCU
spectrum we estimated an energy conversion factor (ECF) for the PCA using the
model folded through the PCA response; the ECF was 
 3.02 10$^{29}$ erg count$^{-1}$ in 1996 and 3.27 10$^{29}$ erg count$^{-1}$ in 1998.  
The spectral sensitivity was not sufficient to
justify extraction of pulse-height spectra for the flaring and quiescent intervals; since
{\it RXTE} is primarily a timing instrument, we concentrate on the
observed light curve.

\subsection{{\it BeppoSAX} Observations}

HR~1099 was observed with {\it BeppoSAX} in 1998 September, from day 6.9 to
10.0 (i.e. slightly more than one orbital period),
using the LECS (0.1--10 keV), MECS (1.6--10 keV) and 
PDS (15--300 keV) detectors. For more information on the {\it BeppoSAX}
satellite refer to \citet{boellaetal1997}. The effective exposure times were 63, 131 and
122 ks in the LECS, MECS and PDS, respectively. The lower exposure times in
the LECS are due to the instrument being operated
only in Earth shadow, thus reducing its observing efficiency.

The LECS and MECS data analysis was based on the linearized and cleaned
event
files obtained from the {\it BeppoSAX} Science Data Center (SDC) on-line
archive. Spectra and light curves were accumulated using the HEASOFT 5.0 
software package, using a circular extraction region
of 8' for the LECS and 4' for the MECS. 
LECS and MECS background spectra accumulated from blank field exposures and
available at the SDC were used for spectral analysis. 
The response matrices released in 2000 January and in 1997 September were used
for the LECS and MECS, respectively.
For the PDS, both the light curve and
the spectrum were retrieved from the SDC on-line archive. The LECS and MECS
light curves show significant variability, with a few small flares. We
therefore extracted separate spectra for the flares and the quiescent
emission. No
significant emission was detected by the PDS instrument: the count rate 
oscillated around zero with a standard deviation of $\sim 0.01$ cts~s$^{-1}$.

In \S~\ref{sec:hr1099sax} we will present the results of the spectral
analysis.
We will use the flux variations from the MECS detector
in \S~\ref{sec:multi} for multi-wavelength comparisons.

\section{Analysis and Results\label{sec:anal}}
The following sections describe in detail the analysis of the large body of data comprising these campaigns.  
One of the main goals of the campaigns was to study flaring, and so it is necessary to describe the means by which
flares are identified. ``Flaring'' can be defined objectively, such as the level where the count rate exceeds 
the average count rate by some factor.  However, solar (and stellar) flares can exist in a range of luminosities
such that small light curve modulations could be evidence of the lower end of the flare luminosity distribution; 
this would hinder our efforts to perform time-resolved spectral analysis of differences between ``flares'' and 
``quiescence''.  
For the purposes of this study, we will use a more subjective way of identifying flares.  Flares are large to
moderate variations
of the count rate above the surrounding count rates that appear to be systematic in time, 
i.e. display a rise, peak, and decay of count rates.

\subsection{Radio Observations}
%\section{Analysis of Radio Observations and Variability \label{sec:radio}}
Radio observations with the VLA and ATCA are a core component of the multi-wavelength
observations.  Data taken with the two arrays offers multi-frequency coverage of
the nonthermal emission from the system during the campaigns and extends the
temporal coverage compared with use of only one of the arrays.  
In these sections we
discuss the observed patterns of radio flux (I) and circular polarization ($\pi_{c}$) in light of 
what we can learn about the emitting properties of HR~1099.
These two 
quantities
can be expressed as \\
\begin{equation}
\pi_{c}(\%) = \frac{V}{I}\times 100 = \frac{RR-LL}{RR+LL} \times 100 \; \; ,
\end{equation}
where V and I are the Stokes parameters for circular polarization and total intensity, respectively,
and RR and LL are the two components of circular polarization: right circularly polarized (RCP)
and left circularly polarized (LCP) emission, respectively.  In radio terminology, X-band refers to 
a wavelength of 3.6 cm (frequency of 8 GHz), C-band refers to a wavelength of 6 cm 
(frequency of 5 GHz), and L-band refers to a wavelength of 20 cm (frequency of 1.4 GHz).

\subsubsection{Light Curve Variability}
\begin{figure}[htbp]
\begin{center}
\figurenum{1a}
\rotatebox{180}{\scalebox{0.8}{
\plotone{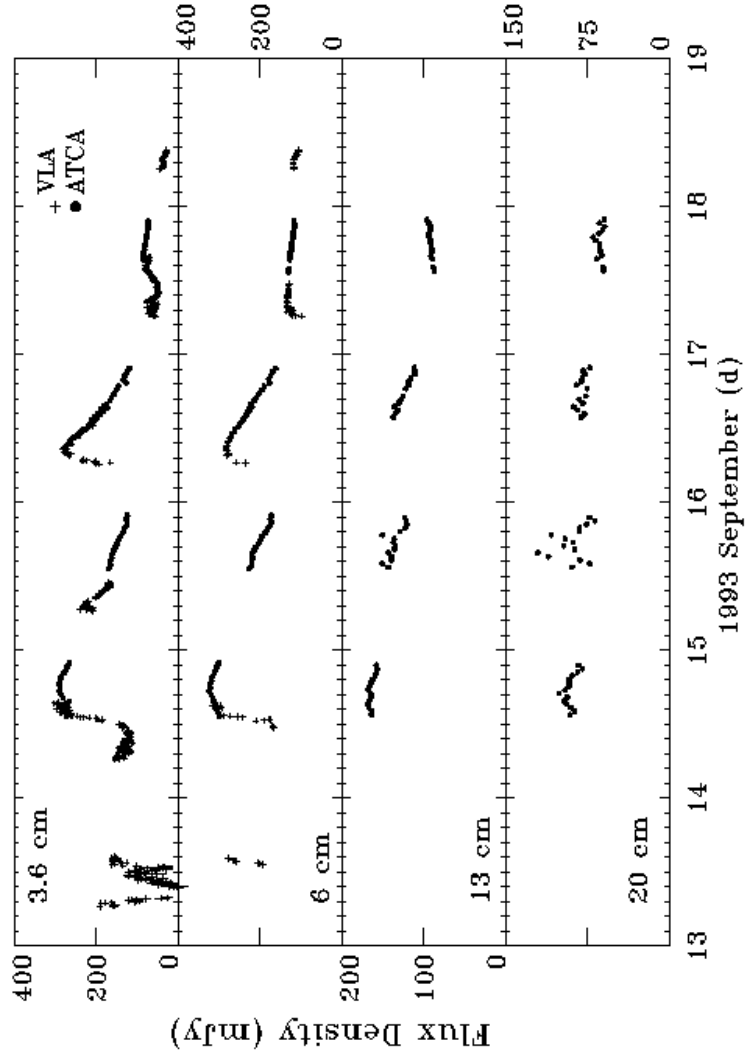}}}
\caption[]{(a)  Flux density variations at 3.6, 6, 13, and 20 cm
during the 1993 observations.  Crosses indicate VLA data; circles denote ATCA observations.  Error bars
are 1 $\sigma$.   
(b) Variations in percent circular polarization at 3.6, 6, and 13 cm during the 1993 observation.
Symbols as in Figure~\ref{1993radioflux}.  Error bars are 1 $\sigma$.
Zero percent circular polarization is shown as a dotted line.
\label{1993radioflux}}
\end{center}
\end{figure}

\begin{figure}[htbp]
\begin{center}
\figurenum{1b}
\rotatebox{180}{\scalebox{0.8}{
\plotone{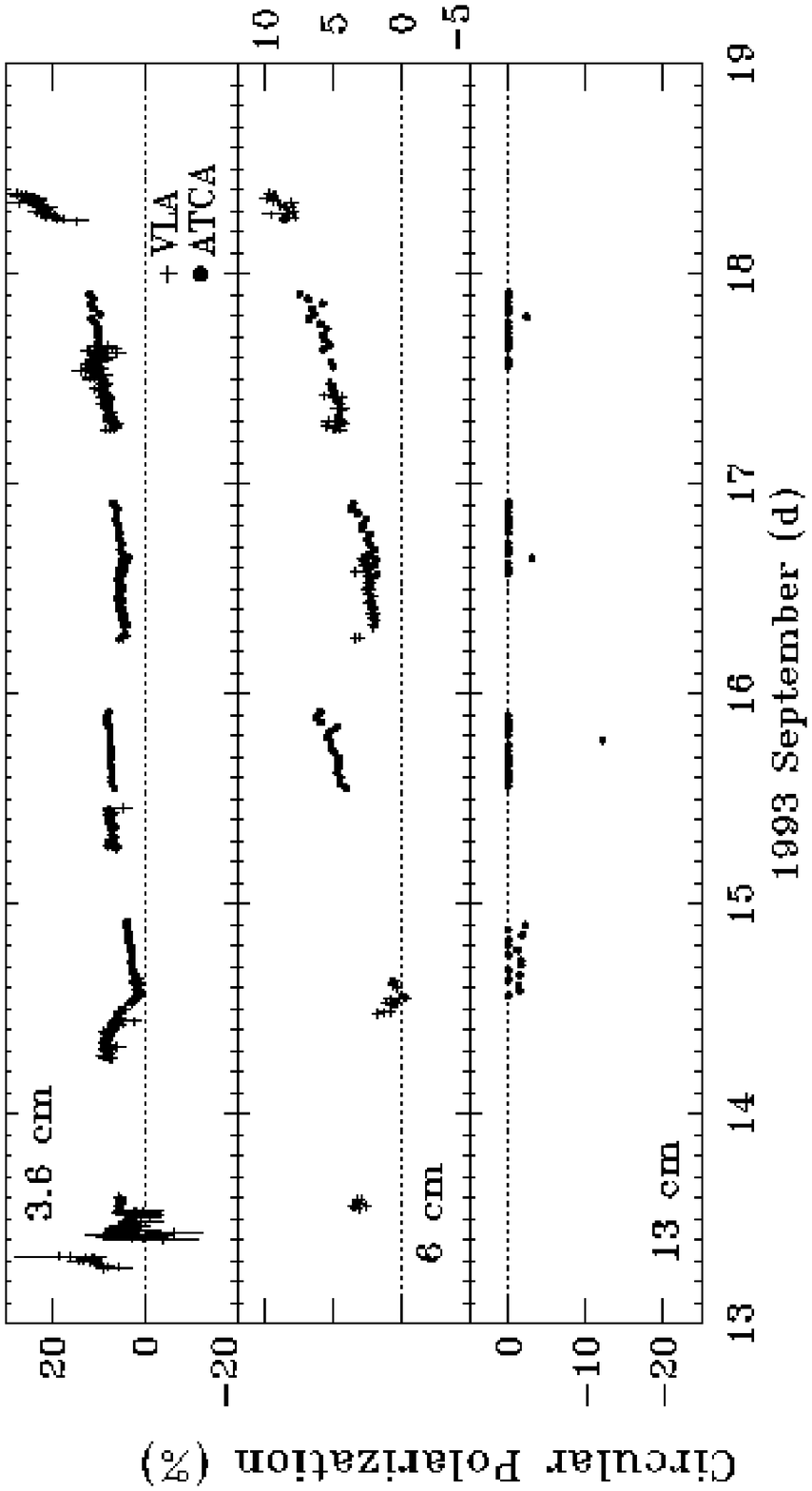}}}
\caption[]{
\label{1993radiopol}}
\end{center}
\end{figure}

\begin{figure}[htbp]
\begin{center}
\figurenum{2a}
\rotatebox{180}{\scalebox{0.8}{
\plotone{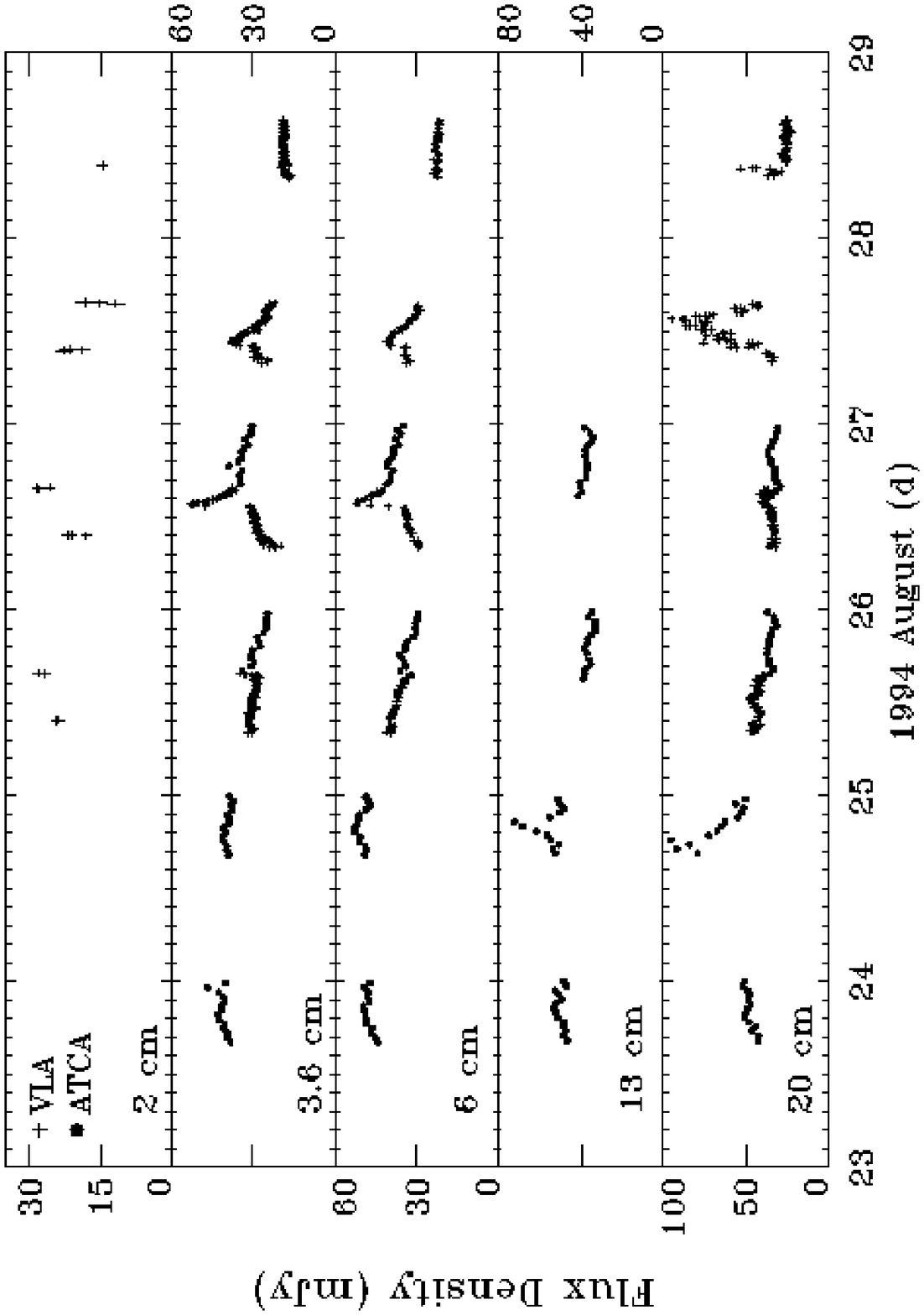}}}
\caption[]{(a)Flux density variations at 2, 3.6, 6, 13, and 20 cm
during the 1994 observations.  Crosses indicated VLA data; circles denote ATCA observations.
Error bars are 1 $\sigma$.  
(b) Variations in percent circular polarization
at 2, 3.6, 6, 13, and 20 cm.  Symbols are as in Figure~\ref{1994radioflux}.
Error bars are 1 $\sigma$.  Zero
percent circular polarization is shown as a dotted line.
\label{1994radioflux}}
\end{center}
\end{figure}

\begin{figure}[htbp]
\begin{center}
\figurenum{2b}
\rotatebox{180}{\scalebox{0.8}{
\plotone{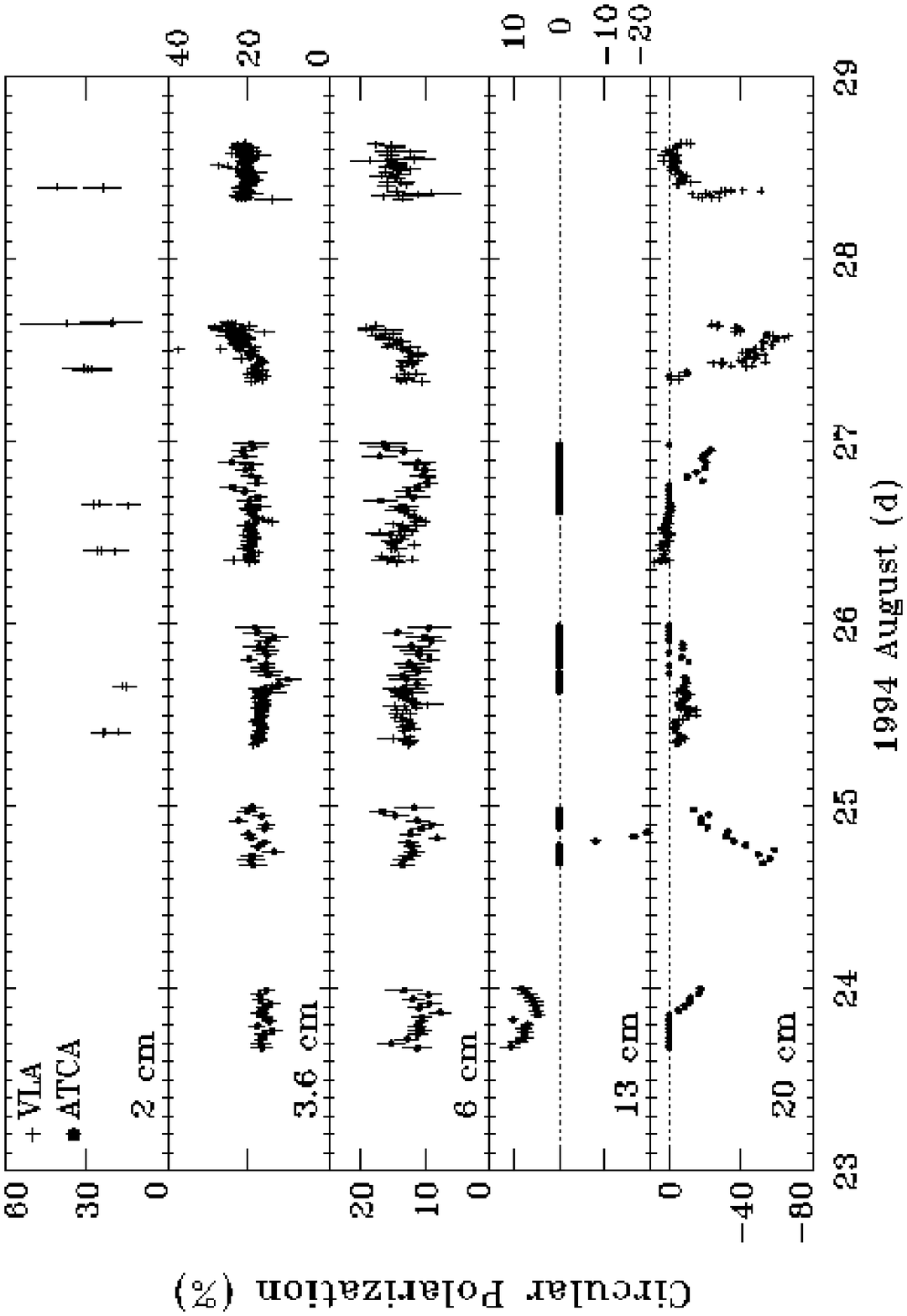}}}
\caption[]{
\label{1994radiopol}}
\end{center}
\end{figure}

\begin{figure}[htbp]
\begin{center}
\figurenum{3a}
\rotatebox{180}{\scalebox{0.8}{
\plotone{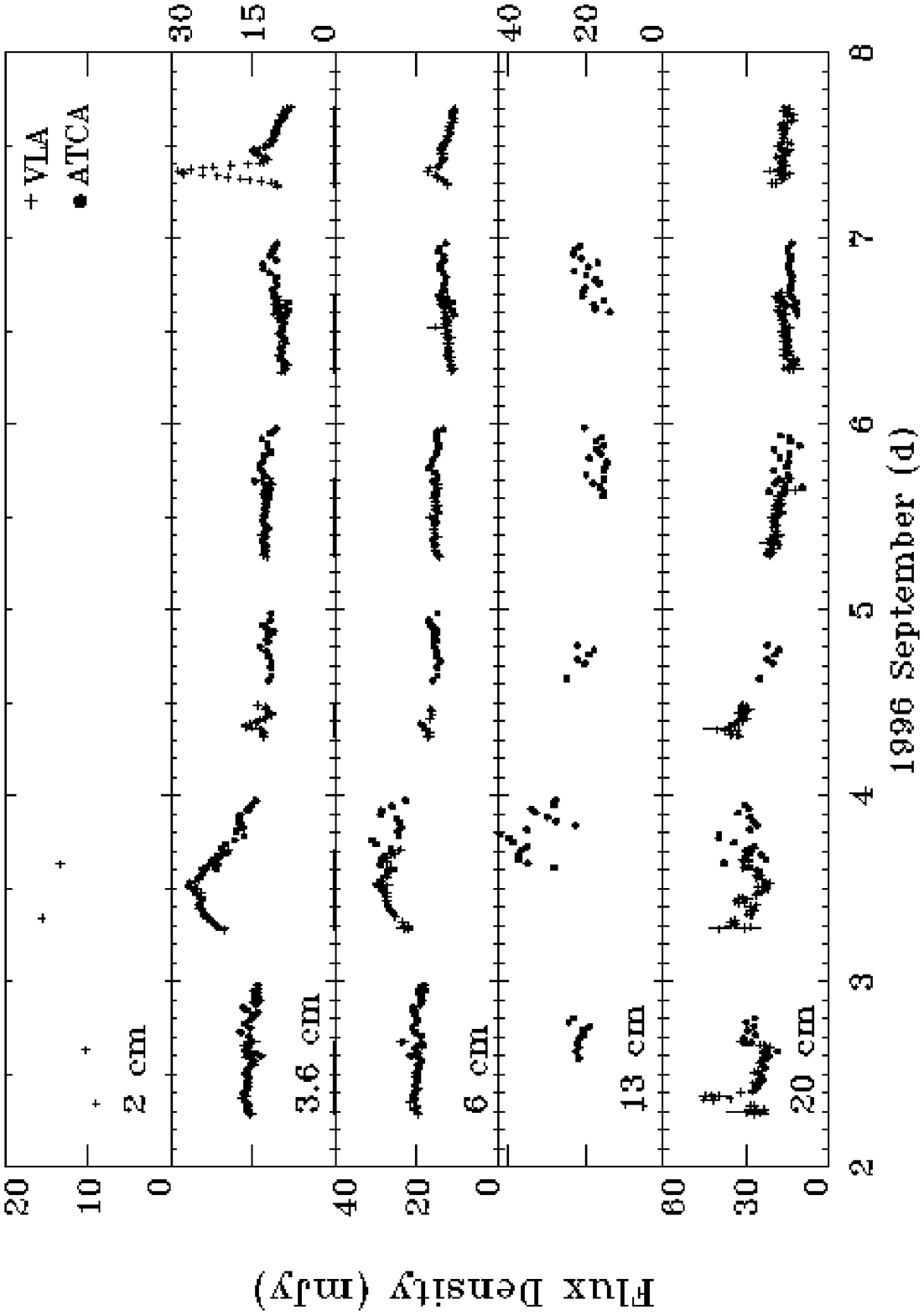}}}
\caption[]{(a) Flux density variations at 2, 3.6, 6, 13, and 20 cm
during the 1996 observations.  Crosses indicate VLA data; circles denote ATCA observations.
Error bars are 1 $\sigma$.  
(b) Variations in percent circular polarization
at 2, 3.6, 6, 13, and 20 cm.  Symbols are as in Figure~\ref{1996radioflux}.  Error bars are 1 $\sigma$.
Zero percent circular polarization is shown as a dotted line.
\label{1996radioflux}}
\end{center}
\end{figure}

\begin{figure}[htbp]
\begin{center}
\figurenum{3b}
\rotatebox{180}{\scalebox{0.8}{
\plotone{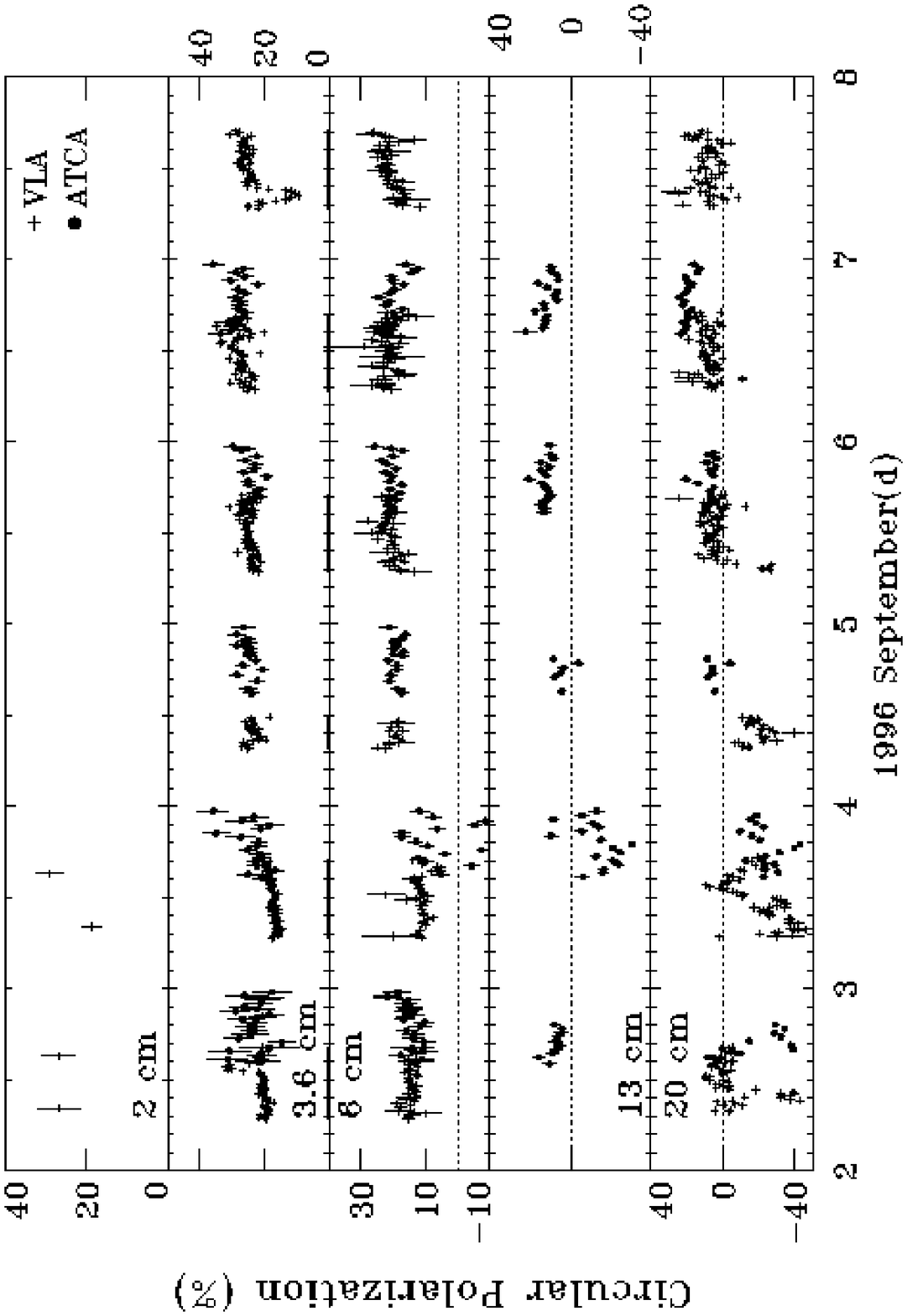}}}
\caption[]{
\label{1996radiopol}}
\end{center}
\end{figure}

\begin{figure}[htbp]
\begin{center}
\figurenum{4a}
\rotatebox{180}{\scalebox{0.8}{
\plotone{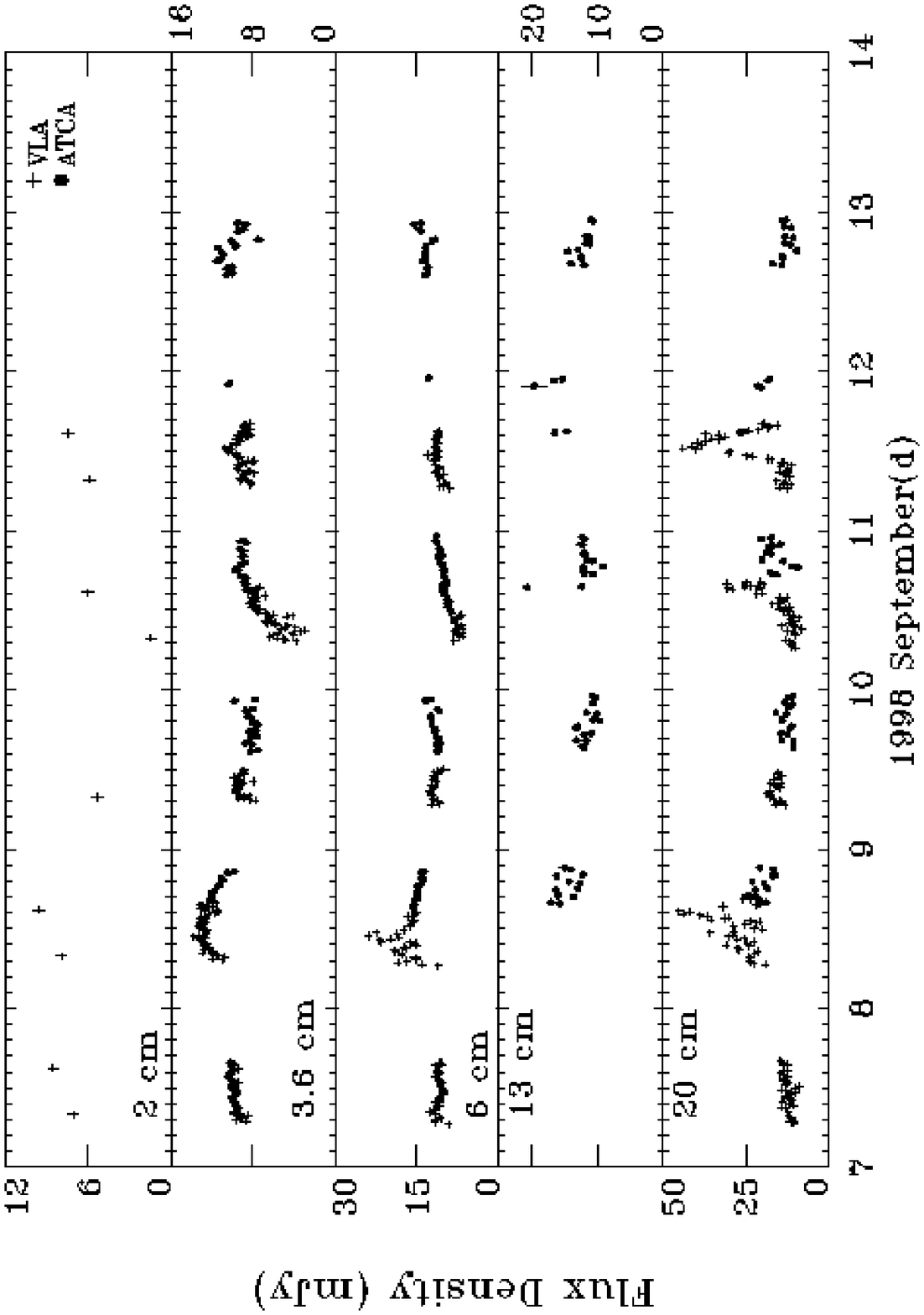}}}
\caption[]{ (a) Flux density variations at 2, 3.6, 6, 13, and 20 cm
during the 1998 observations.  Crosses indicate VLA data; circles denote ATCA observations.
Error bars are 1 $\sigma$.  
(b) Variations in percent circular polarization
at 2, 3.6, 6, 13, and 20 cm.  Symbols are as in Figure~\ref{1998radioflux}.  Error bars are 1 $\sigma$.
Zero percent circular polarization is shown as a dotted line.
\label{1998radioflux}}
\end{center}
\end{figure}

\begin{figure}[htbp]
\begin{center}
\figurenum{4b}
\rotatebox{180}{\scalebox{0.8}{
\plotone{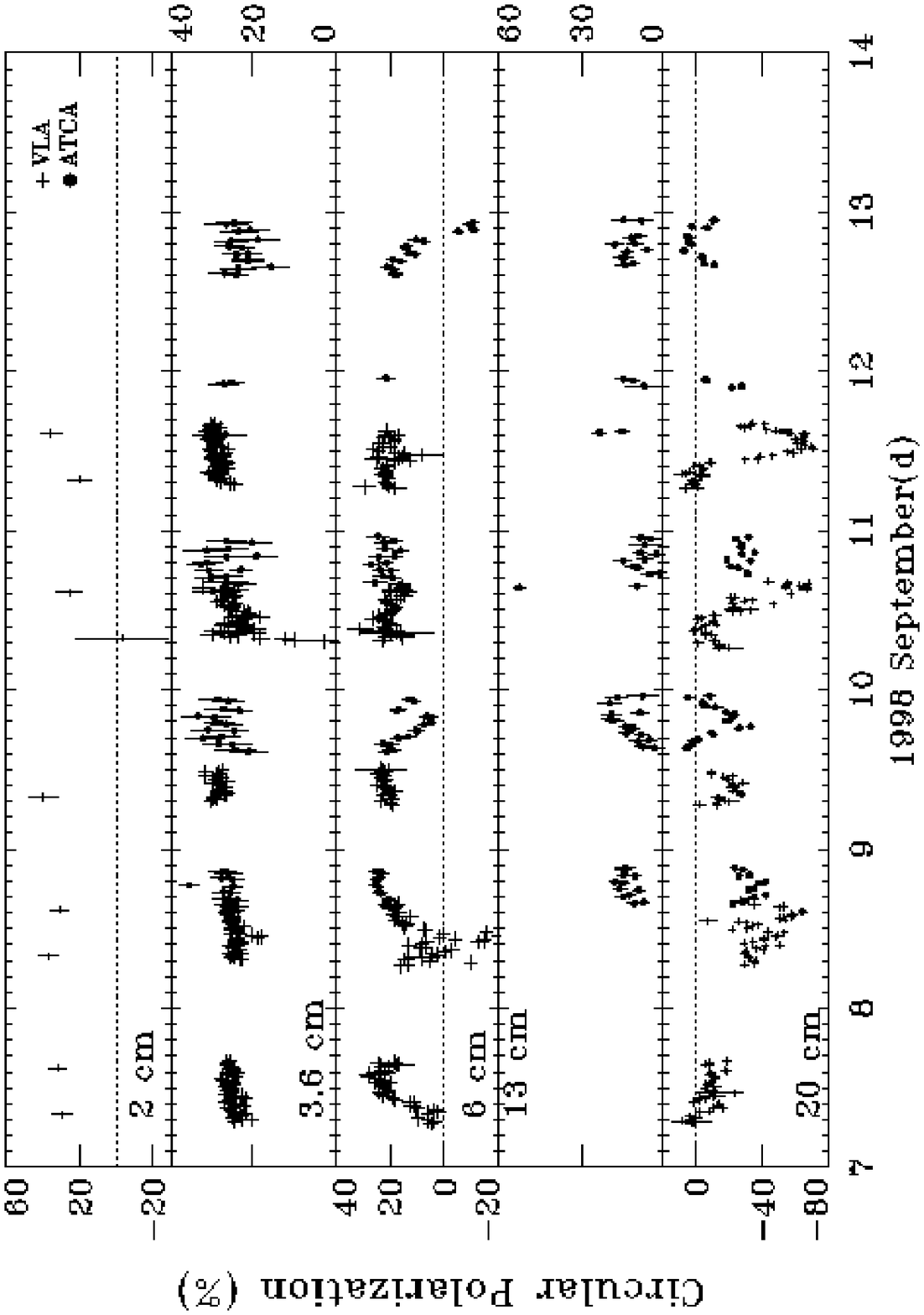}}}
\caption[]{
\label{1998radiopol}}
\end{center}
\end{figure}

Figures~\ref{1993radioflux} through ~\ref{1998radiopol}
show the light curves of flux and polarization variations for the different frequencies
in the four years of observation.  Variations are most obvious at 3.6 and 20 cm, although they
are also evident at other frequencies.  The polarizations of 2, 3.6, and 6 cm 
data are generally positive, i.e., right circularly polarized.  In contrast, the polarizations
at 13 and 20 cm can be RCP, but during some flares the polarization becomes highly negatively (left)
polarized.  We term this behavior, which exhibits large levels of circular polarization and
generally chaotic flux variations with time, as a burst, to distinguish it from flaring behavior seen at higher
radio frequencies.
During two highly polarized events at 13 and 20 cm, the 6 cm emission is also left circularly
polarized.
There is a general trend of declining flux and activity
in later observing campaigns --- the 1993 observations had the largest 3.6 cm flux levels 
and several large flares, but by 1998 the 3.6 cm flux levels are generally much smaller and there
are few to no flares at 3.6 cm.  
These trends can be seen in the average flux and circular polarization values for the four years
at 3.6 cm, listed in Table~\ref{radiotable}. 
The average flux declines and the average polarization increases, from 1993
to 1998; also evident is the decrease in the standard deviation of the average fluxes, indicating
a decreasing amount of flux variability.  
These general trends are consistent with results
reported previously by \citet{muteletal1987} who noted an inverse correlation between flux level and 
percent circular polarization for
RS~CVn systems.  
The decrease in the average flux density 
also implies a decrease in the level of ``quiescence'' during the four years of the
observations.  This is harder to quantify than an average flux density, but an inpsection 
of Figures~\ref{1993radioflux} through ~\ref{1998radioflux} shows that the typical flux densities of times when
no significant variability took place decreased from 1993 -- 1998:  in 1993 the level was 40--50 mJy,
while in 1998 it was of order 8 mJy.
\citet{richardsetal} monitored HR~1099 at 8 and 2.3 GHz using the
Green Bank Interferometer, from 1995--2000, with several observations per day.  While several large
flares (peak fluxes up to 1.4 Jy) were evident on HR~1099 during this interval, none occurred near the times of
our multi-wavelength campaigns in 1996 September and 1998 September, suggesting
that the system was in a relatively inactive state during these times.

%%%%%%%%%%%%%%%%%%%%%%%%%

A closer look at several of the 3.6 cm flares also reveals a 
temporal pattern similar to the broader trend noticed by \citet{muteletal1987}.
Figure~\ref{radioflare}
 show patterns of flux and polarization variations at 3.6 cm
which illustrate this trend.  \citet{muteletal1987}  only used averages from observations, not
a closely spaced temporal sequence.  
There is a hint
in the flare on 1993 September 16 that the circular polarization reaches its minimum value
before the flux level peaks.  This effect described in Figure~\ref{radioflare} for the 3.6 cm data is even clearer in the 20 cm data illustrated in Figure~\ref{radioflare3} .
Inspection of the flares on 1994 August 26 and 1996 September 7
reveals that this is in fact the case:  For the 1996 September 7 flare the flux peaks 
at September 7.361778, and the polarization has its minimum at September 7.35398, $\approx$
11 minutes earlier.  For the 1994 August 26 flare the difference in time between
the peak in the flux and the minimum in the polarization is only $\sim$ 2 minutes, using a data
sampling of 1 minute.

We can investigate the polarization of the flares by assuming a two-component model, composed
of a quiescent source with flux I$_{Q}$ and polarization $\pi_{Q}$, and a flare source with flux
I$_{F}$ and polarization $\pi_{F}$;
the relation between the individual Stokes' parameters and fractional circular polarization
for the quiescent source, flare source, and sum can be expressed as \\
\begin{equation}
I = {\rm total\; flux} = I_{Q} + I_{F} 
\end{equation}
\begin{equation}
V = {\rm total \; circularly\; polarized\; flux} = V_{Q} + V_{F}
\end{equation}
\begin{equation}
\pi = \frac{V}{I}, \; \;\; \pi_{Q}=\frac{V_{Q}}{I_{Q}}, \; \; \; \pi_{F}=\frac{V_{F}}{I_{F}} \; \; \; .
\label{poleqn}
\end{equation}
If the flares are completely unpolarized, then $\pi_{F}=0$. We use values of quiescent flux and
percent circular polarization determined from pre-flare averages, and plot the expected 
variation of total polarization during the flare as the flare flux changes.
This is depicted in the middle panel of Figure~\ref{radioflare} in red crosses and red
circles.  During the flare rise, the observed trends are consistent with
an unpolarized flare emission source.  At later stages in the flare, there is
more observed circular polarization during the flare than is predicted, indicating that a small amount
of flare polarization is present.
Equation~\ref{poleqn} implies 
a small amount of RCP (up to $\approx$ 20\%) during the late stages of the flare.  

We determined estimates of peak luminosities and energies of several radio flares.%, while fitting the light
%curve profiles.  %We assume an exponential rise and decline, of the form: \\
%\begin{eqnarray}
%R(t) & = & R_{0}+ R_{p} e^{-(t-t_{p})/\tau_{r}}\;\;,\;\; t\leq t_{p} \\
     %& = & R_{0} + R_{p} e^{-(t-t_{p})/\tau_{d}}\;\; ,\;\; t\geq t_{p} \nonumber
%\end{eqnarray} 
%where R(t) refers to the flare time evolution, R$_{0}$ is the quiescent flux density, R$_{p}$
%is the peak flux density, $t_{p}$ is the time of peak flux density, and $\tau_{r}$ and $\tau_{d}$
%refer to the exponential rise and decline from maximum flux, respectively.  In a few cases (the flare of 1994
%August 26--27), the flare rise was approximately linear in time, and so we used a linear rise and
%exponential decline to fit this flare.
The flare
energies were determined by integrating the observed flare light curve points, after subtraction of
an estimated constant quiescent flux.  Conversion to cgs units was done using
1 mJy = 10$^{-26}$ erg cm$^{-2}$ s$^{-1}$ Hz$^{-1}$, and the Hipparcos distance to HR~1099 (29 pc).
%The quiescent flux is in general not constant, but varies, and this introduces an uncertainty of up to
%$\sim$30\%(??) into the estimated energies, and $\sim$10\% for the peak luminosities.
Determinations were made for every frequency where the flare was obvious -- typically, X- and C-bands. 
Results are given in Table~\ref{tbl:radfltbl}.  %The flares were typically fit well by 
%exponential rises and decays.
The temporal evolution of the flux density at 3.6 and 6 cm for most of these flares is remarkably similar. 
The monotonic decay of microwave emission during the flare decays is indicative of well-trapped electrons.

Flaring behavior at 20 cm has completely different characteristics than
that displayed at lower wavelengths.
There are flares at 20 cm that display a change in polarization as the flux level
increases, as seen in several 3.6 cm flares.
In contrast with  polarization levels at other
frequencies, the 20 cm polarizations are negative, or left circularly polarized
during these flares.  
Figure~\ref{radioflare3}
shows two examples of 20 cm flares.
%where an increase in flux is met with an 
%increase in the percent LCP, and the percent circular polarization becomes
%less negative as the flux decreases from maximum.
This reversal of sense of polarization, or helicity, was pointed out by
\citet{whitef1995} as evidence of coherent bursts of radiation: an emitting mechanism distinct from
incoherent gyrosynchrotron emission at higher frequencies, which generally keeps the same sense 
of polarization.  Such highly polarized bursts are a relatively common phenomenon on dMe flare
stars \citep{langetal1983,bastianetal1990}, and have been interpreted as either
emission at the plasma frequency or its harmonics, or radiation from an electron cyclotron maser instability,
generally occurring at the cyclotron frequency or its harmonics \citep{stepanov1999,stepanovetal2001,binghametal2001}.

A glance at the polarization light curves (Figures~\ref{1994radiopol},~\ref{1996radiopol},
and ~\ref{1998radiopol}) reveals that for a large amount of the time when HR~1099 was observed at 20 cm,
the emission was highly negatively circularly polarized.  We estimated the fraction of time the system
was exhibiting this behavior, by filtering the polarization data for times when the amount of circular
polarization was less than -20\%.  Roughly 30\% of the time when HR~1099 was observed at 20 cm, the emission
was characterized by moderate to large values of left circular polarization.  The light curves are
averages over typically several minutes, and larger variations on smaller timescales could be masked by 
averaging over large time bins.  Thus this is an estimate, but does point out the apparently common nature
of such events on HR~1099.

%Are the variations seen in these data related to the gyrosynchrotron flares at higher frequencies,
%or could they be related to coherent bursts as have been reported for other active binary 
%systems \citep{whitef1995}?  
We investigated the flare polarizations for these 20 cm flares in the same way as described above
for 3.6 cm flares, by assuming a two-component model of a quiescent source and a flare source contributing
to the total observed flux and polarization.  In contrast to the 3.6 cm flares, 
a large amount of left circular polarization is necessary to reproduce the observed polarization variations
during these flares.  Figure~\ref{radioflare3} shows the expected polarization variations if the 
flare is 100\% LCP in red crosses and circles, which is consistent with the observed polarization levels.  Thus the
total polarization can be described as an equal admixture of quiescent polarization at
a few percent RCP and 100\% left circularly polarized flare emission.  This behavior is consistent with
expected coherent emission, which should be 100\% circularly polarized if it is fundamental plasma emission.

We examined the light curves of flux and polarization changes in Figures~\ref{1993radioflux}
through ~\ref{1998radiopol} to determine if there was a significant degree of correlation between
flares at 20 cm and those at higher frequencies, which are generally attributed to
gyrosynchrotron flares.  We use a relatively lenient restriction,
looking for a close temporal association (within $\sim$ 1 day)
between
3.6 cm flares and either an increase in flux or significant amounts of left circularly
polarized emission at 20 cm.  Table~\ref{tbl:20cm3cmtbl} details the 20 cm 
and 3.6 cm behavior.  There is a low correlation between 3.6 cm flares and 20 cm flux$/$polarization
enhancements; only 3 out of the 12 events show a 3.6 cm flare during the same $\approx$ 17 hours
of each day where radio coverage at both frequencies was obtained.  There are a much larger
number of bursts at 20 cm than at 3.6 cm.  
%This argues against the 20 cm flares
%being caused by the same phenomenon as the 3.6 cm flares:  While there is generally good 
%correlation for the flares and enhancements noted at 3.6 cm and behavior at 6 cm, this
%correlation breaks down at 20 cm.

The ATCA observations at 13 cm also exhibit some of the same behavior as seen at 20 cm:
highly polarized, short duration enhancements.  There are two cases with negative circular polarization
at 6 cm, 1996 September 3--4 and 1998 September 8--9.  The 6 cm data are not simultaneous with the
13/20 cm recordings, but both 6 cm events are accompanied by highly polarized flares at 13 and 20 cm.
Since the ATCA data offer
simultaneous 13 and 20 cm coverage with the same time binning and resolution, we
examined correlations between polarizations at these two frequencies.  The time binning
is rather coarse for investigating what presumably are short duration bursts; the bins
are $\approx$ 15 minutes full width and separated by about 30 minutes.  
There are only two occasions where the 13 cm LCP emission
is accompanied by corresponding LCP 20 cm emission --- 1994 August 24--25 and
1996 September 3--4.  In the former, the 13 cm emission is left circularly polarized
for only $\sim$ 2 hours, whereas the 20 cm data shows LCP for $\sim$ 7 hours.  The
20 cm flux peaks earlier than the 13 cm flux and is declining from maximum when the 13 cm
flux is increasing to its maximum flux.  The 20 cm data reaches its maximum negative
left circular polarization when the 20 cm flux peaks, and likewise with the 13 cm
LCP and flux.  The percent of 20 cm LCP emission is much larger at its peak than the peak of
the 13 cm percent LCP emission by about a factor of three.   This behavior, with a
 temporal difference in peak flux and polarization between 13 and 20 cm, could be attributed to
a frequency drift in coherent emission between these two frequencies, but the frequency difference
($\Delta \nu \sim$ 988 MHz) and temporal difference between flux peaks ($\Delta t \sim$ 8400 seconds)
leads to a frequency drift rate of $\sim$0.12 MHz s$^{-1}$, which is about 3 orders of
magnitude smaller than frequency drift rates seen in solar phenomena \citep[and deduced from 
dMe flare star bursts;][]{gudeletal1989}, and thus the behavior in 
the two frequency bands are probably not related.

The second period of LCP emission at both 13 and 20 cm is 1996 September 3--4.  Here the behavior
is more complicated.  There is a close association in the flux evolution at the two frequencies,
but the variation of the circular polarization shows no apparent correlation.  
The 13 cm and 20 cm circular polarizations are changing rapidly with time in the
half hour between each subsequent scan, suggesting that much finer temporal variations are
taking place.  With such coarse time binning it is difficult to say whether the LCP
behavior at 20 cm is related to the 13 cm behavior.  We note, however, that there are many more
instances of highly left circularly polarized behavior at 20 cm, which implies
that it is a separate phenomenon from the LCP emission at 13 cm, although we cannot rule out
a causal relationship.  There is also 6 cm emission that is left circularly polarized during the general time 
of 13/20 cm bursts --- it is consistent with quiescent emission just before and after the burst, and
a 50--80\% left circularly polarized burst component.

For time periods when the 20 cm light curves show highly LCP emission, we investigated
trends in LL and RR flux with time.  We found in most cases that temporal resolution at
the 10 second level, the time sampling of the VLA data, was
necessary to determine the detailed flux variations; sampling on even shorter
timescales might be necessary, although unobtainable with this dataset.  
Figure~\ref{coherent} illustrates several cases of both long and short duration,
highly polarized bursts.
There are rises and decays which last less than one minute, along
with a complex morphology that might indicate a long series of 
multiple overlapping bursts, each occurring on timescales less than 10 seconds.  For all these
bursts, the LCP flux dominates over the RCP flux, indicating negatively circularly polarized emission.
In contrast to the LCP flux, which shows great variations, the RCP flux is remarkably constant.
We also investigated the difference in temporal evolution of the flux at the two intermediate frequency (IF) bands
used in continuum mode observing at 20 cm: 1440--1490 and 1360--1410 MHz in 1994 August.  
The bottom two panels of Figure~\ref{coherent} 
detail two different kinds of behavior.  In the first case, both IF bands show similar evolution;
and in the second, the flux patterns change with time and frequency.  This latter case is different
from what was observed by \citet{whitef1995} where the LCP flux in the two sidebands was equal and varying in the
same way.

\subsubsection{Flux and Polarization Spectra}

We investigated the trends of flux and polarization with frequency for the multi-frequency
radio datasets.  While the frequency coverage is not strictly simultaneous due to 
frequency switching
or combining coverage with the VLA and ATCA, we 
can obtain contemporaneous multi-frequency coverage of up to five frequencies
spanning an order of magnitude in frequency to within about a half an hour.
During periods
where there is no obvious flaring at any wavelengths, such time intervals should be small enough
to record the general characteristics of the quiescent radio emission.  Problems arise, however,
when variability is present at one or more frequencies, and then we can only diagnose some
``average'' conditions which may or may not represent the actual conditions.

We concentrate on time intervals when 2 cm observations occurred, to extend the frequency coverage,
although for the 1993 observations no recordings at 2 cm were obtained.
We extracted flux and polarization information from the light curve data for each frequency that
was close enough in time to provide adequate temporal and frequency coverage, generally within
about 30 minutes of the 2 cm observation.  Spectral indices were computed for each
successive frequency pair; the spectral index $\alpha$ (where S$_{\nu} \propto \nu^{\alpha}$)
is computed using a least-squares
linear fit to the logarithmic flux and frequency data, and are listed in Table~\ref{tbl:specindex}.
We have sorted the time intervals according to the general light curve behavior the system
was exhibiting:  either quiescent behavior or the  peak$/$decay of a 3.6 cm flare. 
Figure~\ref{fig:radiospec} displays the flux and
polarization spectra typical for these activity levels.

For all the intervals where flux and polarization spectra could be determined, the
polarization shows an increasing trend with frequency, regardless of the 
behavior of the flux spectra.
The characteristics of the flux spectra
are consistent with previously determined behavior on active binary systems.  
The quiescent data generally show a decline in 
flux with increasing frequency. 
The indices for
quiescent observations are flat or slightly negative, $0 \leq \alpha \leq -1$,
with an exception being 1998 Sept. 10.3, which has a 3.6--2 cm spectral index of $-2.32\pm0.39$.
This declining trend of flux with frequency suggests that the emission is optically thin.
The increase in observed
circular polarization with frequency is in contrast to the expected behavior of optically thin
nonthermal gyrosynchrotron radiation from a homogeneous source; however, it is difficult to obtain
large values of percent circular polarization under optically thick conditions \citep{jones1994}.
\citet{whitef1995} have found similar evidence for a rising polarization spectrum with 
frequency being independent of the flux spectrum shape for a sample of active binary systems.

Several flares at 3.6 cm show corresponding intensity variations at 6 cm, and we
investigated these flares for changes in the spectral index $\alpha_{6-3.6}$ during the
course of the flare.  Figure~\ref{radioflare} shows the results for three moderate-sized flares.
In general, there is a positive, linear correlation between the flux and spectral index.
This is in accord with the general trends
noted by \citet{muteletal1987} between the spectral index and luminosity for a sample of active binary 
systems.  
During the rise to peak intensity, the spectral index flattens, becoming less negative, even changing sign and
becoming increasingly positive for two cases (1994 August 26.5 and 1996 September 7.35).  
This may signal a change to optically thick emission at the flare peak, and would explain
the apparent zero polarization level of the flare, in combination with a positive amount of 
quiescent circular polarization. 
The shift from optically thin to optically thick
conditions at the flare peak, and return afterwards, implies that the peak frequency
$\nu_{peak}$ (where $\tau=1$) is $>$ 8.3 GHz during the flare rise and peak, and $<$ 5 GHz during the flare decay.
%The peak frequency (where $\tau=1$) depends most strongly on the total number density of
%nonthermal electrons; this change can be explained qualitatively by  
%an increase in 
%number density of nonthermal electrons during the
%flare, compared with quiescent conditions.

\subsection{Analysis of UV Observations }

\citet{dempseyetal1996} discuss the analysis of the 1993 {\it IUE} observations
and 1993 {\it HST/GHRS} observations. 
\citet{brownetal1994} discuss the analysis of the 1994 {\it IUE} observations.
We will
utilize the temporal variations of the extracted line fluxes of C~IV and Mg~II in our multi-wavelength comparisons in
\S~\ref{sec:multi}.

\subsection{Analysis of {\it EUVE} Observations \label{moreeuv}}

\subsubsection{EUV Light Curve Variability \label{moreeuvlc}}
Figure~\ref{euvelc} shows the light curves in the EUV during the four observations which comprise
this campaign.
During the first three observations of HR~1099 the system was in an extremely active state.
By comparison, during the fourth observation, in 1998, HR~1099 was relatively inactive,
signalling an abrupt change in activity level.
%We estimated the percent of elapsed
%time the system was flaring during the four observations.
%In 1993, for 91\% of the duration of the
%observation the system was flaring.  Similarly active in 1994, 93\% of that
%observation's extent was made up of flares.  During the long observation in 1996, 77\% of the time
%the system was flaring.   Compared to these large flaring rates, the 1998 observation
%signals an abrupt change in activity level --- only one flare was observed, and it represented 
%only 8\% of the elapsed time of the observation.  The exact flaring rate is a subjective
%measurement, dependent as it is on the vagaries of interpretation.  These numbers represent an attempt
%to quantify what the light curves in Figure~\ref{euvelc} illustrate --- active states in 1993, 1994, and 1996,
%and an inactive state in 1998.

We also attempted to quantify and compare the activity of the system by examining the 
average non-flaring count rate during the four observations.  Determining the actual quiescent
level of the system is somewhat difficult, due to the obvious influence of large flares
in the light curve.  The quiescent, or non-flaring level, might also be affected by rotational
modulation of starspots or numerous small-scale flares which occur too close together in time to
resolve with {\it EUVE}'s poor orbital efficiency.  The actual quiescent level may be below
the average count rate outside of flares in these observations.  Nevertheless, we can use these
values as upper limits to the actual level of non-flaring emission.  In 1993, the average
non-flaring luminosity was 9.6$\pm$1.2$\times$10$^{29}$ erg s$^{-1}$; in 1994 it was 5.4$\pm$0.5
$\times$10$^{29}$ ergs s$^{-1}$; in 1996 it was 6.6$\pm$1.0 $\times$10$^{29}$ ergs s$^{-1}$; and in 1998
it was 6.6$\pm$0.9 $\times$10$^{29}$ ergs s$^{-1}$.  The average level in 1993 is 50--100\% higher 
compared with the average levels in 1994, 1996, and 1998.  
The significant variability the system was undergoing during most of the 
observations hampers our efforts to look for evidence of rotational modulation or orbital
phase-dependent effects.  
Also, the observations were spaced far enough apart (generally $\sim$
one or two years) that changes in the ``quiescent'' level of the system,
could be due to evolution in starspot configuration, if these changes are attributable to 
phase-dependent effects from starspots.
%An important caveat in the search for
%rotational modulation is the assumption that outside of large flares the system is in 
%a ``quiescent'' state and any variations can be interpreted as orbital phase-dependent effects
%instead of flaring variability.  Recent studies by \citet{kashyapetal2000} and \citet{audardetal2000}
%have investigated the effect of small-scale flaring on observed EUV light curves of
%main sequence and active stars, and find that it is possible to reproduce {\it EUVE} light curves
%as an assumed constant quiescent state superimposed with numerous small flares.  
%Given the very active nature of the HR~1099 system,
%we cannot make any definite conclusions about time intervals not affected by large, discrete flares
%as being ``quiescent'', and thus can make no statements about rotational modulation.

The top panel of Figure~\ref{cumlum_euve} plots the cumulative luminosity distribution
of events greater than a given L$_{X}$ for all six DS$/$S pointed observations, from 1992, 1993,
1994, 1996, 1998, and 1999, along with the distribution resulting from all the observations taken
together.  For this comparison, the binning of each light curve was fixed, at $\approx$ 3000 seconds.  The
data from the 1992 and 1999 observations, reported in \citet{drakeetal1994} and \citet{ayres2001}, were 
reduced in a similar manner as the data obtained during our 1993, 1994, 1996, and 1998 campaigns.
The behavior of this cumulative distribution varies greatly; only in
1993 and 1996 were there events with luminosities greater than 2 $\times$10$^{30}$ ergs s$^{-1}$.
The flattening of each distribution at low luminosity probably represents the quiescent emission
level, which appears to vary from year to year.  The roll-over at high luminosities
represents the very largest flares, which happen less frequently.  The bottom panel of
Figure~\ref{cumlum_euve} shows the cumulative distribution of the sum of the pointed 
observations, expressed as a percent probability distribution.  Note that even with the 
extremes of variability shown in Figure~\ref{euvelc} the range of luminosities is still
less than a factor of seven between the lowest and highest observed luminosities, the bulk of
the observations occuring over a smaller range of $\approx$ 4 in luminosity.  High 
luminosity events greater than 2 $\times$10$^{30}$ ergs s$^{-1}$ actually comprise about 10\%
of the total number of events in the more than one megasecond of observations from 1992 to 1999.

Figure~\ref{euve_flares1} shows the details of the four largest flares seen with {\it EUVE}.
These flares occurred over the course
of many days, in some cases having durations that equal or exceed the stellar orbital period.
Long duration high-energy flares on RS~CVn binary systems are common 
\citep{graffagninoetal1995,ottmannschmitt1994,kursterschmitt1996}.
Such long timescales require sustained heating; for a $\sim 10^{7}$K plasma with electron
density n$_{e} \sim$ 10$^{8}$--10$^{13}$ cm$^{-3}$, the radiative decay time is
$\sim$ 20--2 $\times$10$^{4}$ s. 
The lack of self-eclipses implies either large sizes or an appropriate location with respect
to the stellar axis and the orbital plane.
%Such long timescales imply that large volumes comparable in size to the orbital separation must be
%involved in the flares, and necessitate an extended emitting region, so the flare
%is not eclipsed by either star due to stellar rotation or orbital motion.  

The flare rises and decays show a remarkable smoothness
in the evolution of the count rate.  In past analyses of EUV light curves 
\citep{ostenbrown1999,ostenetal2000}
we assumed that most EUV and X-ray light curves could be expressed as
a single exponential rise and one or two exponential decays:  
a result of exponential changes of density and$/$or temperature
during the flare.
For the two large flares observed in 1993 September,
this pattern of multiple decay phases appears to hold -- both show two distinct
decay phases, an initial fast decay followed by a second slower decay.
The fast exponential rise and exponential decays plotted in Figure~\ref{euve_flares1}
are taken from \citet{ostenbrown1999}.
Further probing of EUV flare light curves reveals an
increasingly complex flare morphology:  Several of the large flares 
appear to have an inflection in the flare rise with a transition from a slow rise
to fast rise, similar to the two decay phases, as shown in Figure~\ref{euve_flares1}.  
The two flares that occurred
during the 1993 September observation typify this behavior. 
%A long, slow increase in luminosity is
%evident prior to the fast rise of the 1994 August 27 flare, and might also 
%be present before
%the double flare that occurred in 1996 September.  

We examined the significance of these slow rises by comparing how well the data were fit by
an exponential rise and a constant, quiescent rate, using the reduced chi-squared ($\chi^{2}_{\nu}$) statistic.
The exponential fits are shown in Figure~\ref{euve_flares1}.
The constant value was determined by averaging count rates in other temporal regions which
did not seem to be affected by flares.
The slow rise yielded lower $\chi^{2}_{\nu}$ values: For the 1993 September 16 flare, 
1.7 ($\nu=$7) compared with 7.6 ($\nu=$8) for a constant rate; for the 1994 August flare, 1.3 ($\nu=$39) against
25.1 ($\nu=$40) for a straight line; and for the 1996 September large flare, 1.2 ($\nu=$57) versus 
4.7 ($\nu=$58).  Figure~\ref{euve_flares1} illustrates the exponential fits, and delineates
the average count rate for each observation, along with the standard deviation 
of the average count rate.  There is a large amount of variability  present,
even during seemingly quiescent conditions. The increase in luminosity before the
impulsive rise is reminiscent of a flare precursor (destabilization of a filament prior to the main event), 
corresponding to an energy leakage
before the main energy release during the flare, as has been discussed for solar
flares \citep[see discussion in section 6.5 of][]{physicsflares}. 
In the Sun,
pre-flare heating and brightening is due to slow
rearrangement of the magnetic field and subsequent slow energy release before the
impulsive energy release that marks the flare, although not all solar
flares exhibit this behavior.  The slow energy release may reveal itself
as a gradual increase in the plasma radiation.

We compared the similarity of flare light curves by shifting
the flares in time and flux so that the peaks coincide.
The top panel of Figure~\ref{composite_flare} shows the
two flares in 1993 and the large flare at the end of 1994 
superimposed on each other in this way.  There is a striking similarity in the rise phases
of all three flares.  The inflection point in the three flares also appears to occur at roughly
the same time before the peak of the flares.  
The decays of the two
flares in 1993 are remarkably similar; the comparison with the 1994 flare breaks down, however,
as it appears to plateau for $\sim$ 5 hours before undergoing a single decay.
A small flare can be seen peaking on 1994 August 25.8 during the long rise of the 1994
flare shown in Figure~\ref{euve_flares1}.

The middle panel of Figure~\ref{composite_flare} compares the two flares that comprise
the double flare seen in 1996 September.  Here the similarity in the two flares
is even more striking.  Plotted in black is the rise and partial decay of the first flare,
with the partial rise and decay of the second flare in red shifted on top of it.  While
we don't have the complete decay of the first flare nor the complete rise of the second flare,
the parts that do overlap show a surprising amount of symmetry.  We explored this further by
reflecting the second flare about the peak, to compare the rise of the first flare with the decay of the second.  This reflection is shown in the middle panel of Figure~\ref{composite_flare} in green.
Both flares appear to attain a constant value for $\sim$ 8 hours before decaying.
\citet{sf2002} have examined the {\it EUVE} datasets presented here.  They accumulate quiescent and
flare intervals to obtain high S$/$N spectra of quiescence and flaring, and investigate
properties of the emission measure distribution.  They suggest that the double flare
seen in 1996 September is actually one flare that is self-eclipsed.  If this is the case, then
the resulting single flare is of enormous proportions,  lasting $\approx$ 3.5 days after
a possible flare precursor, or 1.2 rotational
and orbital periods.  The symmetry of the rise of the first feature and decay of the second
would then be in line with symmetries found in morphologies of other flares.

We examined the smaller luminosity flares in the same way to look for patterns of similarity.
Most of these six small flares have peaks less than about two times the
quiescent luminosity.  For the most part these flares appear to be symmetrical, with
a long rise and decay.  
Two flares, however, 
show a different morphology which is more akin to the classical picture of stellar flares.
The flare which peaked on 1994 August 24.2 and that which peaked on 1996 September 5.3 
are asymmetrical in the sense that the rise happens very quickly but the decay is long.
%Figure~\ref{composite_flare2} shows these flares shifted and scaled in the same way as the large
%flares.  The four flares shown in the top panel of Figure~\ref{composite_flare2}
%appear to be symmetric about the peak.  The flare in 1998 September appears to be symmetric
%as well but with a more pronounced enhancement between the peak luminosity and quiescence 
%($\approx$ factor of three).  
The two asymmetric flares are remarkably similar as well, when 
shifted and scaled (see Figure~\ref{composite_flare}, bottom panel).

%The identification of slow rises directly preceding large flares in the 1993, 1994, and 1996
%{\it EUVE} light curves, compounded with observations of several smaller enhancement flares
%in these time intervals, results in flares or flare activation taking place almost continually ---
%for many of these flares, there is almost no time between the decay
%of one flare and the rise or activation of another.
%\citet{ostenbrown1999} interpreted such behavior as evidence for sympathetic flares, a phenomenon
%known to occur in the Sun, where a flare in one active region stimulates flaring in distant
%active region by propagation of a wave.  The Extreme Ultraviolet Imaging Telescope (EIT)
%on SOHO and
%the Yohkoh Soft X-ray Telescope have observed coronal 
%disturbances propagating away from the site of a solar flare, 
%and cospatial with 
%the chromospheric Moreton waves \citep{eitmoreton,xraymoreton}.
%In addition, 
%TRACE movies of EUV solar flare behavior 
%also reveal the propagation of a wave through the outer atmosphere and show magnetic loops
%connecting distant active regions.  

The general characteristics of most flares seen here show marked departures from 
the ``canonical'' flare shape --- a fast rise followed by a slow decay.  For many 
of these EUV flares, the evolution of the flare rise and decay is at the same rate.
Only two flares appear to show a fast rise and slow decay.
In the solar scenario, the energy release, particle acceleration
and heating of lower atmosphere plasma to coronal temperatures and subsequent ablation
into coronal loops happens quickly (apart from a possible pre-flare cursor; see above);
this is the impulsive phase.  Perhaps the symmetry between rise and decay
in flares seen on active binary systems reflects the nature of the energy release, which 
happens on longer timescales (gradual instead of explosive evaporation)
or in a larger arcade of loops.  If multiple loops
were involved in producing the flare we see in the light curve, then 
the slow rise could be interpreted as progressive brightenings as subsequent coronal loops
become filled with coronal plasma and radiate.  At what we identify as the peak of 
the flare, the maximum number of loops is filled, or the largest loops in a size distribution
of loops are filled with coronal plasma, the radiation reaches a peak and the plasma cools
via radiation and conduction \citep[and possibly continued heating; see ][]{favata2001}, 
leading to a decline in flux.

\subsubsection{EUV Spectra \label{sec:ch5euvespec}}
We investigated spectral variations in the {\it EUVE} data, dividing the data into temporal segments
corresponding to the general activity in the system.  As described in \S 3.5, the goal of this
paper is to investigate flares on a case-by-case basis.  Based on our analysis of the light
curves in \S 4.3.1, we identified temporal intervals corresponding to quiescence and flares, 
and marked those for subsequent spectral extraction.  
Figure~\ref{euvelc}
shows the light curves and corresponding time intervals in which spectra were accumulated;
`Q' refers to quiescent times, `F' to flare times (often several small-amplitude
flares), and `F1', `F2', etc. to 
discrete large-amplitude flares.   
Table~\ref{table2} lists the exposure time in the SW and MW for the time intervals
indicated in Figure~\ref{euvelc}.
Figures~\ref{euveswspec} and
~\ref{euvemwspec} display the SW and MW spectra for these time intervals, with 
selected emission lines identified.  %Figure~\ref{compositespec} shows spectra from all quiescent
%and flaring intervals accumulated into composite quiescent and composite flare spectra, respectively.
The spectra are dominated by lines of highly ionized
iron --- ionization stages XV--XXIV.  In spite of the
dramatic changes illustrated by the light curves, the corresponding spectral changes
are less overwhelming.  The most noticeable increases that occur during flare segments
compared to quiescent segments are enhancements in the flux from the Fe~XXIII$/$XX $\lambda$ 132.85 
and Fe~XXIV $\lambda \lambda$ 192.02, 255.09.  Other lines appear to be at the same flux level.   
Due to the low sensitivity of the {\it EUVE} spectrometers,
the flare spectra
have not had any estimate of quiescent flux subtracted, and thus represent a combination of 
flare and quiescent plasma.

The assumptions we make in analyzing the spectra are that the emission is from a collisionally
ionized plasma, in equilibrium, and effectively thin.  Under such circumstances,
the plasma is in a steady state between collisional ionization and radiative
and dielectronic recombination;
line formation is a balance between collisional excitation and radiative de-excitation.
The effectively thin assumption implies that all the emitted photons escape; some (typically $1/2$) are
directed outward toward the observer and some are directed inward, possibly to the stellar surface 
(depending on the geometry).
The observed flux can be written as a product of the elemental abundance times an 
integral involving the plasma emissivity and the differential emission 
measure: \\
\begin{equation}
f_{\rm obs} \propto \frac{A}{4 \pi d^2} \int P_{\lambda}(T) n_{e}n_{H} \frac{dV}{d\log T} d\log T
\end{equation}
where A is the elemental abundance, P$_{\lambda}$ is the plasma emissivity.
The quantity \\$n_{e}n_{H} \frac{dV}{d\log T} \equiv \phi(T)$ is the differential emission measure (DEM),
or the emission measure (EM) differential in temperature, \\
\begin{equation}
EM =n_{e}n_{H} dV = \phi(T) d\log T \;\;.
\end{equation}
Bright lines of iron dominate the spectrum and span a relatively
wide temperature range (from Fe~XV at 2 MK to Fe~XXIV at 20 MK); for these reasons we use
these lines to constrain the shape of the coronal differential emission measure distribution.
Since the DEM thus determined is relative to the coronal abundance of iron, this
eliminates the necessity of determining the iron abundance
simultaneously with the DEM.
Previous studies that have attempted to constrain both these quantities
have found that generally two classes of results match the data statistically:
A DEM with a peak between 1 and 30 MK with subsolar iron abundances, or a solar
abundance plasma with a marked peak at 100 MK \citep{schrijveretal1995}.
Restricting the DEM determination by only using iron lines removes some of this
ambiguity.

We measured emission line fluxes of lines in the SW and MW
spectra by generating a linelist from the plasma emissivities taken from \citet{apedref} and assuming
a Gaussian line profile.  These emissivities
have been calculated in the low density limit, and agree with those
of \cite{brs} in their low density limit.  The full width at half maximum of the line
profiles were fixed at the instrumental resolution: 0.38 \AA\ for SW and 1.14 \AA\ for MW.
We corrected for the effects of interstellar absorption in the SW and
MW spectra by using photoabsorption cross section calculations of
\citet{morrisonmccammon1983} and a hydrogen column density of 
1.35 10$^{18}$ cm$^{-2}$ as derived by \citet{piskunovetal1997}.

The analysis of {\it EUVE} spectral data is complicated by the presence of continuum emission.
Although present only at a low flux level, the continuum flux rises to shorter wavelengths,
and can introduce errors in determining the fluxes of emission lines.  Pinning down the
level of continuum emission requires having line-free regions of the spectrum.
The EUV spectral region is littered with numerous weak
emission lines in the atomic codes, and the incompleteness of the codes 
in this region is a recognized problem \citep{jordan1996,beiersdorfer1999}.
%A priori knowledge of the continuum can arise if one knows what the
%coronal abundances are, and can predict the level of continuum emission expected.
%This technique requires analysis of data in another spectral region,
%such as X-ray observations at sufficient spectral resolution to discriminate line from continuum emission.
%The connection of non-simultaneous observations in two disparate spectral regions 
%requires an assumption that the abundance doesn't change, which contradicts
%some findings of variable coronal metallicity during flares \citep{gudeletal1999}.
%The inability of the {\it EUVE} spectral data to assess adequately the continuum flux
%level (and hence elemental abundance) leaves a quandary. 
%Both continuum and emission line fluxes are calculated deterministically
%from the differential emission measure.
%Knowing the continuum level is
%essential to measuring emission line fluxes free from its contaminating influence.
An added complication is the formation of the continuum flux at 
temperatures higher than those diagnosed with the emission lines present in the {\it EUVE} spectrum.
Previous studies of stellar coronae in X-rays have revealed coronal elemental abundances
that are less than solar \citep[e.g.,][]{singhwhitedrake1996} and appear to vary during flares
\citep[e.g.,][]{gudeletal1999}.  This leads to an
increased amount of continuum emission, because the DEM here is computed relative to the
iron abundance.  Decreasing the iron abundance necessitates raising the DEM by the same amount
so that the predicted line flux does not change: A larger DEM consequently produces
more continuum flux.  

We investigated the effect of sub-solar iron abundances in the spectrum.
We performed an initial analysis of the data assuming that the abundance was equal
to the solar photospheric abundances of \citet{andersgrevesse1989}, the minimum
amount of continuum that might exist in the spectrum, and using the method described below
to generate the DEM.
We then predicted what the level of the continuum spectrum should be based on abundances ranging from
solar to one tenth the solar photospheric abundance of iron.  Figure~\ref{contfig} shows
the SW composite quiescent and flare spectra and continuum spectra overlaid.  
The spectra have been corrected for the effect of interstellar absorption.
Abundances $\geq$ 0.2 times the solar value produce a continuum level that can be compatible with the
observed flux levels from both the composite quiescent and flare spectra.
%Extremely sub-solar values (less than 0.2 times the solar value)
%are incompatible with the observed flux levels from both the composite quiescent and flare spectra.  
Table~\ref{tbl:compare_cont} tabulates the discrepancy 
in measuring emission line fluxes between assuming no continuum flux, and
subtracting a continuum calculated with varying abundances.  
%The rise of continuum
%flux towards shorter wavelengths introduces more error into emission line fluxes at the
%short wavelength region of the SW spectrum than at the long wavelength ends.
The discrepancy between line fluxes determined with no continuum subtraction and 
a continuum calculated assuming very sub-solar abundances grows larger with
decreasing abundance.  

We also examined shorter time intervals, in which the line fluxes
 have poorer S$/$N.  In both a shorter quiescent interval ({\bf 1998Q})
and a flare interval ({\bf 1994F}), 
%the spectral data were inconsistent with
%iron abundances $<$0.4.  
the difference in emission line flux introduced 
by not subtracting 
a continuum compared with assuming a continuum with iron abundance $>$0.4 was at the same
level as the photometric errors.  
This is a different result from that obtained for the composite
spectra (where Fe/H $\geq$0.2 solar), due mostly to S/N issues.
In the following analysis of the {\it EUVE} data, which uses the lower S/N spectra to
examine activity-related spectral changes,
we take the coronal iron abundance of HR~1099 to be 0.4 for all time segments, thus assuming
the iron abundance does not vary during flares.  This is a conservative estimate based 
on the fluxes in short time interval spectra:  The {\it EUVE} data can only constrain Fe/H to be
$\geq$ 0.2.

The DEM was determined from the emission lines of iron, as described above; initially no continuum
flux was subtracted before measuring emission lines.  Once the DEM was constrained, 
a continuum was predicted using the DEM and an iron to hydrogen ratio of 0.4 times
the solar photospheric value as given in \citet{andersgrevesse1989} and subtracted from the observed spectrum.  
The emission lines were re-fitted, and the DEM was redetermined.
The method for constraining the DEM is an interactive one; the user chooses a value of
the DEM at each temperature bin ($\Delta \log T=0.1$, from $\log T$=6.2--7.5), and adjusts
the value of the DEM at each temperature
bin until there is suitable agreement between observed line fluxes and those predicted by the
DEM.  Because plasma emissivities in the low density limit are
used, any line exhibiting a departure  from collisional ionization equilibrium
was not used to constrain the DEM.
For temporal intervals where lines of Fe~XV,XVI were not detected, the temperature
coverage available by lines of Fe~XVIII--XXIV is limited to $\log T=$6.7--7.5.
%The assumption of solar photospheric values
%of \citet{andersgrevesse1989} represents the minimum amount of continuum emission to be
%expected, assuming that the continuum does not arise from hotter temperatures than can be
%addressed by the temperature coverage of the emission lines in the EUV spectral range.
Figure~\ref{hr1099qudem} shows the DEMs determined for
quiescent intervals; Figures~\ref{hr1099fldem1} and ~\ref{hr1099fldem2} display
the same quantities for flaring intervals.  Tables~\ref{tbl:euvefluxtbl} and~\ref{table7} list
the emission line fluxes after continuum subtraction.  Figure~\ref{fig:comparedems} compares
the quiescent and flare DEMs for two cases.

After determining the DEM, and subtracting off our estimate of the continuum emission, the
resulting line fits should be due to emission line flux alone.  There are transitions in the EUV where the
assumption of collisional ionization equilibrium breaks down, due to a metastable upper level.  
In these cases the emissivity is proportional to the electron density, not the square.  The ratio of one 
of these lines with a line in collisional ionization equilibrium yields an estimate of the
electron density at the temperature where the lines are emitted.  Two lines of
Fe~XXI, 102.22 \AA\ and 128.73 \AA, satisfy this condition, as do $\lambda$114.41 and 
$\lambda$117.17 of Fe~XXII.  We used the line flux measurements and statistical errors to estimate
the electron densities of Fe~XXI and Fe~XXII.  Table~\ref{tbl:euvdens} lists the
electron densities and errors derived for the different time intervals.  Figure~\ref{hr1099density}
illustrates the variation in electron density of the different time intervals.  
There is no significant difference between electron densities derived from flare intervals
and those derived from quiescent intervals.  This is partly due to the admixture of quiescent flux and flare
flux in the spectra extracted from flare intervals (the flare intervals have been
defined broadly and encompass regions of only moderate enhancement), and partly from the poor S$/$N in flare spectra
due to the shorter duration of the flares compared to quiescent intervals.  Yet, despite these caveats, there
is no evidence for significant density enhancements, such as the factors of $\approx$ 10 or greater as seen
in solar flares \citep{bruner1988}.  Previous investigations of flares on RS~CVn systems with
{\it EUVE} have also shown a lack of conclusive evidence for density enhancements during flare
intervals \citep{sf2002}.

\subsection{Analysis of X-ray Observations \label{morexray}}
Four X-ray observations by three X-ray satellites:
{\it ASCA}, {\it RXTE}, and {\it BeppoSAX}  were obtained as part of the HR~1099 observations
in 1994, 1996, and 1998.
While the wavelength coverage, spectral resolution
and wavelength of peak sensitivity differ, they all diagnose the very hot plasma (T$\geq$ few MK)
present in active binary systems, and are sensitive to much hotter temperatures than
the {\it EUVE} spectra.  
%The full analysis of the {\it BeppoSAX} data are discussed in more detail
%in Tagliaferri et al. (2003).  
In the following we analyze the {\it ASCA},
{\it RXTE} and {\it BeppoSAX} datasets individually.  

\subsubsection{{\it ASCA} \label{sec:hr1099ascasec}}
As seen in Figure~\ref{1994ascalc}, the {\it ASCA} observation revealed a small flare with 
an enhancement of $\approx$ 1.8 in the 0.6--10 keV bandpass over the non-flaring level; 
the flare lasted approximately 8.7 hours.
Because of the variability the system was undergoing during the observation, we extracted
five different time intervals, denoted {\bf Q} for the non-flaring bit, {\bf R} for the rise of
the flare, {\bf P} for the flare peak, and {\bf D1} and {\bf D2} for the two decays of the flare.

The top panel of Figure~\ref{1994ascalc} shows the {\it ASCA} light curve in three different bandpasses, normalized to their
respective average quiescent count rates.  The plot illustrates the
different response of these energy bands (0.6--2 keV, 2--5 keV, 5--10 keV) to the flare.
There is much more flux, relative to the quiescent level, in the higher energy bandpasses.
This indicates the increased
amount of continuum emission present during the flare. 
The bandpass 0.6--2 keV covers the peak in the
observed spectral distribution, which has contributions mainly from iron L shell emission lines.
The 2--5 keV energy range contains emission from hydrogen- and helium-like lines of Ca, Ar, S, and Si,
as well as underlying continuum emission.  The highest energy range, 5--10 keV, contains mainly
continuum emission and the hydrogenic and helium-like lines of iron.  With higher temperatures
come larger amounts of continuum emission, since above temperatures of $\approx$ 15--20 MK 
most of the abundant elements are fully stripped and thus
there are relatively
few line transitions.  

Table~\ref{tbl:ascaflux} gives the
luminosities in these three bandpasses for the five temporal intervals extracted for spectral analysis
(see Figure~\ref{1994ascalc}) along with the total luminosity in the 0.6--10 keV range.  
The luminosity peaks in the {\bf R} and {\bf P} temporal intervals, consistent with 
the light curve.
Using the Hipparcos distance of 29 pc \citep{hipparcos}, 
we subtract the X-ray luminosity during quiescence from the X-ray luminosities during the four subsequent time
intervals to estimate the net amount of energy radiated during the flare.  
The system is bright during
quiescence, with L$_{X}$ of 7.5 10$^{30}$ ergs s$^{-1}$ in 0.6--10 keV and increasing to
1.3 10$^{31}$ ergs s$^{-1}$ at the peak of the flare.
The
total amount of flare radiation is 5.3 10$^{34}$ ergs.  
Table~\ref{tbl:ascaflux} also lists the rise and decay timescales for the flare in 
different energy bands, assuming the count rate increases and decreases exponentially during
the flare rise and decay, respectively.  The higher energy bandpasses display a faster response to
the flare.

The SIS0 spectra for the five time intervals are shown in Fig.~\ref{1994ascaspec}.  The spectra were
fitted with a discrete, two-temperature VMEKAL model; the hydrogen column density, N$_{H}$, was
fixed at 1.35 10$^{18}$ cm$^{-2}$ as determined by \citet{piskunovetal1997}.  The abundances of elements O, Ne, Mg, Si,
S, Ar, Ca, Fe, and Ni were allowed to vary.  Results from spectral fitting to combinations 
of the SIS0, SIS1, and GIS2 spectra are given in Table~\ref{table8}.\footnote{The location
of oxygen lines used in the abundance analysis are near the cutoff at 0.6 keV,
and suffer from calibration problems; see http:$//$asca.gsfc.nasa.gov$/$docs$/$asca$/$watchout.html.} 
We also used SPEX to
determine the emission measure distribution assuming a multi-thermal plasma, with the abundances
fixed at those determined from the 2T fits.  The left panel of Figure~\ref{asca_dem_abundchange} 
shows the resulting emission measures for the five time intervals.  The peaks are consistent with
temperatures derived from 2T fits, and show an increase in temperature during the
flare peak, with a gradual subsidance to the quiescent level as the flare decays.  
The abundances show a statistically significant small increase during the rise of the flare;  iron increases a
maximum of a factor of three from
13\% of the solar photospheric value in quiescence to 32\% during the rise phase, decreasing in 
value over the course of the peak and decay to the value attained during quiescence.
This change can be seen in the evolution of the feature in the {\it ASCA} spectra at 6.7 keV,
due to Fe~XXV.
Any definitive trend for other elemental abundances is marred by the large uncertainties on
these abundances.  The right panel of Fig.~\ref{asca_dem_abundchange} plots the abundances
of Ca, Mg, Fe, Si, S, O, and Ar during the five time intervals against the First Ionization
Potential (FIP).  There does not appear to be any FIP-dependent abundance enhancement.
The abundance of Ne, with a FIP of 21.6 eV, unfortunately suffers from large error bars
and it is difficult to discern any changes in this abundance through the evolution of
the flare.

\subsubsection{{\it RXTE} }
The final, background-subtracted {\it RXTE} light curves are shown in
Fig.~\ref{rxtelc}, binned into 128 second intervals for the 1996 observation
and 256 second intervals for the 1998 observation. The data are 
sparsely sampled in time, but one large flare in 1996 and another in 1998 are
evident.  We plot the light curves for these flares in two energy bandpasses, 
2--5 keV and 5--12 keV, to examine how the flare evolves at different energies.
We see similar behavior to that
in the {\it ASCA} light curve in 1994 --- the higher energy bandpass (here 5--12 keV)
has a much more rapid response to the flare, showing a faster rise and a faster
decay compared with 2--5 keV.  

We can estimate the amount of energy radiated during these flares by determining
a luminosity--count-rate conversion, using the average count rate during the observation,
and converting the integrated flux to luminosity.  The luminosities in 2--5, 5--12 and 2--12 keV
bandpasses were calculated using a 2T MEKAL model applied to the 2--12 keV spectrum accumulated
from the entire observation of each year, and are listed in Table~\ref{tbl:rxtetbl}.
We tabulated the net flare energy, determined by summing the observed
counts during the flare, converting to energy using the luminosity--count-rate conversion, to
obtain the amount of energy radiated during the flare, including the quiescent contribution.
The quiescent contribution to the total flare energy was estimated by taking the
quiescent count rate, multiplying by the flare duration, and converting to energy using the
conversion factor.  This quiescent emission was subtracted to give a net estimate
of the amount of energy radiated during the flare in each bandpass; these values are
listed in Table~\ref{tbl:rxteflares}.  Both flares lasted only a few hours, yet each radiated
a substantial amount of energy ($\sim$ few 10$^{34}$ erg) in the 2--12 keV bandpass.  
The flare energies in 
the 2--5 and 5-12 keV bandpasses also are similar.  

Also listed in Table~\ref{tbl:rxteflares} are the rise ($\tau_{r}$) and decay ($\tau_{d}$) timescales,
computed assuming an exponentially increasing and decreasing time profile of the flare.
Due to the gaps in the timeseries, neither flare was observed
in its entirety; for the 1998 flare we are unable to constrain the rise time due to inadequate
temporal sampling.  A comparison of the flare timescales for the 2--5 and 5--12 keV bandpasses for
both flares reveals smaller timescales, and hence more rapid increases and decreases in flux
changes with time,
for the higher energy bandpass.  This trend was noted by \citet{ostenetal2000} for another
active binary system, $\sigma^{2}$~CrB.

\subsubsection{{\it BeppoSAX} \label{sec:hr1099sax}}

The HR 1099 LECS and MECS light curves are shown in Figure~\ref{1998saxlc}.
A small flare occurred just at the beginning of the observation, followed by
a second one during the decay; a third flare occurred at the end of the
observation. In all three cases the peak count rate in the MECS 
increased by a factor of
$\sim 2-2.5$ with respect to the quiescent one. A few smaller peaks are also
evident in the light curve. Given the variability of the emission, we have
divided the observation into five separate time intervals, covering the
three main flares (labeled {\bf F1}, {\bf F2} and {\bf F3}), the late decay
phase of the first two flares (labeled {\bf D}) and the presumably quiescent
interval before the third flare (labeled {\bf Q}). In Table~\ref{saxlum} we
give the luminosities for the five intervals in the same energy bands as
{\it ASCA} for comparison (see Table~\ref{tbl:ascaflux}). The quiescent
luminosity in the 0.6--10 keV band is 5 10$^{30}$ erg~s$^{-1}$, a factor of
$\sim 1.4$ lower than in the 1994 {\it ASCA} observation, and rises to 7.8
10$^{30}$ erg~s$^{-1}$ during the first two flares. In Table~\ref{saxlum} we
also give the flare timescales in the MECS band (1.6--10 keV) assuming
an exponential rise and decay. While the second and third flare are similar,
with rise and decay timescales of $\sim 2$ and $\sim 4$ hrs, respectively,
the first flare shows a more gradual behaviour, with rise and decay times of
$\sim 11$ and $\sim 13$ hours. We also note that after the second flare
the decay of the emission is slower.

Joint fits of the LECS+MECS spectra have been performed for each time
interval. The LECS data have been fitted in the
$0.1-4$ keV band only due to calibration problems at higher energies
\citep{fioreetal1999}. In order to account for the uncertainties in the
intercalibration of the LECS and MECS instruments, a relative normalization
factor (which was left free to vary) has been included in the fits: the
derived best-fit values are $\sim 0.8$ for all spectra, in agreement with
the allowed range of 0.7--1 \citep{fioreetal1999}. All spectra have been
fitted assuming a 2-temperature MEKAL model with variable global metal
abundances, since the lower energy resolution of {\it BeppoSAX} with respect to
{\it ASCA} does not allow a reliable determination of the individual
abundances. The interstellar column density was kept fixed to the value N$_H
=$ 1.35 10$^{18}$ \citep{piskunovetal1997}.

The best-fit parameters are given in Table~\ref{saxfits} and the spectra are
shown in Figure~\ref{1998saxspec}. There is a very
good agreement with the parameters derived by {\it ASCA}, both for the quiescent
and flaring spectra, with only a slight difference in the quiescent
emission measures, that in {\it BeppoSAX} are lower by a factor of $\sim
1.5$. The temperature and emission measure of the hot component are higher
during the flares, while the cool component remains constant. Contrary to
the {\it ASCA} flare, we do not find any evidence for an increase of the
global abundance, which has a value of $Z \sim 0.3$. This value is
consistent with the individual abundances derived by {\it ASCA}, with the
exception of Ni and Ne.

\section{Multi-Wavelength Comparisons \label{sec:multi}}

Figures~\ref{1993multiw} through ~\ref{1998multiw1} depict the 
portions of the campaigns from different telescopes
that overlap in time. Clear correlations
exist between behavior in different spectral regions. Several large flares were caught with
EUV$/$SXR telescopes, and 3.6 cm radio variability indicates a substantial amount of 
gyrosynchrotron flaring as well.  The data are
sparsely sampled, suffer from gaps due to earth occultations, and incomplete
temporal coverage by ground-based radio telescopes, but several trends shine
through:  Radio gyrosynchrotron and EUV$/$SXR flares generally are correlated, with
radio flares preceding the thermal emission, and UV emissions exhibit 
changes that seem to occur during radio flares.

\citet{gudelbenz} noted an almost linear correlation between X-ray and radio quiescent luminosities for 
stellar coronae on a number of different kinds of stars, from solar flares to dMe flare stars to flares on RS~CVn binary
systems.  For RS~CVn systems containing a subgiant and a main sequence star,
the correlation was $L_{X}/L_{R} =$ 0.17 10$^{15.5\pm0.5}$ Hz.  The observational relation was based on 
measurements of the X-ray luminosity with ROSAT and Einstein, at X-ray energies of 0.2--2.4 keV, typically.
In Table~\ref{tbl:lxlrq}, we list 3.6 cm radio luminosities and simultaneously obtained EUV (80--180\AA)
luminosities from the four different campaigns.  The ratio of the EUV to radio luminosity spans the range
1.2 10$^{13}$ -- 8.2 10$^{13}$ Hz, which is slightly lower than the observed values of \citet{gudelbenz}.
However, if we take into account the relative amount of radiation in the
differing bandpasses  for a thermal plasma with a temperature distributions as determined in Section 6.1,
the numbers are roughly consistent.  Most of the radio and X-ray observations presented in \citet{gudelbenz} 
were obtained at widely different times, and so record only some very average coronal conditions.  By
contrast, our observations are strictly simultaneous, and demonstrate that the ratio of nonthermal
to thermal coronal emission reaches an approximately constant value, albeit with some scatter due to
differing activity states of the star.

Table~\ref{tbl:multiw} summarizes the EUV$/$SXR flares observed, in terms 
of the behavior displayed in the other spectral regions involved in the campaigns.
In addition to the 17 flares summarized there, there were five radio flares that
were not associated with UV or EUV$/$SXR flares:  Three did not have UV or EUV$/$SXR coverage
(1993 September 14.7, 15.3, and 16.4).
During the other two, on 1996 September 3.5 and 1998 September 11.5, there was 
no apparent EUV enhancement.  For three events where a radio flare and EUV/SXR flare were coincident 
(discussed in more detail in \S 6.3), we estimated the peak flare luminosity and flare energies
at 3.6 cm and in the EUV/SXR, by subtracting off an estimate of the quiescent luminosity from the light
curve.  Due to the poor time sampling of some of the data, we fitted functional forms to the pattern of
emission vs time, (e.g. linear rise, exponential decay), and used that representation to estimate the
amount of energy radiated during the flare.  The ratio of peak EUV flare luminosity to peak radio luminosity
is given in Table~\ref{tbl:neupert}, and shows a similar range as the ratios determined from apparent 
quiescence, above.  

Despite the limited coverage of any given flare with the suite of telescopes employed
in these campaigns, we still can investigate rough statistics of correlation in
the radio, UV, EUV, and SXR bands.  There were five EUV$/$SXR flares that had partial UV coverage,
and all showed some kind of enhancement in the UV.  
The response of the Mg~II line fluxes and C~IV line fluxes is similar in three cases;
but for the other two, only the higher temperature C~IV emission shows an increase.
There were five SXR flares
that had EUV coverage, but only two showed an enhancement in the EUV.
Eight EUV$/$SXR flares had radio coverage
during the flare rise; of these, three showed radio bursts (one EUV flare is a relatively small
enhancement above the quiescent count rate, and could be counted as an additional EUV/radio 
flare coincidence).  
There were a total of ten 3.6 cm radio flares: Five were associated with EUV/SXR flares, while
two had EUV/SXR coverage but there was no rise in high energy thermal flux.

The start of the rise phase in EUV$/$SXR flares is generally when the impulsive
UV and 3.6 cm radio flares occur, as demonstrated in 
Figures~\ref{1993multiw}, \ref{1994multiw2}, \ref{1994multiw1}.
The UV and radio radiation show a sudden increase then
decay at this point.  The radio emission originates from gyrosynchrotron emission
from accelerated particles; the abrupt change in behavior suggests an increase in the 
number of radiating particles, increase of magnetic field strength, or possibly a change
 in the distribution of the nonthermal electrons.  The correlation of UV flux with 
radio flux suggests that the transition region (C~IV) and sometimes the chromosphere (Mg~II)
are heated by 
nonthermal particles impinging from the corona.  The trend of Mg~II
fluxes with radio emission is less clear.  For the EUV flare that peaked on 1993 September
17.4, both C~IV and Mg~II showed a peak during the rise of the EUV flare.  In subsequent
EUV$/$SXR
flares, however,  there was no response from Mg~II, in contrast to the behavior illustrated by C~IV.

Radio variability is more common at 20 cm than at higher frequencies.  Section 4.2 showed that
negatively polarized bursting behavior at 20 cm happened about 30\% of the time HR~1099 was observed.
Figures~\ref{1994multiw2}, ~\ref{1996multiw1},~\ref{1998multiw1} compare the flux and polarization behavior
at 20 cm to the variations of the thermal plasma in the EUV and SXR. One might expect a higher association
rate between 20 cm bursts and EUV/SXR flares, since both appear to occur frequently.  Such an assessment 
is difficult to make, due partly to the gaps in time coverage between the two wavelength regions, and
the lack of any structured behavior at 20 cm, compared to the more or less orderly rise and decay of the
largest EUV/SXR flares.  There are several instances when the 20 cm radio flux appears to maintain
a large value of LCP for several hours:  1994 August 24.7--25.0, 1994 August 27.4--27.7,
1996 September 3.3--3.9, 1998 September 8.8.3--8.9, 1998 September 10.5--10.7, and 1998 September
11.4--11.7.  At these times there are few EUV flares, but also many gaps in EUV coverage, so it
is difficult to determine the association rate.  
%As discussed in Section 4.2, these bursts might
%be due to coherent mechanism, such as plasma emission or electron cyclotron maser emission.  On the Sun,
%plasma radiation due to electrom beams propagating through the corona have been used to determine the electron 
%density of the ambient plasma the beam encounters as it propagates away from the site of particle 
%acceleration \citep{aschwandenbenz1997}, and correlations with SXR emission allow a rough determination of
%the physical separation between thermal, SXR-emitting flare loops and the site of flare-associated
%particle acceleration.  Unfortunately, the quality of the observations analyzed here, together with
%an uncertainty as to the physical mechanism producing the coherent bursts, and the low association
%rate between coherent bursts and EUV/SXR flares, means that a similar deduction for the stellar
%case cannot be made.

The DEM distributions for the thermal EUV$/$SXR emission generally has two peaks, 
one at 6--8 MK (0.5 keV) and the other at 20--30 MK (1.7--2.6 keV).  If we make the assumption that
the extent of the 3.6 cm gyrosynchrotron radio emission during these flares is about the size of the primary, as found
by \citet{lestradeetal1984} during a strong radio flare ($\theta=$0.9 mas), we can estimate the 
brightness temperature and energy of the electrons producing the gyrosynchrotron
radiation.  For the four EUV$/$SXR flares accompanied by 3.6 cm radio bursts, the 
peak radio fluxes imply electron energies between 70 and 200 keV, well above the energies of the
thermal electrons.  These electron energies are accessible with the hard X-ray telescopes on board
{\it RXTE} and {\it BeppoSAX}, 
%so it is surprising that 
 but no detectable hard X-ray emission was seen by either satellite.

\section{Discussion \label{sec:ch5conc}}

\subsection{Emission Mechanisms for Radio Bursts\label{sec:radiobursts}}
The 3.6, 6, and 20 cm bands have the most extensive overlap of the five frequencies
used in our radio observations.  Generally, flares are most obvious at 3.6 and 20 cm.
Often, variations at 3.6 cm are repeated by variations at 6 cm, but the bandwidth does not generally
extend to 20 cm.  Emission at 3.6 and 6 cm is consistent with
gyrosynchrotron emission, which is expected to be relatively
broadband:  the approximate expressions in \citet{dulk1985} cover an order of magnitude 
range in harmonic number $s$ of the electron gyrofrequency.  Thus, one would expect
to be able to observe gyrosynchrotron emission from a homogeneous source at 
both 3.6 and 20 cm.  Nondetection of flaring emission at 20 cm could be due to opacity effects, if the
emission were optically thick, or due to a cutoff in the emission due to a high ambient density in the 
flaring region.  Alternatively, 
the lack of correlation between flaring at these two frequencies
could be due to a different emission mechanism operating at lower frequencies.

The persistent nature of gyrosynchrotron emission from accelerated particles
in active binaries in a steady state is puzzling, and investigations have
attempted to explain the observed patterns of flux and polarization with 
various models.   
\citet{chiuderidrago1993} attempted to 
explain the quiescent radio emission as the long-timescale decay of flares, (episodic acceleration of electrons),
in a dipolar magnetic field configuration.
They followed the change in time of the flux spectrum due to decay of a flare, including
collisional and radiative losses of an initially power-law distribution of electrons.
The spectrum at the time of peak flux had a positive slope from 1--5 GHz, becoming negative for larger frequencies.
As losses changed the shape of the electron distribution with energy, the spectrum
evolved to a decreasing function of frequency from 1--30 GHz.  
A three-dimensional dipole field was applied by \citet{jones1994}, who
calculated both the flux and polarization spectra resulting from gyrosynchrotron radiation.
 \citeauthor{jones1994}
found it difficult to reconcile the flat flux spectrum with a changing polarization
spectrum.
Their geometry was a thin shell of radiating electrons of width $\sim$ 0.1 R$_{\star}$
at 1.9--2 R$_{\star}$ in an arcade of magnetic loops, with a continuous
injection of accelerated
particles to maintain the electron distribution against radiative losses.
In order to produce a flat flux spectrum, as was observed from their quiescent
data, the nonthermal electron density must increase
with radius.
\citet{jonesetal1996} examined the VLA and ATCA observations from 1993 and 1994
also presented here.  They interpreted the data in light of the 3D model
presented in \citet{jones1994}.  They argue that the increasing trend of polarization
with frequency is indicative of a change from optically thick gyrosynchrotron emission 
at low frequencies to optically thin radiation 
at high frequencies.  
However, \citet{whitef1995} note in their observations of active binaries that
the increasing trend of polarization with frequency is independent of the shape of the flux spectrum,
contrary to the expectations from homogeneous sources \citep{dulk1985}.  

Figure~\ref{fig:radiospec} displays examples of flux and polarization spectra during the 
decays of high frequency radio flares.  The flux spectrum
is increasing at low frequencies, but turns over at higher
frequencies, with a maximum between 2 and 5 GHz.
This behavior is consistent with optically thick emission at low frequencies and optically thin 
conditions at high frequencies, a pattern that has been seen during the rise and peak
of flares on other active binaries \citep{feldmanetal1978,klein1987,tcetal1998}.
In the limits of small and large optical depth, the frequency dependence of the effective
temperature and absorption coefficient can be combined to estimate the range of spectral indices
$\alpha$ between flux measurements at adjacent frequencies.  For a homogeneous source in a vacuum,
the spectral index $\alpha$ should 
depend on the power-law index of the electron distribution function, $\delta$, as \citep{dulk1985}: \\
\begin{eqnarray}
\alpha_{\tau \gg 1}& =&  2.5 + 0.085\; \delta \\
   2.7\leq \alpha_{\tau \gg 1} \leq 3.1   &  & \rm{for}\; \;  2 \leq \delta \leq 7 \nonumber  \\
\alpha_{\tau \ll 1}& =& 1.2 - 0.895 \; \delta \\
-5.1 \leq \alpha_{\tau \ll 1} \leq -0.6& & \rm{for} \; \;  2 \leq\delta\leq 7 \nonumber .
\end{eqnarray}
The spectral indices listed in Table~\ref{tbl:specindex} in the 1--5 GHz
range are generally flatter than these limits on the spectral index, 
implying that a shallower distribution is required ($\delta \leq$ 2,
and out of the range of validity of the formulae in \citeauthor{dulk1985}),
the source is inhomogeneous, or the magnetic field varies spatially.
\citeauthor{tcetal1998}
used a similar method as \citet{chiuderidrago1993} to model the flux spectrum, but examined the rising phase
of flares, 
and determined that
a continuous supply of relativistic electrons and loss mechanisms was necessary.  

The radio flares at 3.6 and 20 cm generally have quite different characteristics:
20 cm bursts are very highly polarized, and display rapid flux changes ($<$60 s),
while 3.6 cm bursts vary over longer timescales, and are consistent with little to no 
intrinsic polarization.
This suggests a different mechanism at 20 cm, bringing us to the question of what 
kind of process might be operating to produce the 20 cm flares.
The quantities necessary to discriminate different radio emission mechanisms
are: (1) time duration of the event; (2) brightness temperature T$_{b}$; (3) 
sense and amount of circular polarization; (4) bandwidth $\Delta \nu$.  
The time duration is limited by the temporal sampling of the VLA data to be $\geq$ 10 seconds.
There are large LCP fluxes that persist for 
long timescales, $\approx$ 5.6 hours, as well as fluctuations on smaller timescales.
Often increases, peaks, and declines in flux were observed to occur with a duration $\leq$ 60 seconds.
Knowledge of the 
brightness temperature involves an estimate of the emitting source size, which we cannot
unambiguously deduce 
from these data.  We can place a lower limit on T$_{b}$ by assuming that the radio source
is $\approx$ the size of the binary orbit: consistent with previous VLBI observations
of quiescence in active binaries at 
higher frequencies \citep{muteletal1985}, and constrained by
a light travel time of $\sim$ 60s (the approximate duration of the shortest individual bursts).
The brightness temperature can be
determined from the observed flux S$_{\nu}$, at a wavelength of 20 cm, assuming a source
size $\theta_{\rm mas} =$2.5 using orbital parameters from \citet{cabs} and references therein, \\
\begin{equation}
T_{b}(K) \geq 2.2 \times 10^{8}\;\; S_{\nu}(mJy) \;\;.
\end{equation}
The brightness temperatures of the 20 cm bursts was in the range 1--4 10$^{10}$ K, pointing to
a nonthermal process, suggesting electron energies of a few tens of MeV\footnote{If the emission mechanism
is coherent, with particles radiating in phase, the average energy implied by the brightness
temperature can be greater than the energy of a single particle.  Thus this may be an upper limit
to the actual electron energy.}.
The high degrees of circular polarization  argue for a relatively localized phenomenon,
so this is a lower limit to the brightness temperature.
The observed circular polarizations are very large and negative 
and consistent with 100\% LCP emission 
assuming a two-component quiescent and flaring model.  
The polarization at higher frequencies, where the spectral indices indicate optically thin conditions,
suggests X-mode emission.  Since the low frequency emission has the opposite helicity, it appears to be o-mode
emission, which is consistent with the hypothesis of plasma emission.
O-mode plasma emission at the fundamental can be explained by the lower cut-off
frequency for o-mode emission compared to x-mode \citep{benzbook}; polarizations
can reach quite high levels. 
We can place rough limits on the bandwidth
by comparing the behavior at 20 cm with that at 13 or 6 cm.
From a lack of correlation with higher frequency data, the bandwidth must be smaller than $\approx$ 988 MHz.
The difference in flux behavior in the two intermediate frequencies of the VLA L band receiver also constrains the
bandwidth of the phenomenon.  The two IFs span 50 MHz centered
on 1385 and 1465 MHz respectively.
The bottom two panels of Figure~\ref{coherent} 
compare the behavior in these two IFs during some 20 cm bursts.  
From 1994 August 27.525--27.533, the higher frequencies dominated in flux, although the temporal
pattern of both IFs is similar, suggesting $\Delta \nu \geq$ 130 MHz.  
On 1994 August 28.372--28.380 the lower IF dominates in flux
initially, followed by a change to higher frequency dominating at the end of the burst event,
suggesting that the bandwidth was smaller than 130 MHz.
The differential frequency information
available in the two IFs from these continuum observations is not enough to pin down the bandwidth; 
finer spectral binning must be used, such as available with the Arecibo
telescope, the Green Bank Telescope, or the expanded VLA.

Highly polarized, short time duration events have been observed on the Sun, although generally
at lower frequencies than seen here.  The metric and decimetric solar radio bursting and spiky emission
is usually attributed to plasma radiation at the fundamental or second harmonic 
of the plasma frequency ($\nu \sim s \nu_{p}$, s$=$1,2).
This emission is associated with electron beams, and results from the conversion of electron
beam-excited coronal plasma waves into plasma radiation.
Table~1 of \citet{dulk1985} summarizes the characteristics of
solar radio bursts.  
Some bursts display analogous behavior to
that seen here.
Type~I bursts have brightness temperatures $>$ 10$^{10}$ K and are
highly polarized in the sense of the o-mode.  
However, they tend to occur at much lower frequencies ($\nu \leq$ few hundred MHz), 
and can have very small bandwidths $\Delta \nu \approx$ 1 MHz and last for
one second at most.  Solar radio continuum emissions have been observed 
which are relatively long-lived, highly polarized, and
show large polarizations in the sense of the o-mode; an example is type~IV storm
continuum.  
\citet{islikerbenz1994}
have catalogued solar flare radio emission from 1--3 GHz, and give examples of type~IV
bursts which last for minutes, modulated on timescales less than 10 seconds, and are
strongly polarized.  

The two main mechanisms which have been proposed to explain similar behavior in dMe
flare stars are plasma radiation or emission from an electron-cyclotron maser instability.
%O-mode plasma emission at the fundamental frequency ($\nu_{p} \approx 9 \sqrt{n_{e} (cm^{-3})} (GHz)$)
%can explain the large values of percent left circular polarization.  
Both interpretations suffer from uncertainties in the model parameters, and imprecise
knowledge of stellar atmospheric structure.  Plasma emission can occur at the
fundamental or second harmonic of the plasma frequency ($\nu_{p} \approx 9 \sqrt{n_{e} (cm^{-3})} (GHz)$),
while emission from an electron cyclotron maser instability generally occurs at the fundamental or
harmonics of the electron cyclotron frequency ($\nu_{B} \sim 2.8 B (G)$ MHz).  
These parameters depend on the electron density $n_{e}$ and magnetic field strength $B$,
whose variation with height (for the case of a homogeneous, stratified atmosphere)
or relative changes in inhomogeneous structures are not known.  
Often only a single
coronal electron density or photospheric magnetic field strength measurement exists;
it is difficult to infer whether spatial regimes in the stellar upper atmosphere
favoring either mechanism exist.
Additional
complicating factors include the propagation and escape of the radiation once it has been
generated: for example, o-mode plasma emission at the fundamental frequency can explain
the large values of percent left circular polarization, but this radiation experiences strong
free-free absorption due to the high inferred electron densities.  
Similarly, fundamental radiation from electron-cyclotron maser emission
experiences large amounts of gyroresonant absorption, due to the high inferred magnetic field strengths.
A coronal loop provides a natural set-up for a magnetic mirror and resulting loss-cone distribution; 
however, recent research on auroral kilometric radiation (AKR) which 
invokes a cyclotron maser emission mechanism has shown that a shell maser configuration
is more appropriate in this situation \citep{ergunetal2000}.  A shell maser in a stellar corona
would allow the source to be located higher in the corona due to a more moderate requirement of the magnetic mirror
ratio.
In addition to these two possibilites, \citet{bastian1996} has proposed some exotic mechanisms
to explain the extremely high brightness temperature phenomena observed in dMe flare star
coronae.

Without better temporal and frequency resolution, it is impossible to
discriminate amongst these two coherent emission mechanisms in the observational data. The hours-long 
highly LCP emission could be an envelope of thousands of short duration bursts ($\leq$ 10 s)
overlapping each other, or it could be diffuse continuum emission.  
\citet{whitef1995} suggested the solar analogy of type~I noise storms \citep{kai1985}
to explain this phenomena:  a superposition
of type~I bursts with bandwidths of about 100 MHz is consistent with the data here.
Similarly, continuous pumping of free energy into the maser from a parallel electric field
could explain the long duration of large negative polarization. The smaller timescale bursts
could be due to pulses from the maser.
The high coronal temperatures (T$\geq$ 10 MK) found in HR~1099 and other active binary systems could reduce the
free-free absorption that would result from fundamental plasma emission (free-free absorption coefficient
$\sim$ n$_{e}^{2}$/T$^{1/2}$) and explain the higher frequencies where the stellar emission 
is found, compared to the lower temperature solar corona.  The evidence from high energy spectra for
high coronal electron densities (n$_{e} >$ 10$^{12}$ cm$^{-3}$) suggests there must be extreme density
inhomogeneities, however; otherwise the high inferred electron densities would obliterate any effect of
high temperatures in reducing the free-free opacity.
One of the requirements
for generating the electron-cyclotron maser is the condition of a low density plasma,
$\nu_{p}/\nu_{B} \leq$ 1.   We do not have exact knowledge of the 
spatial variations of the electron density or magnetic field in the coronal
environment of HR~1099.  The situation is made more complicated by the presence of
two stars, and possible intrabinary emission. 
We can make estimates for
plausibility arguments.
Taking a base coronal electron density of 10$^{12}$
cm$^{-3}$ (see \S~\ref{sec:ch5euvespec}), a base magnetic field of $\leq$ 1000 G as determined from Zeeman Doppler
imaging \citep{donatietal1990}, and estimates for the density scale height and magnetic scale
height using the parameters for the K1~IV primary, the  plasma frequency always exceeds the
electron gyrofrequency: \\
\begin{equation}
\frac{\nu_{p}}{\nu_{B}} = (4\pi m_{e} c^{2})^{1/2} \frac{n_{0}^{1/2}}{B_{0}} e^{r(\frac{1}{H_{B}}-\frac{1}{2H_{n}})}
\end{equation}
assuming $n(r) = n_{0} e^{-r/H_{n}}$ and $B(r) = B_{0} e^{-r/H_{B}}$, for H$_{B} \leq$ H$_{n}$;
the density scale height H$_{n}$ at 10$^{7}$ K is 5.6 10$^{11}$ cm, or approximately twice the radius
of the K1 subgiant.
This assumes a homogeneous coronal plasma: The existence of extreme density inhomogeneities
could make conditions more favorable
for the electron-cyclotron maser.  

\subsection{DEM Analysis and Abundances\label{sec:DEMS}}

The temperatures and abundances of the coronal plasma in HR~1099 can be diagnosed
in both the EUV ({\it EUVE}) and X-ray ({\it ASCA, RXTE, BeppoSAX}).  The EUV spectra permit identification
of individual emission lines, while X-ray spectra show only broad ``humps'' of emission line complexes
and the underlying continuum.  Here we discuss the different analyses and their 
different views of the coronal structure in HR~1099.

Figure~\ref{fig:comparedems} compares the differential emission measures derived 
from {\it EUVE} spectra for
quiescent and flaring conditions in 1993 and 1996, to discern plasma changes
before and during a flare.  The same basic shape exists in all the DEMs at temperatures
hotter than $\approx$ 5 MK.  The flare intervals are
integrated over changing plasma conditions $n_{e}$, $T_{e}$, and $V$ during the flare itself, so
a comparison with the quiescent DEM only yields information about the ``average''
DEM during a flare.  There is also a contribution from quiescent emission to these DEMs,
since no estimate of the quiescent spectrum was subtracted from the
flare spectra. 
The flare DEMs plotted in Figure~\ref{fig:comparedems}
have had the quiescent DEM subtracted from them, to get a sense of the changes that
occur in the temperature distribution during the flare.
During the flares there is a large amount of
hot plasma, T$\geq$ 15 MK, over and above the amount present outside of flares.
There is also a significant enhancement in plasma at 6--9 MK during flares.
For time intervals where Fe~XV and ~XVI are
detected, there is a large drop from the DEM at $T \geq$ 5 MK to
lower temperatures, showing a minimum around $T=$ 3 MK and rising to 
a small maximum around $T=$ 1.6 MK.  
The DEM shows evidence of
increasing at high temperatures, T$\geq$ 30 MK, beyond which the line diagnostics in the {\it EUVE}
spectra are not sensitive.

Portions of the {\it EUVE} data have been examined by others.  \citet{griffithsjordan1998}
performed an analysis of the 1993 and 1994 spectral datasets, without
performing time-resolved spectroscopy, despite the fact that several large flares
occurred during these observations.  
They determined that
bremsstrahlung continuum was minimal, and determined the fluxes of emission lines in the
spectra by using a ``local continuum'' level.  \citeauthor{griffithsjordan1998}
calculated the emission measure distribution using temperature bins of $\Delta \log T =$0.3,
and emissivities calculated at varying electron densities.  They find a peak at $\log T=$6.9, 
a decrease in the EM and subsequent increase to hotter temperatures for $\log T \geq$ 7.1.
For comparison with X-ray spectra, which have overlapping temperature sensitivities,
\citeauthor{griffithsjordan1998} examined EXOSAT medium energy proportional counter
and low energy imaging telescope data, fitted using one or two discrete temperatures.
The maxima in the {\it EUVE} emission measure distributions match those found from
the 1T or 2T X-ray fits.  \citet{sf2002}  also examined the {\it EUVE} spectral data of 
all observations from 1992--1999,
accumulating one quiescent and one flare spectrum from all individual quiescent and flare
time segments, and derive emission measure distributions for these and a spectrum summed  from all
{\it EUVE} exposures of HR~1099, regardless of activity.  In their spectral analysis, they
subtract a local continuum determined by visual inspection from the spectral data.
The general results of \citeauthor{sf2002} are consistent with those found in \citet{griffithsjordan1998}
and also found here for the separate time intervals:  There is a small peak at $\log T=$6.3
(due to emission from Fe~XV and XVI) and another, larger peak at $\log T=$6.9.  For hotter
temperatures, the emission measure decreases and subsequently rises to hotter
temperatures up to $\log T=$7.5, above which {\it EUVE} spectral data become
insensitive.

%The difference between the present analysis and these previous ones is our
%framework of studying individual flares, the spectral changes
%that occur, and possible variations from flare to flare.  In contrast to
%the analysis of \citet{griffithsjordan1998}, we do not lump together flaring and
%quiescent intervals, neither do we only construct a composite quiescent and composite flare
%spectra as \citet{sf2002} do, resulting in a spectrum which may not represent actual conditions. 
%We have also attempted to be more consistent in our treatment of line and continuum emission,
%especially given the importance stellar coronal abundance measurements have received
%in recent years with the advent of the {\it Chandra} and {\it XMM-Newton} spectrographs.  

The fundamental problem
in dealing with {\it EUVE} spectra is the lack of sufficient information to constrain all
the quantities that characterize the coronal plasma:  The key amongst them being elemental
abundances, and hence continuum emission in the spectrum.  The general trends reported by
the above authors are consistent with the distribution of plasma with temperature
derived here from {\it EUVE}.

The {\it ASCA} analysis has used the MEKAL plasma code, allowing the abundance of each element 
(O, Ne, Mg, Si, S, Ar, Ca, Fe and Ni)
having
strong emission lines in the 0.6--10 keV region to vary independently of each other.
Due to the low spectral resolution of the CCDs, some of these
abundances are compromised because of severe line blending (the oxygen abundance is suspect
due to calibration problems).  
The analysis of {\it BeppoSAX} spectra is simplified to discrete multiple temperatures,
and an abundance scaling $Z$.  The presence of emission lines from multiple ionic species,
and the high energy sensitivity, allows the X-ray spectra to constrain temperatures,
abundances, and the continuum level; in practice, due to the low resolution, some of these
quantities are entangled.
%The results during
%quiescence indicate very sub-solar abundances.  

The low iron abundance during the quiescent {\it ASCA} observation (Fe/H$\sim$0.12)
is apparently incompatible
with the observed {\it EUVE} flux obtained simultaneously; it predicts a continuum
level that is larger than the observed EUV fluxes.  We took the DEM 
and abundances derived from the quiescent
{\it ASCA} spectrum (described in \S 4.3.1) and synthesized the EUV spectrum.  We compare the
synthesized EUV spectrum with the composite quiescent EUV spectrum.  A comparison of the DEMs
derived from the {\it ASCA} and {\it EUVE} analyses shows that there are hotter temperatures
present in the {\it ASCA} DEM compared with the one derived from {\it EUVE} data. The 
hotter plasma in the {\it ASCA} DEM reveals itself in the {\it EUVE} spectrum primarily
through the continuum, and in order to match the observed EUV iron line strengths, the iron
abundance can be at 0.1 times the solar abundance.  This difference with the {\it EUVE} 
analysis lies in the temperature sensitivity; when there are hotter temperatures, the intrinsic
continuum level is higher, and therefore the line-to-continuum ratio can be lower.  
Disregarding the hotter temperatures artificially depresses the estimated continuum level,
necessitating a higher iron abundance.  
The low Z ($\sim$0.3) derived from the {\it BeppoSAX} X-ray spectra is consistent with the
constraint on abundance determined from the composite {\it EUVE} spectra, where Fe/H $\geq$ 0.2;
the wavelength overlap between these two instruments ({\it BeppoSAX}: 1.2--120 \AA; {\it EUVE}:
80--370 \AA) ensures that a consistent picture
can be obtained.
Analysis of other low resolution X-ray spectra
yield similar results to those found here:  \citet{drakeetal1996} investigated Einstein, EXOSAT,
and ROSAT spectra with 2T variable abundances 
and found a low temperature
component at $\sim$ 0.7 keV and another between 2--4 keV, with some variability.
The abundance Z was always subsolar, between 0.4--$\leq$0.9 solar, and variable.  
\citet{drakekashyap1998}
investigated the {\it ASCA} data presented here; however, they did not do a time-resolved spectral
analysis, and kept the abundances scaled to a common Z. Their 2T fit agrees with our results,
but their Z is $\approx$ 0.4 solar; their data, however, include quiescent and
flare intervals (as discussed in \S4.3.1). This
could account for why their Z is higher than the individual abundances derived here for
quiescent and flaring intervals. 
We mention that the abundances derived by {\it ASCA} and {\it BeppoSAX} are
in good agreement with the iron photospheric abundance of $\sim 0.2$ solar
derived for the K star by \citet{randichetal1993}. 

In more recent years, HR~1099 has been observed by {\it Chandra} and {\it XMM-Newton} 
\citep[][respectively]{drake2001,audard2001a}.  The {\it XMM-Newton} observation included a large flare.
The analysis by \citeauthor{audard2001a} of the 7--35 \AA\ {\it XMM-Newton} spectral region in
quiescence utilized a similar technique to that performed here:  They used
four discrete temperature components (compared to the two in the present analysis
of {\it ASCA} data) using the plasma code SPEX. 
The abundances derived during the quiescent interval are remarkably similar
to those found for the quiescent interval of the present analysis:  The abundance of iron
relative to the solar photospheric abundance of \citet{andersgrevesse1989} is 0.15$\pm$0.02, 
very close to our derived values from analyses of SIS0, SIS1, and GIS2 spectra of
0.12$\pm$0.04 from Table~\ref{table8}.  Abundances of other elements are similar
in the {\it ASCA} and {\it XMM-Newton} observation, but due to the differing wavelength coverage
and sensitivity, the iron abundance is the more important to compare since in principle 
it can be determined more accurately.
The large flare captured by the {\it XMM-Newton} 
observation also revealed an enhancement in the iron abundance, about a factor of three
compared to the quiescent value. 
The continuous emission measures \citeauthor{audard2001a} determined using the
coronal abundances from discrete temperature fitting imply a broad plateau of emission
between 6--30 MK during quiescence, and during the flare there is a high temperature
peak at 40--50 MK, along with a secondary peak at $\approx$ 6 MK as in quiescence.

In contrast to the varying behavior seen in the {\it XMM-Newton} observation, the {\it Chandra} HETG observation
showed no evidence for flares \citep{ayres2001}.  \citet{drake2001} analyzed the
spectrum to determine DEM and abundances.  The DEM showed a broad peak
from $\log T=$6.8--7.4, with a minimum at $\log T=$6.6 and rise towards
cooler temperatures.
They determined abundances of
O, Ne, Mg, Si, S, and Ar relative to the iron abundance, expressed
in relation to the solar photospheric abundances
of \citet{andersgrevesse1989} for the elements O, Ne, Mg, Si, S, and Ar, and 
the photospheric iron abundance of \citet{holweger1991}.  The use of a different value
for the solar photospheric iron abundance makes a cross-comparison difficult, but they
determined that the iron abundance was depleted in the coronal spectrum, by about a third
relative to the solar photospheric abundance; the coronal abundances of
O, Mg, Si, and S are slightly subsolar, whereas the noble gases Ne and Ar are enhanced
by factors of several over the solar photospheric values.

\subsection{The Neupert Effect\label{sec:neupert}}
The time delays observed between radio and EUV$/$SXR flares are in the same sense as
expected from the solar analogy, but the scale of the delays is much larger (hours compared with
minutes in the solar case).
There is an inherent difficulty in causally connecting behaviors based solely on temporal
associations.  We can take the solar case as an example, and use expectations from 
solar observations coupled with a theoretical physical connection between emitting mechanisms.

The Neupert effect is the observed correlation in solar flares between the
time-integrated radio$/$HXR light curve and the SXR light curve during the
rising phase of the latter \citep{neupert1968}; alternately, between the radio$/$HXR light curve
and the time derivative of the SXR light curve.  This observational association can be explained
using the chromospheric evaporation theory, in which the HXR emission is thick-target nonthermal
electron-ion bremsstrahlung \citep{lietal1993} and the
microwave radiation arises from gyrosynchrotron emission from electrons trapped in 
a magnetic loop.
Most of the energy contained in the
nonthermal particles is lost by Coulomb collisions with the ambient plasma in the chromosphere.  The
chromospheric material at the particle impact site undergoes a radiative instability
and can be heated to coronal temperatures (T$>$ 10$^{6}$ K) on a timescale short compared with the
hydrodynamic expansion time.  The resulting overpressure of the plasma causes it to
fill the coronal loop where it radiates thermal soft X-ray emission.
This is the ``explosive evaporation'' model of \citet{fisheretal1985}.

There have been numerous studies of the Neupert effect in solar flares, and the results are contradictory.
\citet{denniszarro1993} found that the Neupert effect relationship
between the soft X-ray time derivative and the hard X-ray emission holds for 80\% of impulsive solar flares;
events that vary more gradually show a poorer correlation.  \citet{mctiernanetal1999} demonstrated
from an analysis of the DEM during solar flares that the Neupert effect is more likely to be seen
in higher temperature plasma than at low temperatures, suggesting that the bandpasses of the soft X-ray
observations may skew  sensitivity to this effect.  
\citet{hawleyetal1995}
observed the flare star AD~Leo using {\it EUVE} and optical observations, and found a
correlation between the two suggestive of a Neupert effect.  The optical emission, which serves
as a proxy for nonthermal radiation due to electron beam impingement on the stellar photosphere,
reaches a peak during the rise of the EUV flare: The time integral of the U band
luminosity matches the evolution in EUV count rate during the rise phase.
\citet{gudeletal1996} investigated radio
and X-ray observations of flares on a dMe flare star, finding evidence for a Neupert effect;
and most
recently, \citet{gudeletal2002} found evidence of the Neupert effect in the active binary
$\sigma$~Gem.

We investigated evidence for the Neupert effect in the cases where a radio burst occurred
during the rise phase of an EUV$/$SXR flare.  
%Because of the complex dependencies of the general
%formulation of the Neupert effect on the plasma quantities $n$ and $T$, and their changes with time
%through the EUV$/$SXR flare rise, we concentrated on the simple formulation of the relationship 
%between the thermal energy content and the radio flux.  Since we are examining luminosities,
%the classical Neupert effect should be stated as $L(t)\propto \int_{t0}^{t}F_{r}(u)du$.
\citet{gudeletal1996} constructed a simple model, which we follow here, for the rate of change of the energy content of a volume $V$ of
thermal plasma at temperature $T$ and density n$_{e}$: a source of energy is provided by heating, presumed
to originate from the nonthermal energy input as diagnosed by the radio emission, and energy is lost by 
radiation.  This equation can be written \\
\begin{equation}
\frac{d}{dt} (3n_{e} kTV) = \alpha F_{R} - n_{e}^{2}V P_\lambda (T)
\end{equation}
\citep[equation 1][]{gudeletal1996} where k is Boltzmann's constant, $\alpha F_{R}$ gives the
amount of nonthermal energy deposition, and $P_\lambda (T)$ is the radiative loss function for optically thin,
high
temperature thermal plasmas
\citep{meweetal1985}.
Using a constant radiative time scale $\tau$=$\frac{3kT}{n_{e}P_\lambda (T)}$, and neglecting any initial energy content,
an analytical solution is :\\
\begin{equation}
E(t)= \alpha \int_{t_{0}}^{t} F_{R}(u)e^{-(t-u)/\tau}du
\end{equation}
and the observed luminosity $L(t)=E(t)/\tau$.  The equivalent expression for the classical Neupert effect
is \\
\begin{equation}
E(t)=\alpha \int_{t_{0}}^{t} F_{R}(u) du \;\;\; ,
\end{equation}
which is the first solution, in the limit of infinite $\tau$.  
The analyses of the data 
(Sections~\ref{moreeuv} and \ref{morexray}) have been able to constrain density and temperature,
but only in a general way; i.e., we can characterize the overall quiescent-flare 
and flare-flare changes of the plasma, but not a detailed characterization of the
plasma evolution during the flare. 
In reality, the density and temperature
will be changing during the course of the flare.
%The time sampling of the radio and
%EUV$/$SXR points is insufficient to perform a derivative-based analysis, so we investigated
%time integrals of the radio flux.
In addition, two of the large EUV flares which had accompanying radio flares during
their rise (1994 Aug. 26.5--28, 1996 Sept. 7--11) had complex morphologies during the
flare rises and decays (for 1996 Sept. 7--11 the radio coverage extended over only 
$\approx$ 10 hours, compared with the 83.6 hour duration of  the EUV flare),
making an interpretation in terms of the Neupert effect difficult.  The {\it ASCA}/{\it EUVE}
flare on 1994 August 25.6--26.0, however, has a relatively simple shape, and we investigated it
in detail.  Figure~\ref{fig:neupert} shows the flare variation of the thermal plasma in the
SXR and EUV bands, along with the radio flare.  Estimates of the quiescent emission have been subtracted
from all three light curves.  We fitted the radio light curve with a linear rise and exponential decay,
then used that function to perform the integrations discussed above. The radiative decay timescale $\tau$
of 6000 seconds was chosen to match roughly the observed soft X-ray flare decay: this timescale is
appropriate for a temperature of 10$^{7}$K and n$_{e} \sim$ 10$^{11}$ cm$^{-3}$.  The agreement with the
observed soft X-ray profile is decent, indicating that the chromospheric evaporation scenario is
applicable to stellar flares.  Table~\ref{tbl:neupert} lists the peak luminosities, integrated energies, and flare durations, for
the three EUV$/$SXR flares which showed a radio burst during the flare rise. The L$_{X}$/L$_{R}$
(and L$_{EUV}$/L$_{R}$) ratios have a similar range to those tabulated during quiescent intervals
in Table~\ref{tbl:lxlrq}, and are comparable in magnitude to the results from \citet{gudelbenz}
for quiescent X-ray and radio emission from a sample of active stars and solar flares.  This could suggest
a common heating mechanism for quiescent and flaring coronal plasma.

We also call attention to the EUV flare of 1993 September 17.4, where there was a gap in the
radio coverage but UV coverage was available from {\it IUE}.  The UV fluxes of C~IV and Mg~II
peak near the start of the EUV flare rise, in a pattern similar to that seen with several
radio flares.  Solar flare UV radiation shows a close temporal and spatial association with
hard X-ray emission \citep{cheng1999} ---
The interpretation is that the chromospheric and
transition region material is directly heated by the nonthermal particles, so that the increase
in intensity of these lines traces the nonthermal energy input.  
The close association between
UV and radio fluxes is clearer in the flare on 1994 August 27.0, where both are
seen to be rising at the start of the EUV flare rise.  

%\citet{fisheretal1985} noted from their models that
%the expansion velocity of explosively evaporated chromospheric material could not
%exceed 2.35 c$_{s}$, where c$_{s}$ is the sound speed in the evaporated material.
%If we take the characteristic temperature of the explosively evaporated chromospheric material
%to be one of the two temperatures where the DEM appears to peak, e.g., 
%$\sim$ 6--8 MK and 20--30 MK, compute the isothermal sound speed,
%and equate the time delay between
%radio and EUV$/$SXR fluxes with the duration of the material expansions,
%we obtain an upper limit to the size scale involved
%in the flare expansion.  For the flares listed in Table~\ref{tbl:neupert},
%the long time delays imply large size scales, on the order of the binary separation.

\subsection{Flare frequency\label{sec:flarec}}
Several flares were detected with multiple EUV$/$SXR telescopes, and show a general trend
in the amount of contrast between the flare peak and surrounding ``quiescent'' level:
A larger response is seen at the harder energies.
In several 
cases, the enhancement in the EUV light curve is consistent with the level of
quiescent variations, and only by examining the flare in a multi-wavelength context can hints
of flare emission in the EUV be recognized.  The salient flares are those
on 1994 August 25.76, 1996 September 4.8, 1998 September 7.9, and 1998 September 9.8.
While the SXR light curves exhibit enhancements of 2--3 over the level outside of the flare,
the EUV light curve typically shows variations at the level of 20--30\%.
This suggests that lower energy bandpasses will be insensitive to all but the most
energetic flares. Unfortunately, none of the large EUV flares were observed in the SXR;
we expect the flare contrast would have been huge.  

In \citet{ostenbrown1999}
the flare frequency was determined for a sample of active binary systems that were observed with
{\it EUVE}.  After consideration of the poor flare contrast in the EUV, it is possible that those flare
frequencies were underestimated.  The most extreme example of this is illustrated by the {\it EUVE}
observations in 1998.  The light curve of EUV variations in Figure~\ref{euvelc} shows only one moderate flare
noticeable during an observation spanning 8.7 days.  We noted in Section~\ref{moreeuvlc}
that there was a sudden shift in activity from the behavior displayed in the 1993--1996
observations and that illustrated in 1998.  Inspection of the {\it BeppoSAX} and {\it RXTE} light curves
in Figure~\ref{1998multiw1} show that there are roughly five more flares, 
in addition to the one obvious in the {\it EUVE} light curve.  Based on the {\it EUVE} light curve,
we would have concluded that the flare frequency was $\sim$ 0.1 day$^{-1}$; using the multi-wavelength
light curves, we would infer a much higher rate, $\sim$ 0.7 day$^{-1}$.
\citet{feldmanetal1995} examined a 
sample of solar flares spanning four orders of magnitude in peak emission measure, and found an almost power-law 
dependence between the peak emission measure ($\sim$ peak luminosity) and the peak flare temperature, with
a steep dependence (EM$\propto$T$^{5}$).  \citeauthor{feldmanetal1995} extrapolated the results to a few
of the largest stellar flares which had been reported in the literature.  This relation implies that larger
flares are also hotter, which would skew the flare spectral energy distribution to harder photon energies, and
explain the apparent discrepancy between the {\it EUVE} flare enhancements and the {\it RXTE}/{\it BeppoSAX}
flare enhancements.

\citet{gudeletal2003} examined the flare rate distribution on the M dwarf flare star AD~Leo using simultaneous
{\it EUVE} and {\it BeppoSAX} LECS and MECS light curves in combination with model-flare-synthesized light
curves generated assuming a power law distribution of flares with energy.  They found a systematic flattening
of the derived distributions generated using the harder energy MECS data, compared with the {\it EUVE} and LECS
detector data. They interpreted this skewing as due to the harder energy range of the MECS detector, and the
increasing peak temperatures of more luminous flares.  The softer bandpasses are more sensitive to quiescent emission.
Such a statistical treatment of flare rate distributions will be less biased than simply
counting individual large flares.  Harder X-ray bandpasses are more favorable, however, for
investigating the temperatures involved in large stellar flares.

\section{Conclusions}
There are several conclusions to be drawn from such a broad
investigation of flaring on the active binary system HR~1099.

One of the surprising results obtained from radio observations is how common
bursts at low frequencies are on HR~1099 -- occurring roughly one third
of the time. We have investigated the nature
of L-band bursts, and find that they are 
characterized by emission that can achieve large values of
percent left circular polarization ($\rightarrow$ 100\%), is highly variable (possibly varying on timescales
 less than the 10 second integration time of the VLA observations) and long-lasting
(attaining high flux levels and high degrees of left percent circular
polarization for hours). 
Due to an inability to constrain the source size of 
these bursts, only a moderate lower limit on the brightness temperature, T$_{b} \geq$
10$^{10}$K, can be determined.  Such brightness temperatures could be consistent with
incoherent gyrosynchrotron emission; coupled with large values of circular polarization,
short timescales and apparent lack of similarity with higher frequency
variations, however, it is more likely that this phenomenon is an example of a coherent
mechanism.  Determining what mechanism might be operating in the coronae of HR~1099
requires observations with better time resolution (to constrain the source size),
and larger frequency coverage (to determine the bandwidth of the emission).  

%large values of circular polarization (approaching 100\%), extremely high brightness temperatures
%($T_{b} \geq$ 10$^{16}$K), and fast variations ($\Delta t \leq$ 20 milliseconds) 
%have been observed \citep{abadasimon1997a,bastianetal1990}.  
The characteristics of X-ray and microwave emission from active binary systems 
and dMe stars share similarities.
The 
thermal coronal emission from both is characterized by 
electron temperatures T$_{e}$ of 5--30 MK, high densities (n$_{e} \approx$ 10$^{12}$ cm$^{-3}$),
and coronal abundances that are subsolar in quiescence and appear to vary during flares.
Centimeter-wavelength radio emission is generally attributed to nonthermal gyrosynchrotron radiation
outside of flares \citep{gudelbenz1996}.
The phenomena observed on HR~1099 at 20 cm reported here, and on other active binary
systems reported by others, has parallels with behavior on dMe stars: 
in both cases, there is evidence for highly polarized and time variable emission
(dMe star behavior shows evidence for fast fluctuations [$\Delta t \leq$ 20 milliseconds]
and extremely high brightness
temperatues, [$T_{b} \geq$ 10$^{16}$K]).
Constraining the observational
properties of these low-frequency bursts on active binary systems 
will reveal the extent of the apparent similarity in dynamics of
these two kinds of coronal environments.  

Radio observations show several large flares at 3.6 cm. 
The increase of flux
and decrease of polarization during 3.6 cm flares is consistent with a constant
quiescent level in intensity and polarization and a flare which increases in flux but remains
unpolarized.  The 6--3.6 cm spectral indices during the flares show an increase
during the flare rise, attaining the largest value
as the flux reaches its maximum value \citep[see also][]{richardsetal}.  This behavior is consistent with
an increase in optical depth during the flare rise, and also consistent with little or no
circular polarization during the flare; it is difficult to obtain large values of circular polarization
under optically thick conditions.  
The behavior of the radio flux spectra
illustrate the general trends of active binary gyrosynchrotron emission:  Spectral
indices are flat or slightly negative during quiescence, and show a peak between 2 and 5 GHz
during flares.
The observed polarizations always increase with frequency, regardless
of the behavior of the flux spectra with frequency \citep[as also discussed in][]{whitef1995}.  

There were no definite detections of HXR emission at energies larger than $>$ 12 keV during
or outside of flares.  Yet the radio fluxes indicate the existence (and persistence) of 
accelerated particles.  To date, there have been only a few detections of hard X-ray (20--50 kev)
emission by {\it BeppoSAX} at the peak of very strong flares
\citep{favataschmitt1999,maggioetal2000,pallavic2001,francioetal2001}. This HXR emission
however has been interpreted as thermal emission due to the very high
temperatures (T $\sim 10^8$ K) reached in these flares. Our {\it BeppoSAX}
PDS observation of HR 1099 gives only an upper limit to hard X-ray emission
which is two orders of magnitude lower than the soft X-ray emission.
%and the interpretation as to the thermal/nonthermal nature has been ambiguous.  
The fact that no HXR emission has been detected by either {\it RXTE} or {\it BeppoSAX}
is due to the fact that the temperatures reached by the observed X-ray flares
on HR~1099 are much lower than those needed (about 100 MK) for thermal hard X-ray emission; 
and secondly, to the fact that the nonthermal electrons indicated by the radio observations 
are apparently insufficient to produce detectable non-thermal hard X-ray emission (as is 
also true for the Sun where the ratio of non-thermal to thermal hard X-ray emission in impulsive
solar flares is $\geq$ 10$^{-5}$).

Flares lasting several days appear to be a common feature on active stars.  Such 
long-duration flares have now been detected not just on active binary systems 
\citep{kursterschmitt1996}, but on active evolved single giant stars \citep{ayresetal1999,ayresetal2001}
as well as higher gravity dMe flare stars \citep{cullyetal1994}.  A feature common to many of
these flares is the presence of a change in the light curve decay phase, from an initially
fast decay to a slower decay.  Several flares on HR~1099 discussed in this paper also show evidence
for such decay morphologies.  Unfortunately, the sensitivities of the EUV and SXR telescopes which 
observed these flares are not sufficient to determine the change of plasma parameters (n$_{e}$, T$_{e}$)
with time during the decay phase; usually only a gross estimate of the total flare changes with 
respect to quiescent intervals is possible.  This renders an interpretation of the underlying
cause of the observed decay change difficult:  a change to low-density structures during the late
stages of the decay could explain the long decay timescales.  Another viable explanation is
the presence of continued heating during the decay \citep[e.g.][]{favataschmitt1999}.  
A possible clue may come from studying
solar flares; a few long-duration solar events \citep[e.g.][]{feldman1995} appear to possess a break during 
the decay, similar to
the behavior seen in stellar long-duration events.

{\it EUVE} observations reveal many flares; HR~1099 was in a high state of activity for three out of the four
years it was observed during these campaigns.  There is suggestive evidence of a flare
precursor  in four large EUV flares, consisting of a slow increase in flux before
the flare rise.  The dynamic range of events exhibited by HR~1099 was less than a factor of 7
in the EUV.  Most flares in the EUV show a remarkable amount of symmetry between the flare
rise and decay, an observation that is at odds with the fast rise and
slow decay typical of chromospheric evaporation in solar flares.  An investigation of the
time-resolved spectral variations confirms the creation of hot plasma during flare intervals.
The existence of many weak lines blended with continuum radiation in the EUV hinders spectral
interpretation, particularly with regard to elemental abundances and activity-related abundance
changes.  This also limits the hottest temperatures to which EUVE spectra are sensitive.
Densities 
determined from {\it EUVE} spectra indicate high values (n$_{e}$ $\sim$ few 10$^{12}$ cm$^{-3}$)
during quiescence, but no evidence for a statistically significant enhancement during flares.

{\it ASCA} observations of a small flare indicate an enhancement of the iron abundances
by about a factor of three between quiescence and the peak of the flare.  The emission measure distribution
shows the same general structure during quiescence and flares:
A peak at 6--10 MK and another between 20 and 30 MK.  During flares, the high temperature
peak becomes more prominent and moves to hotter temperatures.  
The temperatures derived from {\it BeppoSAX} and {\it ASCA} spectra are generally consistent with
each other; however, the small flares observed with {\it BeppoSAX} did not show any evidence for
abundance changes.
The behavior in different SXR bands as investigated with {\it ASCA} and {\it RXTE} reveal faster
changes at the harder energies, indicative of hotter temperature plasma.

The multi-wavelength flare data illustrate fair correlation between
EUV$/$SXR flares and radio/UV bursts. This can
be interpreted in the light of solar flare mechanisms using the Neupert effect, 
where the time integral of the nonthermal
radiation is proportional to the rise phase of the EUV$/$SXR light curve.
The correspondence of radio and UV light curves suggests that the UV line fluxes also can be
used as a proxy for the the nonthermal radiation impinging on the chromosphere.
Three EUV$/$SXR flares showed a radio burst during the flare rise, with time 
delays between peak radio and EUV$/$SXR of 2.5--30 hours. These values are
much larger than typical values for solar flares, where the delay is on the order of
minutes.
The highly polarized emission at 20 cm shows a low degree of correlation with
gyrosynchrotron flaring activity at 3.6 cm.
The ratios of flare luminosities in the EUV and 3.6 cm bandpasses
indicate a remarkable amount of similarity with ratios derived from simultaneous
observations of quiescence in the two bandpasses.  This suggests that a similar mechanism which
forms the quiescent thermal/nonthermal radiation is also present, at some level, in producing the time-varying
flare emission.
The radio and EUV flares considered here last orders of magnitude longer than
typical solar flares, where nonthermal radio radiation lasts $\sim$ minutes, and
EUV/SXR thermal radiation lasts $\sim$ hours. Yet, the ratios of EUV-to-radio flare duration
for the three flares discussed are in agreement with the solar and stellar flares discussed by
\citet{gudeletal1996}.

There is a noticeably greater enhancement in the flare luminosity compared with
the quiescent luminosity for higher energy bandpasses.  For times when EUV and X-ray observations
are simultaneous, often flares are obvious in the X-ray light curve but only appear as 
``quiescent'' modulations in the EUV.
More luminous flares tend to be hotter, and thus the flare spectral energy distribution
will be shifted to higher energies, and involve predominantly continuum emission; 
flares will then favor the X-ray spectral regions over the EUV.
This can lead to a bias that lowers the observed flare frequency, and
affects the distribution of flares with energy.  

This work was supported by NASA grants NGT5-50241, NAG5-7020, NAG5-3226,
NAG5-2259, NSG5-4589, NAG5-2530, NAG5-7398, and NSF grant AST-0206367
to the University of Colorado. 
RAO is grateful for the support of a GSRP fellowship.  
This represents the results of VLA projects AB691, AB719, AB793, and AB874, and ATCA
projects C302, C370, and C546.
 EF, RP and GT acknowledge partial support from the Italian Space Agency
(ASI).  We gratefully acknowledge Keith Jones (University of Queensland, Australia; retired)
for his effort in obtaining the early ATCA observations, and Bryan Deeney for his contribution in
the IUE data analysis.  We thank the referee for a careful reading of this lengthy paper.

%\bibliographystyle{natbib}
%\bibliography{/users/rosten/thesis/refs1,/users/rosten/thesis/refs2,/users/rosten/thesis/refs3,/users/rosten/thesis/refs4,/users/rosten/thesis/refs5,/users/rosten/thesis/refs6,/users/rosten/thesis/refs7,/users/rosten/papers/hr1099/elena}

\clearpage
%%%%%%%%%%%%%%%%%%%%%%%%%%%%%
%PUT TABLES HERE
%Table 1 goes here
\begin{deluxetable}{lllll}
\rotate
\tablewidth{0pt}
\tablenum{1}
\tablecolumns{5}
\tablecaption{Summary of Multi-wavelength Observations \label{table1}}
\tablehead{ \colhead{Year} & \colhead{Radio} & \colhead{UV} & \colhead{EUV} & \colhead{X-ray} }
\startdata
1993 & VLA\tablenotemark{a}, ATCA\tablenotemark{b} & {\it IUE}\tablenotemark{c}, {\it HST$/$GHRS}\tablenotemark{d} & {\it EUVE}\tablenotemark{e} & \ldots \\
     & Sept 13.6--18.4, Sept. 14.5--17.9 & Sept. 16.4--19.2, Sept. 15.0--19.7 & Sept. 16.4--21.6 & \\
1994 & VLA, ATCA & {\it IUE} & {\it EUVE} & {\it ASCA}\tablenotemark{f} \\
     & Aug. 25.3--28.6, Aug. 23.7--27.0 & Aug. 23.5--28.7 & Aug. 24.0--28.0 & Aug. 25.2--26.1 \\
1996 & VLA, ATCA & \ldots & {\it EUVE} & {\it RXTE}\tablenotemark{g} \\
     & Sept. 2.3--7.7, Sept. 2.6--7.0 &   &  Sept. 1.5--11.1 & Sept. 7.6--11.2 \\
1998 & VLA, ATCA & \ldots & {\it EUVE} & {\it BeppoSAX}\tablenotemark{h}, {\it RXTE} \\
     & Sept. 7.3--11.7, Sept. 8.6--12.9 &   & Sept. 3.0--11.7 & Sept. 6.9--10.0, Sept. 7.6--11.2 \\
\enddata
\tablenotetext{a}{VLA observations are at 3.6,6,20 cm (1993) and at 2,3.6,6,20 cm (1994,1996,1998)}
\tablenotetext{b}{ATCA observations are at 3.6,6,13,20 cm} 
\tablenotetext{c}{{\it IUE} observations C~IV, Mg~II} 
\tablenotetext{d}{{\it HST$/$GHRS} observations C~IV} 
\tablenotetext{e}{{\it EUVE} 80--380 \AA\ }
\tablenotetext{f}{{\it ASCA} 0.6--10 keV} 
\tablenotetext{g}{{\it RXTE} 2--12 keV }
\tablenotetext{h}{{\it BeppoSAX} 1.5--10 keV } 
\end{deluxetable}     
%%%%%%%%%%%%%%%%%%%%%%%%%%%%%%%%%%%%%%%%%%%%%%%%%%%%%%

%Table 2
%%%%%%%%%%%%%%%%%%%%%%%%%%
\begin{deluxetable}{lll}
\tablewidth{0pt}
\tablenum{2}
\tablecolumns{3}
\tablecaption{Average 3.6 cm fluxes
for HR~1099 \label{radiotable}}
\tablehead{\colhead{Year} & \colhead{Avg. Flux} & \colhead{Avg. Pol} \\
\colhead{} & \colhead{(mJy)} & \colhead{(\%)} } 
\startdata
1993 Sept. 13.6--17.9 & 143.5$\pm$80.6 & 7.3$\pm$5.6 \\
1994 Aug. 23.7--27.0& 28.7$\pm$7.3 & 19.0$\pm$2.6 \\
1996 Sept. 2.3--7.7& 14.7$\pm$4.8 & 23.4$\pm$4.5 \\
1998 Sept. 7.3--12.9& 9.1$\pm$2.4 & 26.0$\pm$5.2 \\
\enddata
\end{deluxetable}

%Table 3
\begin{deluxetable}{lllll}
\tabletypesize{\footnotesize}
\tablewidth{0pt}
\tablenum{3}
\rotate
\tablecolumns{5}
\tablecaption{Radio Flare Properties\label{tbl:radfltbl}}
\tablehead{ \colhead{Wavelength} & \colhead{L$_{\rm pk}$} & \colhead{E$_{\rm int}$} & \colhead{t$_{p}$\tablenotemark{1}} &
 \colhead{$\Delta$t\tablenotemark{2}} \\
\colhead{(cm)} & \colhead{(erg s$^{-1}$ Hz$^{-1}$)} & \colhead{(erg Hz$^{-1}$)} & \colhead{(d)} & \colhead{(d)} }
\startdata
\multicolumn{5}{c}{\it 1993 September 16--17} \\
3.6 & 2.2 10$^{17}$ & 7.4 10$^{21}$ & 16.3561$\pm$0.0003 &  0.64 \\
6 & 1.6 10$^{17}$ & 5.7 10$^{21}$ & 16.3948$\pm$0.0005 &  0.64 \\
\multicolumn{5}{c}{\it 1994 August 25--26} \\
3.6 & 1.2 10$^{16}$ & 3.8 10$^{19}$ & 25.6618$\pm$0.0002 & 0.08 \\
\multicolumn{5}{c}{\it 1994 August 26--27} \\
3.6 & 2.6 10$^{16}$ & 1.4 10$^{20}$ & 26.56354$\pm$0.00007 & 0.15 \\
6 & 2.0 10$^{16}$ & 1.3 10$^{20}$ & 26.5658$\pm$0.0002 &  0.15 \\
20 & 1.0 10$^{16}$ & 5.8 10$^{19}$ & 26.5687$\pm$0.001 &  0.15\\
\multicolumn{5}{c}{\it 1994 August 27--28} \\
3.6 & 1.1 10$^{16}$ & 3.8 10$^{19}$ & 27.4359$\pm$0.0004 & 0.17 \\
6 & 9.6 10$^{15}$ & 4.1 10$^{19}$ &27.440$\pm$0.002 &  0.17  \\
\multicolumn{5}{c}{\it 1996 September 4--5} \\
3.6 & 3.6 10$^{15}$ & 9.8 10$^{18}$ & 4.3759$\pm$0.0006 & 0.09 \\
6 & 3.8 10$^{15}$ & 1.8 10$^{19}$ &4.380$\pm$0.002 &  0.09\\
\multicolumn{5}{c}{\it 1996 September 7a} \\
3.6 & 1.9 10$^{16}$ & 1.1 10$^{20}$ & 7.3630$\pm$0.0002 &  0.13\\
6 & 5.2 10$^{15}$ & 2.8 10$^{19}$ & 7.360$\pm$0.001 &  0.13\\
\multicolumn{5}{c}{\it 1996 September 7b} \\
3.6 & 5.5 10$^{15}$ & 2.9 10$^{19}$ & 7.4738$\pm$0.0008  & 0.09\\
\multicolumn{5}{c}{\it 1998 September 11--12} \\
3.6 & 1.8 10$^{15}$ & 7.6 10$^{18}$ & 11.510$\pm$0.002 &  0.10 \\
\enddata
\tablenotetext{1}{t$_{p}$: time of flare peak from profile fitting}
\tablenotetext{2}{$\Delta$t: flare duration}
\end{deluxetable}

%Table 3
\begin{deluxetable}{lll}
\tablewidth{0pt}
\tablenum{4}
\tablecolumns{3}
\tablecaption{Correlations between 20 cm and 3.6 cm flares \label{tbl:20cm3cmtbl}}
\tablehead{ \colhead{Day} & \colhead{20 cm behavior} & \colhead{3.6 cm behavior} } 
\startdata
1994 Aug. 24--25 & flux $\Uparrow$, large LCP & no variation \\
1994 Aug. 26--27 & no change in flux, & moderate flare \\
                 & moderate LCP & \\
1994 Aug. 27--28 & flux $\Uparrow$, large  LCP & small flare \\
1994 Aug. 28--29 & flux $\Uparrow$, large LCP & no variation \\
1996 Sept. 2--3 & flux $\Uparrow$, moderate LCP & no variation \\
1996 Sept. 3--4 & no change in flux, & small enhancement \\
               & moderate LCP & \\
1996 Sept. 4--5 & flux $\Downarrow$, moderate LCP & small flare \\
1996 Sept. 7--8 & no variation & 2 moderately large flares \\
1998 Sept. 8--9 & flux $\Uparrow$, large LCP & slow modulation \\
1998 Sept. 9--10 & no change in flux, & no variation \\
                 & moderate LCP & \\
1998 Sept. 10--11 & flux $\Uparrow$, large LCP & slow increase in flux \\
1998 Sept. 11--12 & flux $\Uparrow$, large LCP & no variation \\
\enddata
\end{deluxetable}

%Table 4
\begin{deluxetable}{llccccc}
\tablewidth{0pt}
\tablenum{5}
\tablecolumns{7}
\tablecaption{Radio spectral indices for HR~1099 \label{tbl:specindex}}
\tablehead{ \colhead{Year} & \colhead{Day} & \colhead{$\alpha_{3.6-2}$} & \colhead{$\alpha_{6-3.6}$} & \colhead{$\alpha_{13-6}$} 
& \colhead{$\alpha_{20-13}$} & \colhead{$\alpha_{20-6}$} } 
\startdata
            \multicolumn{7}{c}{Quiescence} \\
\cline{1-7}
1993 & Sept. 17.7 & \ldots & $-0.40$ & 0.30 & 0.90 &\ldots \\
1994 & Aug. 25.4 & $-0.46\pm0.06$ & $-0.48\pm0.02$ & \ldots & \ldots & $-0.04\pm0.01$ \\
1994 & Aug. 26.4 & $-0.45\pm0.08$ & $-0.34\pm0.03$ & \ldots & \ldots & $0.01\pm0.02$ \\
1996 & Sept. 2.6 & $-0.74\pm0.06$ & $-0.46\pm0.07$ & $-0.08\pm0.06$ & $0\pm0.08$ & \ldots \\
1998 & Sept. 7.3 & $-0.43\pm0.09$ & $-0.49\pm0.03$ & \ldots & \ldots & $0\pm0.02$ \\
1998 & Sept. 7.6 & $-0.16\pm0.08$ & $-0.31\pm0.06$ & \ldots & \ldots & $-0.07\pm0.03$ \\
1998 & Sept. 9.3 & $-0.84\pm0.12$ & $-0.64\pm0.03$ & \ldots & \ldots & $-0.30\pm0.04$ \\
1998 & Sept. 10.3 & $-2.32\pm0.39$ & $-0.45\pm0.09$ & \ldots & \ldots & $-0.49\pm0.04$ \\
1998 & Sept. 11.3 & $-0.66\pm0.10$ & $-0.40\pm0.04$ & \ldots & \ldots & $-0.32\pm0.03$ \\
\cline{1-7}
                  \multicolumn{7}{c}{Flare decays$/$peaks} \\
\cline{1-7}
1993 & Sept. 14.6 & \ldots & 0.16 & 0.64 & 1.51 & \ldots \\
1993 & Sept. 15.9 & \ldots & $-0.25$ & 0.28 & 1.29 & \ldots \\
1993 & Sept. 16.6 & \ldots & $-0.11$ & 0.49 & 1.09 & \ldots \\
1994 & Aug. 25.6 & $-0.36\pm0.07$ & $0.08\pm0.06$ & $-0.23\pm0.05$ & $0.21\pm0.07$ & \ldots \\
1994 & Aug. 26.6 & $-0.48\pm0.06$ & $-0.24\pm0.04$ & $0.06\pm0.04$ & $0.77\pm0.08$ & \ldots \\
%\cline{1-7}
    %&           & \multicolumn{5}{c}{20 cm polarized bursts} \\
%\cline{1-7}
%1994 & Aug. 27.4 & $-0.75\pm0.13$ & $-0.32\pm0.02$ & \ldots & \ldots & $-0.39\pm0.01$ \\
%1994 & Aug. 27.6 & $-0.74\pm0.18$ & $-0.44\pm0.04$ & \ldots & \ldots & $-0.36\pm0.02$ \\
%1994 & Aug. 28.4 & $-0.29\pm0.11$ & $-0.60\pm0.04$ & \ldots & \ldots & $-0.52\pm0.02$ \\
%1996 & Sept. 2.3 & $-0.93\pm0.08$ &$-0.48\pm0.08$ & \ldots & \ldots & $-0.28\pm0.04$ \\
%1996 & Sept. 3.3 & $-0.67\pm0.05$ & $-0.16\pm0.02$ & \ldots & \ldots & $-0.24\pm0.02$ \\
%1996 & Sept. 3.6 & $-0.90\pm0.07$ & $-0.36\pm0.03$ & $-0.31\pm0.03$ & $-0.19\pm0.07$ & \ldots \\
%1998 & Sept. 8.3 & $-0.60\pm0.08$ & $-0.88\pm0.03$ & \ldots & \ldots & $-0.24\pm0.02$ \\
%1998 & Sept. 8.6 & $-0.46\pm0.08$ & $-0.43\pm0.04$ & $0\pm0.05$ & $-0.61\pm0.08$ & \ldots \\
%1998 & Sept. 10.6 & $-0.55\pm0.12$ & $-0.36\pm0.07$ & $-0.93\pm0.05$ & $-0.03\pm0.09$ & \ldots \\
%1998 & Sept. 11.6 & $-0.36\pm0.10$ & $-0.32\pm0.07$ & $-0.53\pm0.05$ & $-1.11\pm0.09$ & \ldots \\
\enddata
\end{deluxetable}

%Table 5
\begin{deluxetable}{lll}
\tablewidth{0pt}
\tablenum{6}
\tablecolumns{3}
\tablecaption{Summary of {\it EUVE} Exposure Times \label{table2}\tablenotemark{1}}
\tablehead{ \colhead{Time Interval} & \colhead{SW} & \colhead{MW} \\
\colhead{} & \colhead{(s)} & \colhead{(s)} }
\startdata
1993 {\bf Q} & 16524.5 & 16064.3 \\
1993 {\bf F1} & 24453.7 & 22652. \\
1993 {\bf F2} & 20427.6 & 19229.6 \\
1994 {\bf Q} & 5687.0\tablenotemark{a} & 5678.7\tablenotemark{a} \\
1994 {\bf F} &49484.8 &49255.1  \\
1996 {\bf Q} &43742.4  &40766.8  \\
1996 {\bf F} &44229.0  &41090.9  \\
1996 {\bf F1} &67882.2  &66570.9  \\
1996 {\bf F2} & 36778.7 & 36596.3 \\
1998 {\bf Q} & 141476. & 118332.9 \\
1998 {\bf F} & 12457.\tablenotemark{a} & 10547.3\tablenotemark{a} \\
sumQ & 402273. & 370690. \\
sumF & 266539. & 257924. \\
\enddata
\tablenotetext{1}{See Section~\ref{sec:ch5euvespec} for explication of how time intervals were determined; Figure~\ref{euvelc} depicts time intervals.}
\tablenotetext{a}{Spectrum was not detected.} 
\end{deluxetable}
%%%%%%%%%%%%%%%%%%%%%%%%%

%Table 6
\begin{deluxetable}{lllll}
\rotate
\tablewidth{0pt}
\tablenum{7}
\tablecolumns{5}
\tablecaption{
Comparison of continuum effect on measured line fluxes \label{tbl:compare_cont}}
\tablehead{ \colhead{}
                          & \multicolumn{2}{c}{ composite quiescent spectrum} & 
\multicolumn{2}{c}{composite flare spectrum} \\
\colhead{Fe$/$H $/$(Fe$/$H)$_{\odot}$ } & \colhead{Fe~XVIII $\lambda$ 93.92\tablenotemark{a} } 
& \colhead{Fe~XXII $\lambda$ 135.78\tablenotemark{a} } 
& \colhead{Fe~XVIII $\lambda$ 93.92\tablenotemark{a}} & \colhead{Fe~XXII $\lambda$ 135.78\tablenotemark{a} } }
\startdata
1.0 &0.94 & 0.88 & 0.91 & 0.92 \\
0.9 & 0.93 & 0.88 & 0.90 & 0.92 \\
0.8 & 0.92 & 0.87 & 0.89 & 0.91 \\
0.7 &  0.92 & 0.87 & 0.87 & 0.91\\
0.6 &  0.90 & 0.86 & 0.85 & 0.90\\
0.5 &  0.89 & 0.86 & 0.83 & 0.89\\
0.4 &  0.86 & 0.84 & 0.78 & 0.88\\
0.3 &  0.81 & 0.83 & 0.72 & 0.86\\
0.2 &  0.69 & 0.80 & 0.58 & 0.81\\
0.1 &  0.46 & 0.71 & 0.16 & 0.67\\
\enddata
\tablenotetext{a}{ Measured emission line flux with continuum subtracted at 
given abundance 
divided by emission line flux with no continuum subtraction} 
\end{deluxetable}

\clearpage
\thispagestyle{empty}
%%%%%%%%%%%%%%%%%%%%%%%%%%%%%%
%Table 7
\begin{deluxetable}{lllllllllll}
\tabletypesize{\tiny}
\rotate
\tablewidth{0pt}
\tablenum{8}
\tablecolumns{10}
\tablecaption{{\it EUVE} Line Fluxes for HR~1099 \label{tbl:euvefluxtbl}} 
\tablehead{ \colhead{\small Line} & \colhead{\small  {\bf 1993Q}} &\colhead{\small {\bf 1993F1}} &
\colhead{\small {\bf 1993F2}} &\colhead{\small {\bf 1994F}} &\colhead{\small {\bf 1996Q}} &\colhead{\small {\bf 1996F}} &
\colhead{\small {\bf 1996F1}} & \colhead{\small {\bf 1996F2}} & \colhead{\small {\bf 1998Q}} \\
\colhead{} &   \multicolumn{9}{c}{\small (10$^{-13}$ erg cm$^{-2}$ s$^{-1}$) } } 
\startdata
\small Fe~XVIII $\lambda$93.92 & \small2.00$\pm$0.33 & \small 2.32$\pm$0.44 & \small 1.53$\pm$0.42 & \small 1.31$\pm$0.26 & \small 1.01$\pm$0.23 &
\small 0.99$\pm$0.22 & \small 0.87$\pm$0.19 & \small 1.83$\pm$0.34 & \small 1.09$\pm$0.12   \\
\small Fe~XXI $\lambda$97.88 & \small $<$0.7 & \small $<$1.1 & \small $<$1.1 & \small $<$0.6 & \small $<$0.5 &\small  $<$0.6 &\small  0.61$\pm$0.18 &\small  0.90$\pm$0.28 &
\small 0.64$\pm$0.10 \\
\small Fe~XIX $\lambda$101.55 &\small  $<$0.7 &\small  1.28$\pm$0.36 &\small  $<$1.0 &\small  0.58$\pm$0.17 &\small  0.59$\pm$0.18 &\small  $<$0.5 &\small  0.81$\pm$0.16 & 
\small 1.28$\pm$0.29 &\small  0.40$\pm$0.08 \\
\small Fe~XXI $\lambda$102.22 &\small  1.20$\pm$0.26 &\small  1.51$\pm$0.39 &\small  1.26$\pm$0.36 &\small  0.82$\pm$0.18 &\small  0.82$\pm$0.19 &
\small $<$0.5& 1.02$\pm$0.17 &\small  1.91$\pm$0.31 & \small 0.67$\pm$0.09 \\
\small Fe~XVIII $\lambda$103.937 &\small  0.65$\pm$0.21 &\small  0.98$\pm$0.32 &\small  $<$1.0 &\small  $<$0.52 &\small  0.48$\pm$0.15 &\small  
0.69$\pm$0.19 &\small  1.00$\pm$0.17 &\small  1.30$\pm$0.28 & \small 0.49$\pm$0.08 \\
\small Fe~XIX $\lambda$108.37 &\small  1.03$\pm$0.23 &\small  2.53$\pm$0.44 & \small 1.16$\pm$0.38 &\small  1.48$\pm$0.22 &\small  0.58$\pm$0.18 &\small  
0.96$\pm$0.20 &\small  1.37$\pm$0.18 &\small  1.46$\pm$0.28 &\small  0.67$\pm$0.10 \\
\small Fe~XIX $\lambda$109.97 &\small  $<$0.6 &\small  $<$1.0 &\small  $<$1.0 &\small  $<$0.4 &\small  $<$0.5 &\small  $<$0.4 &\small  0.60$\pm$0.14 &
\small 1.01$\pm$0.27 &\small  0.34$\pm$0.08 \\
\small Fe~XIX $\lambda$111.70 &\small  $<$0.6 & \small 1.00$\pm$0.31 &\small  $<$1.0 &\small  0.76$\pm$0.18 &\small  $<$0.4 &\small  $<$0.4 &\small  0.49$\pm$0.14 &
\small 1.31$\pm$0.26 &\small  0.32$\pm$0.07 \\
\small Fe~XXII $\lambda$114.41 &\small  $<$0.6 &\small  1.18$\pm$0.33 &\small  $<$1.1 &\small  $<$0.5 &\small  $<$0.5 &\small  $<$0.5 &\small  0.87$\pm$0.16 &
\small 1.49$\pm$0.29 &\small  0.29$\pm$0.08 \\
\small Fe~XXII $\lambda$117.17 &\small  2.03$\pm$0.31 &\small  3.24$\pm$0.48 &\small  2.53$\pm$0.48 &\small  2.03$\pm$0.27 &\small  1.30$\pm$0.22 &
\small 1.78$\pm$0.24 &\small  2.09$\pm$0.22 &\small  3.35$\pm$0.36 &\small  0.93$\pm$0.11\\
\small Fe~XX $\lambda$118.66 &\small  0.72$\pm$0.24 & \small 1.19$\pm$0.37 &\small  $<$1.1 &\small  0.80$\pm$0.20 &\small  0.56$\pm$0.18 &
\small 0.78$\pm$0.19 &\small  1.25$\pm$0.17 &\small  1.01$\pm$0.25 &\small  0.54$\pm$0.09 \\
\small Fe~XIX $\lambda$120.00 &\small  $<$0.7 &\small  1.34$\pm$0.38 &\small  $<$1.0 &\small  $<$0.5 &\small  $<$0.5 &\small  $<$0.5 &\small  0.61$\pm$0.15 &
\small $<$0.7 & $<$0.2 \\
\small Fe~XX $\lambda$121.83 & \small 0.92$\pm$0.26 &\small  1.43$\pm$0.40 &\small  $<$1.2 &\small  0.72$\pm$0.19 &\small  0.90$\pm$0.20 &\small  
1.35$\pm$0.23 &\small  1.05$\pm$0.18 &\small  1.64$\pm$0.30 &\small  0.73$\pm$0.09 \\
\small Fe~XXI $\lambda$128.73 &\small  2.54$\pm$0.36 &\small  3.02$\pm$0.52 &\small  2.49$\pm$0.54 &\small  1.98$\pm$0.26 &\small  1.55$\pm$0.25 &
\small 1.97$\pm$0.28 &\small  2.16$\pm$0.22 &\small  3.20$\pm$0.38 & \small 1.41$\pm$0.13 \\
\small Fe~XX,XXIII $\lambda$132.85 &\small  6.13$\pm$0.60 &\small  19.10$\pm$1.14 &\small  12.2$\pm$1.03 &\small  5.35$\pm$0.45 &\small  5.70$\pm$0.48 &\small  
5.86$\pm$0.51 &\small  6.31$\pm$0.41 & \small 11.90$\pm$0.74 & \small 4.73$\pm$0.24 \\
\small Fe~XXII $\lambda$135.78 &\small  1.04$\pm$0.34 & \small 3.98$\pm$0.59 &\small  2.35$\pm$0.60 &\small  1.68$\pm$0.30 &\small  1.36$\pm$0.27 &
\small 1.18$\pm$0.26 &\small  1.42$\pm$0.24 &\small  2.08$\pm$0.37 & \small 0.97$\pm$0.13 \\
\small Fe~XXIV $\lambda$192.02 &\small  4.34$\pm$0.92 &\small  11.30$\pm$1.58 &\small  8.64$\pm$1.57 &\small  5.02$\pm$0.67 &\small  3.85$\pm$0.66 &
\small 3.32$\pm$0.66 &\small  6.27$\pm$0.62 &\small  7.28$\pm$0.90 &\small  2.33$\pm$0.36 \\
\small Fe~XXIV $\lambda$255.09 &\small  $<$3.0 &\small  7.16$\pm$1.51 &\small  6.82$\pm$1.76 &\small  3.88$\pm$0.80 &\small  $<$2.5 &\small  3.89$\pm$0.84 &\small  
3.47$\pm$0.64 &\small  6.10$\pm$1.09 &\small  1.40$\pm$0.43 \\
\small Fe~XV $\lambda$284.16 & \small $<$2.7 &\small  $<$3.4 &\small  5.10$\pm$1.53 &\small  $<$1.7 &\small  $<$2.0 &\small  $<$2.0 &\small  1.41$\pm$0.44 &\small  
2.38$\pm$0.74 &\small  1.10$\pm$0.37 \\
\small Fe~XVI $\lambda$335.41 &\small  $<$2.6 & \small $<$3.4 &\small  $<$4.3 &\small  $<$1.8 &\small  $<$1.9 &\small  $<$1.9 &\small  1.56$\pm$0.46 &\small  2.56$\pm$0.71 &
\small $<$1.1 \\
\small Fe~XVI $\lambda$360.76 &\small  $<$2.4 & \small $<$3.2 & \small $<$4.1 &\small  $<$1.7 &\small  $<$1.9 & \small $<$1.8 &\small  $<$1.3 &\small  $<$2.1 &\small  1.11$\pm$0.34 \\
\enddata
\end{deluxetable}

\clearpage

%%%%%%%%%%%%%%%%%%%%%%%%%%%%%%
%Table 8
\begin{deluxetable}{llll}
\tablewidth{0pt}
\tablenum{9}
\tablecolumns{4}
\tablecaption{ {\it EUVE} Line Fluxes from composite quiescent and flare spectra \label{table7}} 
\tablehead{\colhead{Line} & \colhead{$\lambda$} &\colhead{   {\bf sumQ}}  & \colhead{{\bf sumF}} \\
\colhead{} & \colhead{(\AA)} &  \multicolumn{2}{c}{(10$^{-13}$ erg cm$^{-2}$ s$^{-1}$)}  }
\startdata
Fe~XVIII & 93.92 & 1.19$\pm$0.97 & 1.27$\pm$0.10 \\
Fe~XXI & 97.88 & 0.59$\pm$0.06 & 0.60$\pm$0.09\\
Fe~XIX & 101.55 & 0.61$\pm$0.06 & 0.73$\pm$0.08\\
Fe~XXI & 102.22 & 0.75$\pm$0.06 & 1.00$\pm$0.08\\
Fe~XVIII & 103.94 & 0.51$\pm$0.05 & 0.86$\pm$0.08\\
Fe~XIX & 108.37 & 0.84$\pm$0.06 & 1.20$\pm$0.08\\
Fe~XIX & 109.97 & 0.36$\pm$0.05 & 0.46$\pm$0.07\\
Fe~XIX & 111.70 & 0.35$\pm$0.05 & 0.59$\pm$0.07\\
Fe~XXII & 114.41 & 0.48$\pm$0.05 & 0.76$\pm$0.08\\
Fe~XXII & 117.17 & 1.53$\pm$0.08 & 2.13$\pm$0.11\\
Fe~XX & 118.66 & 0.61$\pm$0.06 & 0.90$\pm$0.08 \\
Fe~XIX & 120.00 & 0.34$\pm$0.05 & 0.42$\pm$0.07\\
Fe~XX & 121.83 & 0.95$\pm$0.07 & 1.09$\pm$0.09\\
Fe~XXI & 128.73 & 1.74$\pm$0.09 & 2.07$\pm$0.11\\
Fe~XX,XXIII & 132.85 & 5.41$\pm$0.16 & 6.82$\pm$0.22\\
Fe~XXII & 135.78 & 1.48$\pm$0.09 & 1.68$\pm$0.12\\
Fe~XXIV & 192.02 & 3.50$\pm$0.22 & 6.73$\pm$0.34\\
Fe~XXIV & 255.09 &  2.21$\pm$0.25 & 4.66$\pm$0.35\\
Fe~XV & 284.16 & 1.30$\pm$0.21 & 2.26$\pm$0.24 \\
Fe~XVI & 335.41 &1.21$\pm$0.21 & 2.11$\pm$0.24 \\
Fe~XVI & 360.76 & 0.95$\pm$0.20 & 1.25$\pm$0.22\\
\enddata
\end{deluxetable}

\clearpage

%Table 9
\begin{deluxetable}{ccccc}
\rotate
\tablewidth{0pt}
\tablenum{10}
\tablecolumns{5}
\tablecaption{Electron densities
in {\it EUVE} spectra
from Fe~XXI and Fe~XXII line ratios \label{tbl:euvdens}}
\tablehead{
 \colhead{Time interval}  & \multicolumn{2}{c}{Fe~XXI 128.73$/$102.22} & \multicolumn{2}{c}{Fe~XXII 117.17$/$114.41} \\
\colhead{}& \colhead{ratio$\pm$ 1$\sigma$ error} & \colhead{n$_{e}$ (cm$^{-3}$) }& \colhead{ratio$\pm$ 1$\sigma$ error} & \colhead{n$_{e}$ (cm$^{-3}$)} }
\startdata
1993Q & 2.1$\pm$0.5 & 3.0$\times$10$^{12}$ (1.5$\times$10$^{12}$ -- 6$\times$ 10$^{12}$) & 4.0$\pm$1.8 & \ldots \\
1993F1 & 2.0$\pm$0.6 & 3.2$\times$10$^{12}$ (1.6$\times$10$^{12}$--8.7$\times$10$^{12}$)& 2.7$\pm$0.9 & $>$4.7 10$^{12}$ \\
1993F2 & 2.0$\pm$0.7 & 3.2$\times$$^{12}$(1.5$\times$10$^{12}$ -- 10$^{13}$) & 3.5$\pm$1.9 & 5.1$\times$10$^{12}$(6.8$\times$10$^{11}$--10$^{13}$) \\
1994F &2.4$\pm$0.6 & 2.0$\times$10$^{12}$(10$^{12}$ -- 4.3$\times$10$^{12}$) & 4.7$\pm$1.9 & \ldots \\
1996Q & 1.9$\pm$0.5 & 3.7$\times$10$^{12}$(1.9$\times$10$^{12}$--9.1$\times$10$^{12}$) & 2.8$\pm$1.1 & $>$3.4 10$^{12}$ \\
1996F & 5.1$\pm$2.3 & $<$1.4$\times$10$^{12}$ & 4.0$\pm$1.6 & \ldots \\
1996F1 & 2.1$\pm$0.4 & 2.8$\times$10$^{12}$ (1.7$\times$10$^{12}$--4.9$\times$10$^{12}$)& 2.4$\pm$0.5 & \ldots \\
1996F2 & 1.7$\pm$0.3 & 5.1$\times$10$^{12}$ (3.2$\times$10$^{12}$--9.3$\times$10$^{12}$) & 2.2$\pm$0.5 & \ldots \\
1998 & 2.1$\pm$0.3 & 3.0$\times$10$^{12}$ (1.9$\times$10$^{12}$ -- 4.5$\times$10$^{12}$)& 3.2$\pm$0.9 & 7.2$\times$10$^{12}$ (2.8$\times$10$^{12}$--10$^{13}$) \\
sumF & 2.1$\pm$0.2 & 3.0$\times$10$^{12}$ (2.3$\times$10$^{12}$--3.8$\times$10$^{12}$)& 2.8$\pm$0.3 & $>$8.1 10$^{12}$ \\
sumQ & 2.3$\pm$0.2 & 2.2$\times$10$^{12}$ (1.8$\times$10$^{12}$--2.8$\times$10$^{12}$)& 3.2$\pm$0.4 & 7.4$\times$10$^{12}$(4.9$\times$10$^{12}$--10$^{13}$) \\
\enddata
\end{deluxetable}

%Table 10
\begin{deluxetable}{llllll}
\tablewidth{0pt}
\tablenum{11}
\tablecolumns{6}
\tablecaption{Luminosities and timescales
during the
1994 {\it ASCA} observation \label{tbl:ascaflux}}
\tablehead{ \colhead{Energy Range} & \colhead{{\bf Q}\tablenotemark{1}} & \colhead{{\bf R/Q}} & \colhead{{\bf P/Q}} & \colhead{{\bf D1/Q}} & 
\colhead{{\bf D2/Q}} \\
\colhead{(keV)} &\colhead{   (ergs s$^{-1}$)} &\colhead{}         & \colhead{}         & \colhead{}           &  \colhead{} } 
\startdata
0.6--2 & 5.2 10$^{30}$ & 1.4 & 1.6 & 1.3 & 1.2 \\
2--5 & 1.9 10$^{30}$ & 2.2 & 2.0 & 1.4 & 1.1 \\
5--10 & 4.4 10$^{29}$ & 3.6 & 2.5 & 1.5 & 1.2  \\
0.6--10 & 7.5 10$^{30}$ & 1.7 & 1.7 & 1.3 & 1.1 \\
\cline{1-6}
      &     \multicolumn{2}{c}{$\tau_{r}$ (hrs)} & \multicolumn{3}{c}{$\tau_{d}$ (hrs)} \\
\cline{1-6} 
0.6--2 & \multicolumn{2}{c}{4.3$\pm$0.7} & \multicolumn{3}{c}{8.4$\pm$2.0}  \\
2--5 & \multicolumn{2}{c}{2.2$\pm$0.3} & \multicolumn{3}{c}{4.8$\pm$1.5} \\
5--10 & \multicolumn{2}{c}{0.9$\pm$0.2} & \multicolumn{3}{c}{2.8$\pm$1.3} \\
0.6--10 & \multicolumn{2}{c}{3.8$\pm$0.5} & \multicolumn{3}{c}{7.6$\pm$1.6} \\
\enddata
\tablenotetext{1}{N$_{H}$=1.35 10$^{18}$ cm$^{-2}$ was used to estimate the luminosities.}
\end{deluxetable}

%Table 11
\begin{deluxetable}{lrrrrrr}
\tablewidth{0pt}
\tablenum{12}
\tablecolumns{6}
\tablecaption{VMEKAL 2-Temperature spectral fits to flaring$/$quiescent {\it ASCA} spectra \label{table8}} 
\tablehead{
\colhead{}  & \colhead{{\bf Q}\tablenotemark{1}} &  \colhead{{\bf R}\tablenotemark{1}} &  \colhead{{\bf P}\tablenotemark{1}} &  \colhead{{\bf D1}\tablenotemark{1}} &  \colhead{{\bf D2}\tablenotemark{1}} }
\startdata
 & \multicolumn{5}{c}{-- SIS0 0.6 - 10 keV --} \\
\cline{2-6} \\
kT$_{1}$ (keV) & 0.7$^{+0.02}_{-0.03}$ & 0.72$\pm$0.06 & 0.67$\pm$0.05 & 0.65$^{+0.04}_{-0.06}$ & 0.71$\pm$0.05 \\
kT$_{2}$ (keV) & 2.21$^{+0.13}_{-0.09}$ & 2.86$^{+0.19}_{-0.18}$ & 2.39$^{+0.15}_{-0.12}$ & 2.21$^{+0.11}_{-0.10}$ & 2.46$^{+0.50}_{-0.43}$ \\
$\log$ EM$_{1}$ (cm$^{-3}$) & 53.54$^{+0.12}_{-0.09}$ & 53.23$^{+0.14}_{-0.12}$ & 53.42$^{+0.10}_{-0.12}$ & 53.42$^{+0.12}_{-0.10}$ & 53.84$^{+0.16}_{-0.39}$ \\
$\log$ EM$_{2}$ (cm$^{-3}$) & 53.74$^{+0.02}_{-0.03}$ & 53.97$\pm$0.04 & 54.05$\pm$0.04 & 53.87$\pm$0.03 & 53.74$\pm$0.10 \\
O & 0.2$\pm$0.1 & 0.54$^{+0.36}_{-0.29}$ & 0$^{+0.19}_{-0}$ & 0.27$^{+0.19}_{-0.15}$ & 0.13$^{+0.32}_{-0.11}$ \\
Ne & 0.97$^{+0.19}_{-0.20}$ & 1.57$^{+0.57}_{-0.51}$ & 1.16$^{+0.37}_{-0.30}$ & 1.59$^{+0.35}_{-0.30}$ & 0.73$^{+0.64}_{-0.22}$ \\
Mg & 0.27$^{+0.11}_{-0.09}$ & 0.25$^{+0.39}_{-0.25}$ & 0.26$^{+0.28}_{-0.23}$ & 0.51$^{+0.24}_{-0.20}$ & 0.17$^{+0.31}_{-0.10}$ \\
Si & 0.23$\pm$0.07 & 0.68$^{+0.35}_{-0.29}$ & 0.33$^{+0.22}_{-0.17}$ & 0.40$^{+0.17}_{-0.14}$ & 0.15$^{+0.14}_{-0.07}$ \\
S & 0.27$\pm$0.10 & 0.25$^{+0.35}_{-0.25}$ & 0.22$^{+0.240}_{-0.216}$ & 0.34$^{+0.20}_{-0.18}$ & 0.21$^{+0.18}_{-0.17}$ \\
Ar & 0.44$^{+0.31}_{-0.30}$ & 1.27$^{+0.94}_{-0.90}$ & 0$^{+0.23}_{-0}$ & 0.88$^{+0.54}_{-0.51}$ & 0.31$^{+0.65}_{-0.31}$ \\
Ca & 0.56$^{+0.45}_{-0.44}$ & 0.73$^{+1.05}_{-0.73}$ & 0.92$^{+0.85}_{-0.84}$ & 0.12$^{+0.70}_{-0.12}$ & 0.002$^{+0.628}_{-0.002}$ \\
Fe & 0.12$\pm$0.04 & 0.33$\pm$0.11 & 0.22$^{+0.09}_{-0.06}$ & 0.20$\pm$0.06 & 0.05$^{+0.12}_{-0.02}$ \\
Ni & 1.14$^{+0.40}_{-0.35}$ & 1.88$^{+1.57}_{-1.32}$ & 1.48$^{+1.11}_{-0.90}$ & 1.12$^{+0.80}_{-0.84}$ & 0.75$^{+0.75}_{-0.39}$ \\
$\chi^{2}/$dof & 156.2$/$160 & 143.8$/$128 & 91.0$/$121 & 122.5$/$130 & 130.6$/$121 \\
\cline{2-6}
 & \multicolumn{5}{c}{-- SIS0 + SIS1 0.6 - 10 keV --} \\
\cline{2-6} \\
kT$_{1}$ (keV) & 0.70$\pm$0.02 & 0.71$\pm$0.05 & 0.73$^{+0.05}_{-0.04}$ & 0.66$\pm$0.03 & 0.73$^{+0.04}_{-0.03}$ \\
kT$_{2}$ (keV) & 2.22$^{+0.11}_{-0.07}$ & 2.82$^{+0.15}_{-0.13}$ & 2.44$^{+0.13}_{-0.10}$ & 2.20$^{+0.09}_{-0.08}$ & 2.20$^{+0.38}_{-0.11}$ \\
$\log $EM$_{1}$ (cm$^{-3}$) & 53.55$^{+0.09}_{-0.08}$ & 53.26$^{+0.11}_{-0.09}$ & 53.48$^{+0.14}_{-0.10}$ & 53.40$^{+0.08}_{-0.07}$ & 53.58$^{+0.25}_{-0.11}$ \\
$\log$ EM$_{2}$ (cm$^{-3}$) & 53.73$\pm$0.02 & 53.97$\pm$0.03 & 54.05$\pm$0.04 & 53.87$\pm$0.02 & 53.78$^{+0.03}_{-0.09}$ \\
O & 0.21$\pm$0.08 & 0.47$^{+0.25}_{-0.21}$ & 0.12$^{+0.16}_{-0.12}$ & 0.27$^{+0.14}_{-0.11}$ & 0.30$^{+0.14}_{-0.18}$ \\
Ne & 1.04$^{+0.15}_{-0.16}$ & 1.58$^{+0.42}_{-0.37}$ & 1.10$^{+0.29}_{-0.27}$ & 1.60$^{+0.26}_{-0.24}$ & 1.21$^{+0.24}_{-0.41}$ \\
Mg & 0.29$^{+0.09}_{-0.07}$ & 0.30$^{+0.28}_{-0.26}$ & 0.36$^{+0.21}_{-0.19}$ & 0.53$^{+0.17}_{-0.16}$ & 0.32$^{+0.14}_{-0.17}$ \\
Si & 0.26$^{+0.05}_{-0.06}$ & 0.62$^{+0.25}_{-0.21}$ & 0.25$^{+0.14}_{-0.13}$ & 0.40$^{+0.12}_{-0.11}$ & 0.22$^{+0.10}_{-0.08}$ \\
S & 0.31$\pm$0.08 & 0.24$^{+0.25}_{-0.24}$ & 0.25$^{+0.18}_{-0.17}$ & 0.23$^{+0.14}_{-0.13}$ & 0.29$^{+0.14}_{-0.13}$ \\
Ar & 0.48$^{+0.23}_{-0.24}$ & 0.84$^{+0.64}_{-0.67}$ & 0$^{+0.23}_{-0}$ & 0.45$^{+0.38}_{-0.37}$ & 0.28$^{+0.41}_{-0.28}$ \\
Ca & 0.60$\pm$0.34 & 0.98$^{+0.80}_{-0.84}$ & 0.84$^{+0.65}_{-0.66}$ & 0.31$^{+0.54}_{-0.31}$ & 0.009$^{+0.481}_{-0.009}$ \\
Fe & 0.12$\pm$0.03 & 0.30$\pm$0.08 & 0.20$^{+0.07}_{-0.06}$ & 0.21$\pm$0.05 & 0.14$^{+0.05}_{-0.08}$ \\
Ni & 1.05$^{+0.28}_{-0.27}$ & 1.92$^{+1.14}_{-0.88}$ & 1.72$^{+0.84}_{-0.75}$ & 1.28$^{+0.64}_{-0.54}$ & 0.82$^{+0.49}_{-0.46}$ \\
$\chi^{2}/$dof & 355.8$/$325 & 264.8$/$253 & 253.7$/$244 & 260.4$/$259 & 263.0$/$243 \\
\cline{2-6} \\
 & \multicolumn{5}{c}{-- SIS0 + GIS2 0.6 - 10 keV --} \\
\cline{2-6} \\
kT$_{1}$ (keV) & 0.69$^{+0.03}_{-0.02}$ & 0.72$\pm$0.06 & 0.68$^{+0.06}_{-0.05}$ & 0.65$^{+0.04}_{-0.05}$ & 0.70$\pm$0.05 \\
kT$_{2}$ (keV) & 2.17$^{+0.09}_{-0.07}$ & 2.86$^{+0.05}_{-0.04}$ & 2.36$^{+0.11}_{-0.10}$ & 2.25$^{+0.09}_{-0.10}$ & 2.31$^{+0.47}_{-0.25}$ \\
$\log$ EM$_{1}$ (cm$^{-3}$) & 53.53$^{+0.09}_{-0.07}$ & 53.24$^{+0.13}_{-0.11}$ & 53.46$\pm$0.11 & 53.42$^{+0.11}_{-0.09}$ & 53.75$^{+0.21}_{-0.27}$ \\
$\log$ EM$_{2}$ (cm$^{-3}$) & 53.74$\pm$0.02 & 53.97$\pm$0.03 & 54.05$\pm$0.04 & 53.86$\pm$0.03 & 53.76$^{+0.11}_{-0.07}$ \\
O & 0.19$^{+0.09}_{-0.08}$ & 0.56$^{+0.31}_{-0.27}$ & 0$^{+0.16}_{-0}$ & 0.30$^{+0.17}_{-0.15}$ & 0.18$^{+0.35}_{-0.16}$ \\
Ne & 1.02$^{+0.16}_{-0.17}$ & 1.55$^{+0.51}_{-0.49}$ & 1.16$^{+0.37}_{-0.28}$ & 1.61$\pm$0.32 & 0.84$^{+0.48}_{-0.31}$ \\
Mg & 0.29$^{+0.10}_{-0.08}$ & 0.30$^{+0.35}_{-0.30}$ & 0.28$^{+0.26}_{-0.20}$ & 0.51$^{+0.23}_{-0.18}$ & 0.20$^{+0.23}_{-0.13}$ \\
Si & 0.23$^{+0.06}_{-0.05}$ & 0.54$^{+0.27}_{-0.22}$ & 0.28$^{+0.18}_{-0.13}$ & 0.37$^{+0.14}_{-0.12}$ & 0.11$^{+0.09}_{-0.08}$ \\
S & 0.22$\pm$0.08 & 0.28$^{+0.29}_{-0.26}$ & 0.08$^{+0.22}_{-0.08}$ & 0.32$^{+0.16}_{-0.15}$ & 0.24$^{+0.18}_{-0.15}$ \\
Ar & 0.68$\pm$0.26 & 0.24$^{+0.70}_{-0.24}$ & 0$^{+0.16}_{-0}$ & 0.77$\pm$0.44 & 0.42$^{+0.63}_{-0.42}$ \\
Ca & 0.87$^{+0.40}_{-0.39}$ & 0.58$^{+0.95}_{-0.58}$ & 1.30$^{+0.64}_{-0.65}$ & 0$^{+0.71}_{-0}$ & 0.13$^{+0.73}_{-0.13}$ \\
Fe & 0.13$\pm$0.03 & 0.32$\pm$0.09 & 0.22$^{+0.07}_{-0.06}$ & 0.20$^{+0.07}_{-0.05}$ & 0.07$^{+0.09}_{-0.04}$ \\
Ni & 1.08$^{+0.35}_{-0.32}$ & 2.14$^{+1.33}_{-1.17}$ & 1.20$^{+0.96}_{-0.78}$ & 1.14$^{+0.74}_{-0.67}$ & 0.77$^{+0.62}_{-0.45}$ \\
$\chi^{2}/$dof & 529.0$/$490 & 338.8$/$321 & 285.4$/$313 & 337.4$/$350 & 314.1$/$316 \\
\enddata
\tablenotetext{1}{N$_{H}$ fixed at 1.35 10$^{18}$ cm$^{-2}$.  Uncertainties are 90\% confidence intervals.}
\end{deluxetable}

%Table 12
\begin{deluxetable}{lll}
\tablewidth{0pt}
\tablenum{13}
\tablecolumns{3}
\tablecaption{Luminosity in {\it RXTE} observations \label{tbl:rxtetbl}}
\tablehead{ \colhead{Energy Range} & \colhead{1996} & \colhead{1998} \\
\colhead{(keV)} & \multicolumn{2}{c}{ (erg s$^{-1}$)} }
\startdata
2--5  & 2.6 10$^{30}$ & 1.7 10$^{30}$ \\
5--12  & 7.7 10$^{29}$ & 3.8 10$^{29}$ \\
2--12 & 3.4 10$^{30}$ & 2.1 10$^{30}$ \\
\enddata
\end{deluxetable}

%Table 13
\begin{deluxetable}{llllr}
\tablewidth{0pt}
\tablenum{14}
\tablecolumns{5}
\tablecaption{Flare energies in {\it RXTE} observations \label{tbl:rxteflares}}
\tablehead{ \colhead{Energy Range} & \colhead{$\tau_{r}$} & \colhead{$\tau_{d}$} & \colhead{$\Delta$ t} & \colhead{Energy} \\
\colhead{(keV)} &\colhead{ (hrs)} &\colhead{ (hrs)} &\colhead{ (hrs)} &\colhead{(ergs)} } 
\startdata
\multicolumn{5}{c}{1996 Flare on September 4.8} \\
\hline
2--5 & 1.66$\pm$0.11 & 3.85$\pm$0.54 & 2.6 & 1.6 10$^{34}$ \\
5--12 & 1.13$\pm$0.09 & 2.24$\pm$0.35 & 2.6 & 8 10$^{33}$ \\
2--12 &1.38$\pm$0.09 & 2.63$\pm$0.24 & 2.6 & 2.7 10$^{34}$\\
\hline
\multicolumn{5}{c}{1998 Flare on September 9.8} \\
\hline
2--5 & \ldots\tablenotemark{a} & 3.88$\pm$0.66 & 3.4 & 1.3 10$^{34}$ \\
5--12 & \ldots\tablenotemark{a} & 1.89$\pm$0.34 & 3.4 & 3 10$^{33}$ \\
2--12 & \ldots \tablenotemark{a} & 2.61$\pm$0.37 & 3.4 & 1.6 10$^{34}$ \\
\enddata
\tablenotetext{a}{Flare rise was not observed.} 
\end{deluxetable}

\begin{deluxetable}{llllll}
\tablewidth{0pt}
\tablenum{15}
\tablecolumns{6}
\tablecaption{Luminosities and MECS timescales
during the 1998 {\it BeppoSAX} observation \label{saxlum}}
\tablehead{
\colhead{Energy Range}& \colhead{{\bf Q}}& \colhead{{\bf F1}}&
\colhead{{\bf F2}}& \colhead{{\bf D}}& \colhead{{\bf F3}} \\
\colhead{(keV)}& \colhead{(erg s$^{-1}$)}&  \colhead{(erg s$^{-1}$)}&
\colhead{(erg s$^{-1}$)}& \colhead{(erg s$^{-1}$)}& \colhead{(erg s$^{-1}$)}
}
\startdata
0.6--2  & 3.7 10$^{30}$& 5.0 10$^{30}$& 5.1 10$^{30}$& 4.1 10$^{30}$& 4.4
10$^{30}$ \\
2--5    & 1.1 10$^{30}$& 2.2 10$^{30}$& 2.1 10$^{30}$& 1.3 10$^{30}$& 1.5
10$^{30}$ \\
5--10   & 2.3 10$^{29}$& 6.0 10$^{29}$& 6.1 10$^{29}$& 2.8 10$^{29}$& 3.9
10$^{29}$ \\
0.6--10 & 5.1 10$^{30}$& 7.8 10$^{30}$& 7.8 10$^{30}$& 5.7 10$^{30}$& 6.4
10$^{30}$ \\
\cline{1-6}
%Timescales& {\bf F1}& {\bf F2}& {\bf F3}& & \\
%\cline{1-6}
$\tau_{r}$ (hrs)& & 11.1$\pm$0.7& 2.1$\pm$0.3& & 2.3$\pm$0.3 \\
$\tau_{d}$ (hrs)& & 13.2$\pm$3.8& 3.7$\pm$0.2& & 4.5$\pm$0.4 \\
\enddata
\end{deluxetable}

\begin{deluxetable}{llllll}
\tablewidth{0pt}
\tablenum{16}
\tablecolumns{6}
\tablecaption{MEKAL 2-Temperature spectral fits to flaring$/$quiescent {\it
BeppoSAX} LECS + MECS spectra \label{saxfits}}
\tablehead{
\colhead{}& \colhead{{\bf F1}}& \colhead{{\bf F2}}& \colhead{{\bf D}}&
\colhead{{\bf Q}} & \colhead{{\bf F3}} }
\startdata
kT$_{1}$ (keV)& 0.90$^{+0.09}_{-0.09}$& 0.83$^{+0.13}_{-0.11}$&
0.77$^{+0.07}_{-0.07}$& 0.81$^{+0.04}_{-0.05}$& 0.78$^{+0.11}_{-0.09}$ \\
kT$_{2}$ (keV)& 2.41$^{+0.18}_{-0.15}$& 2.42$^{+0.34}_{-0.22}$&
1.98$^{+0.13}_{-0.11}$& 1.98$^{+0.10}_{-0.08}$& 2.24$^{+0.31}_{-0.18}$ \\
$\log$ EM$_{1}$ (cm$^{-3}$)& 53.42$^{+0.15}_{-0.16}$&
53.35$^{+0.24}_{-0.23}$& 53.34$^{+0.12}_{-0.13}$& 53.35$^{+0.09}_{-0.08}$&
53.40$^{+0.20}_{-0.18}$ \\
$\log$ EM$_{2}$ (cm$^{-3}$)& 53.76$^{+0.04}_{-0.05}$&
53.75$^{+0.05}_{-0.07}$& 53.63$^{+0.04}_{-0.04}$& 53.55$^{+0.03}_{-0.04}$&
53.64$^{+0.05}_{-0.08}$ \\
Z& 0.27$^{+0.07}_{-0.05}$& 0.34$^{+0.14}_{-0.10}$& 0.30$^{+0.07}_{-0.05}$&
0.29$^{+0.04}_{-0.03}$& 0.30$^{+0.10}_{-0.08}$ \\
$\chi^{2}/$dof& 179.2/172& 113.7/115& 170.5/177& 329.8/287& 121.4/130 \\
\enddata
\end{deluxetable}

%Table 14
\begin{deluxetable}{llll}
\tablewidth{0pt}
\tablenum{17}
\tablecolumns{4}
\tablecaption{Radio, EUV Quiescent Emission Properties\label{tbl:lxlrq} }
\tablehead{ \colhead{Dates} & \colhead{L$_{\rm 3.6}$} & \colhead{L$_{\rm EUV}$}  & 
\colhead{$\log \frac{L_{EUV}}{L_{3.6}}$}  \\
\colhead{} & \colhead{ (erg s$^{-1}$ Hz$^{-1}$) } & \colhead{ (erg
s$^{-1}$) } & \colhead{ (Hz) } }
\startdata
1993 Sept. 17.7--17.9 & 9.5 10$^{16}$ & 1.2 10$^{30}$ & 13.1\\
1994 Aug. 25.3--25.6 & 3 10$^{16}$ & 6.6 10$^{29}$ &13.3\\
1994 Aug. 26.4--26.5 & 2.9 10$^{16}$ & 8.4 10$^{29}$&13.5 \\
1996 Sept. 2.3--2.8 & 1.9 10$^{16}$ & 6.0 10$^{29}$ & 13.5\\
1996 Sept. 6.3--7.0 & 9.9 10$^{15}$ & 7.2 10$^{29}$ & 13.9\\
1998 Sept. 7.4--7.6 & 9.9 10$^{15}$ & 7.3 10$^{29}$& 13.9 \\
1998 Sept. 9.6--9.8 & 8.0 10$^{15}$ & 6.6 10$^{29}$& 13.9 \\
\enddata
\end{deluxetable}

%Table 16
\begin{deluxetable}{lll}
\rotate
\tablewidth{0pt}
\tablenum{18}
\tablecolumns{3}
\tablecaption{EUV$/$SXR flares on HR~1099
and multi-wavelength behavior \label{tbl:multiw}}
\tablehead{ \colhead{Time of peak flux} & \colhead{EUV$/$SXR telescope}   & \colhead{comments} \\
             \colhead{}    & \colhead{that saw flare} &\colhead{} }
\startdata
1993 Sept. 17.4 & {\it EUVE} & C~IV$/$Mg~II flux rises, peaks $\approx$ 4 hrs before EUV peak \\
                &	     & secondary maximum in Mg~II flux after EUV peak \\
               &              & gap in radio coverage during EUV flare rise \\
1993 Sept. 19.2 & {\it EUVE} & C~IV flux rise and peak $\approx$ 3 hrs before EUV flare peak \\
                &             &Mg~II flux increases but does not reach a peak \\
               &              &no radio coverage\\
1994 Aug. 24.2 & {\it EUVE} & flare seen in Mg~II, C~IV; peak is missed \\
                &          &gap in radio coverage  \\
1994 Aug. 25.76 & {\it EUVE}, {\it ASCA} & radio flare peaks $\sim$ 2.5 hrs before EUV$/$SXR maxima \\
                &                  & no response of Mg~II to flare, slight increase in C~IV \\
1994 Aug. 27.00 & {\it EUVE} & radio flare accompanied by increase in C~IV flux; both peak $\approx$ \\
                &            & 4.8 hrs before the end of the EUV rise\\
                &            & no response from Mg~II \\
1996 Sept. 2.0 & {\it EUVE} & no radio, SXR coverage \\
1996 Sept. 2.9 & {\it RXTE} & no radio flare, no EUV enhancement \\
1996 Sept. 4.4 & {\it EUVE} & small radio flare peaks $\sim$ 0.5 hr before small EUV enhancement, gap in SXR coverage \\
1996 Sept. 4.8 & {\it EUVE}, {\it RXTE} & possibly small radio enhancement\\
1996 Sept. 5.3 & {\it EUVE} & gaps in SXR coverage; radio coverage limited to after EUV peak \\
1996 Sept. 8.7--9.6 & {\it EUVE} & 2 radio flares during EUV rise, peaking $\sim$ 30 hrs before 1st EUV peak\\ 	
                 &          &radio coverage limited to beginning of EUV flare; gap in SXR coverage \\
1998 Sept. 5.9 & {\it EUVE} & no radio, SXR coverage \\
1998 Sept. 7.1 & {\it BeppoSAX} & gap in EUV, radio coverage \\
1998 Sept. 7.3 & {\it BeppoSAX} & limited EUV, radio coverage; no increase in EUV or radio flux \\
1998 Sept. 7.9 & {\it BeppoSAX}, {\it RXTE} & small enhancement in {\it BeppoSAX} light curve, rise in {\it RXTE} \\
               &                     &no EUV flare, gap in radio coverage \\
1998 Sept. 8.4 & {\it BeppoSAX} & small enhancement in {\it BeppoSAX} light curve \\
               &                &no EUV enhancements, no radio flare \\
1998 Sept. 9.8 & {\it BeppoSAX}, {\it RXTE} & no EUV or radio enhancement \\
\enddata
\end{deluxetable}

%Table 17
\begin{deluxetable}{llll}
\tablewidth{0pt}
\tablenum{19}
\tablecolumns{4}
\tablecaption{Comparison
of radio, EUV$/$SXR flare properties \label{tbl:neupert}}
\tablehead{ \colhead{Date} & \colhead{1994 Aug. 25.76} & \colhead{1994 Aug. 27.0} &  \colhead{1996 Sept. 7}\\ }
\startdata
EUV$/$SXR peak Lum. (erg s$^{-1}$) & 2.4 $\times$10$^{29}$ (EUV) & 1.2 $\times$10$^{30}$ &  1.9 $\times$10$^{30}$  \\
                                   & 5.2 $\times$10$^{30}$ (SXR) &                                          &              \\
3.6 cm peak Lum. (erg s$^{-1}$ Hz$^{-1}$) & 1.2 $\times$10$^{16}$ & 2.6 $\times$10$^{16}$ &  1.9 $\times$10$^{16}$ \\
ratio L$_{EUV,SXR}$/L$_{R}$ (Hz)			& 2.0 $\times$10$^{13}$ (EUV) & 4.6 $\times$10$^{13}$ &      1.0 $\times$10$^{14}$ \\
				& 4.3 $\times$10$^{14}$ (SXR) & 				&		\\
\cline{1-4} \\
EUV$/$SXR energy (erg) & 1.8 $\times$10$^{33}$ (EUV) &8.1 $\times$10$^{34}$  & 3.5 $\times$10$^{35}$ \\
			& 5.3 $\times$10$^{34}$(SXR) &	&		 		\\
3.6 cm energy (erg Hz$^{-1}$)   & 3.8 $\times$10$^{19}$ & 1.4 $\times$10$^{20}$  & 1.1 $\times$10$^{20}$ \\
ratio E$_{EUV,SXR}$/E$_{R}$ (Hz) &  4.7 $\times$10$^{13}$ (EUV)& 5.8 $\times$10$^{14}$ & 3.2 $\times$10$^{15}$ \\
           &   1.4 $\times$10$^{15}$ (SXR) &              &                              \\
\cline{1-4} \\
EUV$/$SXR duration $\Delta t_{EUV,SXR}$ (hr) & 9.4 & 30.1 &  83.6 \\
3.6 cm duration $\Delta t_{R}$ (hr) & 2  & 2.3 &  10.2 \\
ratio    $\Delta t_{EUV,SXR}$/$\Delta t_{R}$            & 4.5 & 13.1 &  8.2 \\
t$_{peak,EUV(SXR)}$-t$_{peak,R}$ (hr) & 2.5 & 4.8 &  30 \\
 &                   &    &      \\
\enddata
\end{deluxetable}

\clearpage
%%%%%%%%%%%%%%%%%%%%%%%%%%%%%%%
%PUT FIGURES HERE
%%%%%%%%%%%%%%%%%%%%%%%%%%%%%%%%%%%%%%%%%%
\clearpage
%Figure 2a

\clearpage
%Figure 2b
%%%%%%%%%%%%%%%%%%
%Figure 1993radiopol

\clearpage
%Figure 3a
%%%%%%%%%%%%%%%%%%%%%%

\clearpage
%Figure 3b
%%%%%%%%%%%%%%%%%%%%%%%
%Figure 1994radiopol

\clearpage
%Figure 4a
%%%%%%%%%%%%%%%%%%%%%%%%%
%Figure 1996radioflux

\clearpage
%Figure 4b
%%%%%%%%%%%%%%%%%%%%%%%%%
%Figure 1996radiopol

\clearpage
%Figure 5a
%%%%%%%%%%%%%%%%%%%%%%%%%%
%Figure 1998radioflux

\clearpage
%Figure 5b
%%%%%%%%%%%%%%%%%%%%%%%%

\addtocounter{figure}{4}
%Figure 6a
\clearpage
%%%%%%%%%%%%%%%%%%%%%%%%%%%%%%%%%%%%%%%%%%%%%%%%%%%%%%%%
\begin{figure}[htbp]
\begin{center}
\scalebox{0.8}{\rotatebox{90}{
\plotone{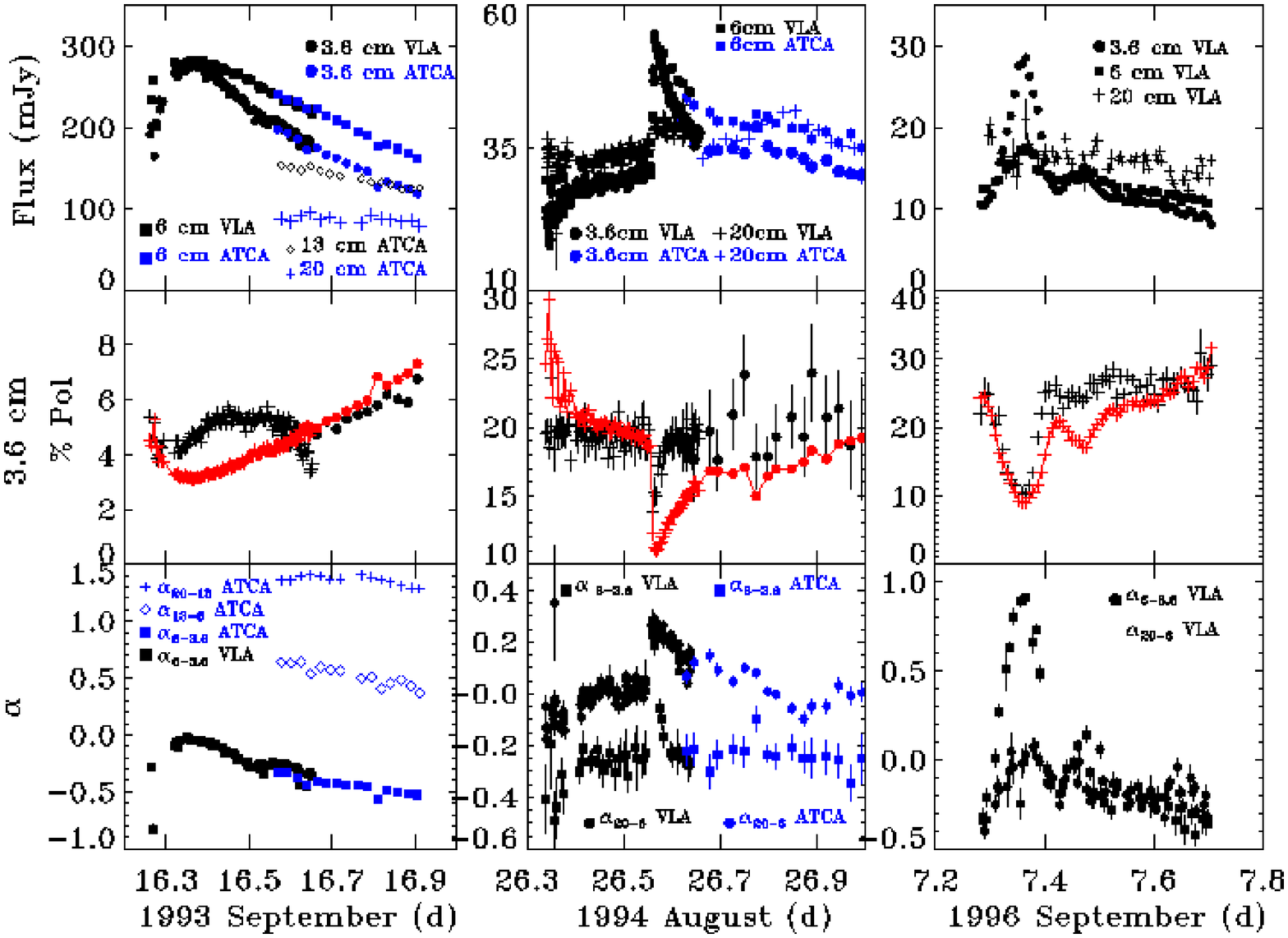}}}
\caption[]{ 
Flux, polarization and spectral index variations 
of three radio flares. { \bf (top row)}:  Flux variations
at 3.6, and 6 cm; 13 and 20 cm flux variations are plotted where there is coverage. 
The VLA data has a temporal sampling of 5 minutes,
while the ATCA data is binned at $\sim$ half an hour.  
{\bf (middle row)}: Observed 3.6 cm circular polarization variations.  An increase in flux (as seen in 
the top row) is
accompanied by a marked decrease in percent circular polarization; the polarization
minima occur at the flux maximuma for these flares.
The red line
indicates the polarization variations from a combination of an unpolarized flare
and steady, polarized quiescent emission; see text for details.
{\bf (bottom row)}: Observed 6--3.6 cm, 13--6, 20--6, and 20--13 spectral indices during the flares.
The maximum spectral index is reached at the time of the flare peak flux.
The flare spectral indices are consistent with an increase in optical depth during the flare
rise.
\label{radioflare}}
\end{center}
\end{figure}

\clearpage
%Figure 7
\begin{figure}[htbp]
\begin{center}
\scalebox{0.8}{
\plotone{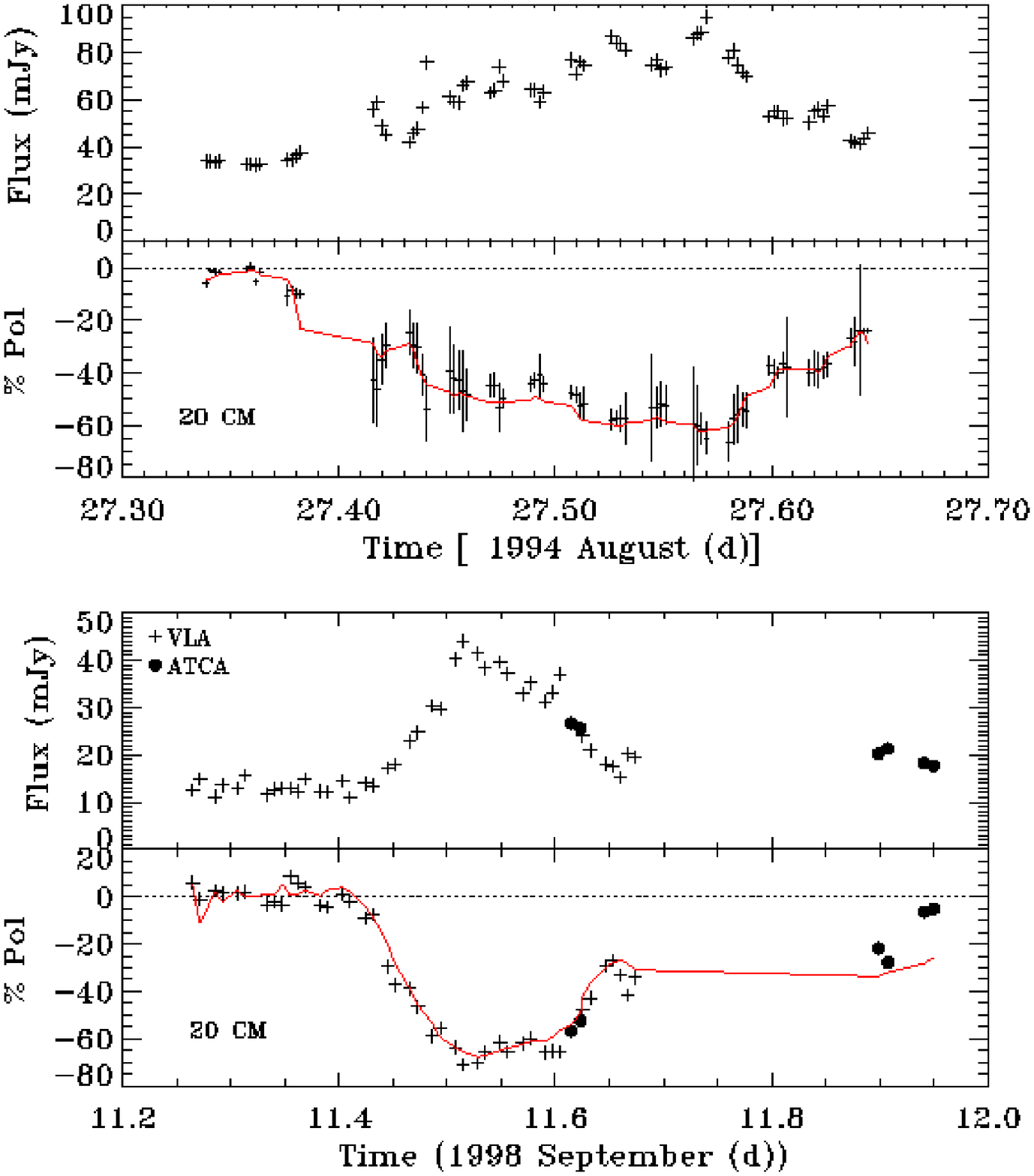}}
\caption[]{
Flux and polarization variations of two flares seen at 20 cm in 1994 and 1998.
Time bins are 300 seconds in length.
Crosses indicate VLA data; circles indicate ATCA data.  
A noticeable decrease in polarization accompanies increase in flux during a flare, and
the polarization increases as the flux declines.  Red line
indicates the amount of polarization expected if the flare is 100\% left circularly polarized; see
text for details.
\label{radioflare3}}
\end{center}
\end{figure}

\clearpage
%Figure 8
\begin{figure}[htbp]
\begin{center}
\scalebox{0.8}{
\plotone{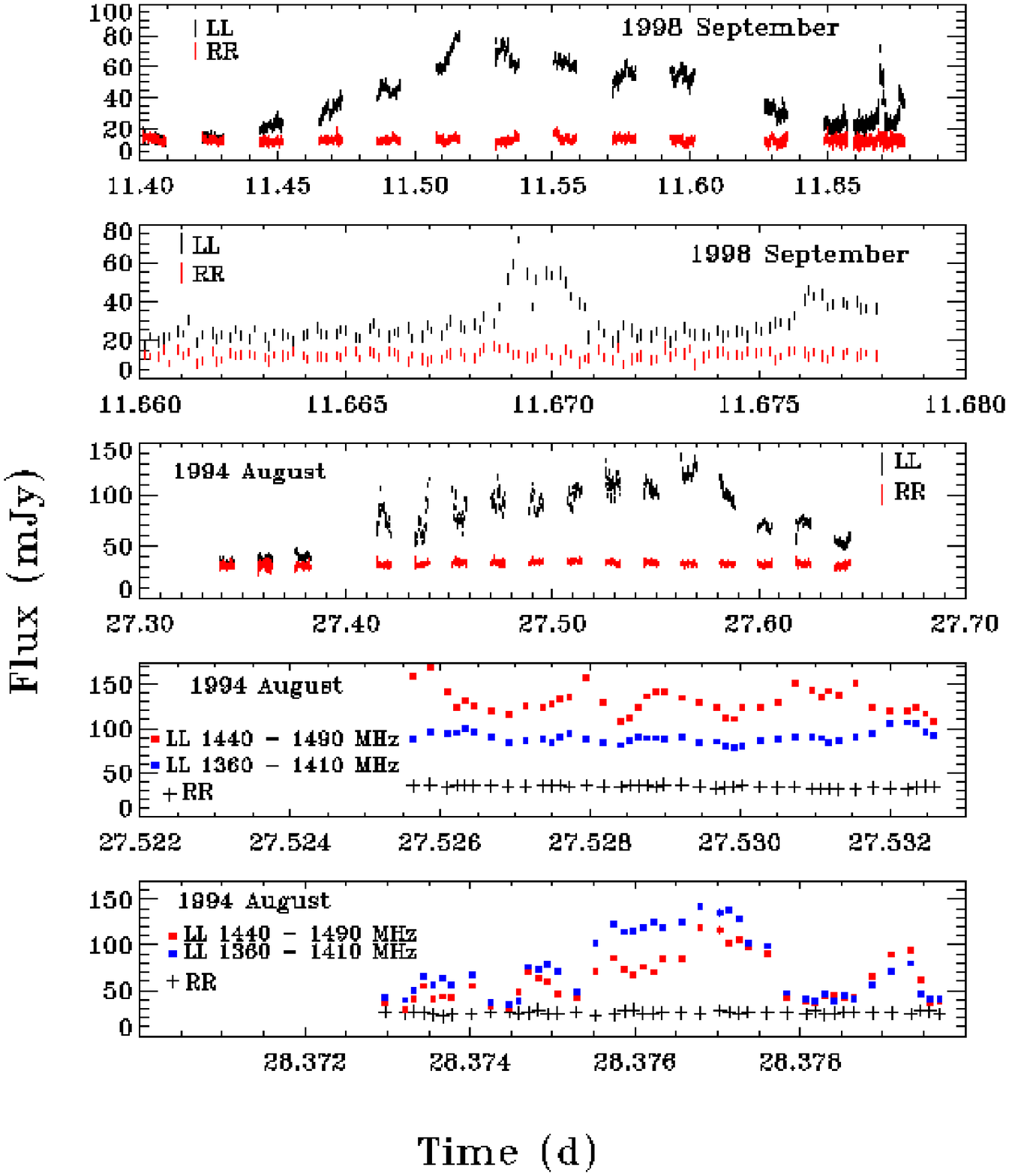}}
\caption[]{
Examples of L-band highly polarized emissions (plotted at the maximum time 
resolution of 10 seconds).  Top panel displays LL, RR emission over
5--6 hours in 1998; subsequent panel shows a short duration ($\approx$ 80 seconds) burst.
Another example of long time duration, highly polarized fluxes in 1994 is shown.  The bottom
two panels compare RCP flux with LCP flux in the two frequency bands used in the observation; 
there is some evidence for different flux density variations at the two bands which comprise the 20 cm
continuum observation.
\label{coherent}}
\end{center}
\end{figure}

\clearpage
%Figure 9
\begin{figure}[htbp]
\begin{center}
\scalebox{0.8}{
\plotone{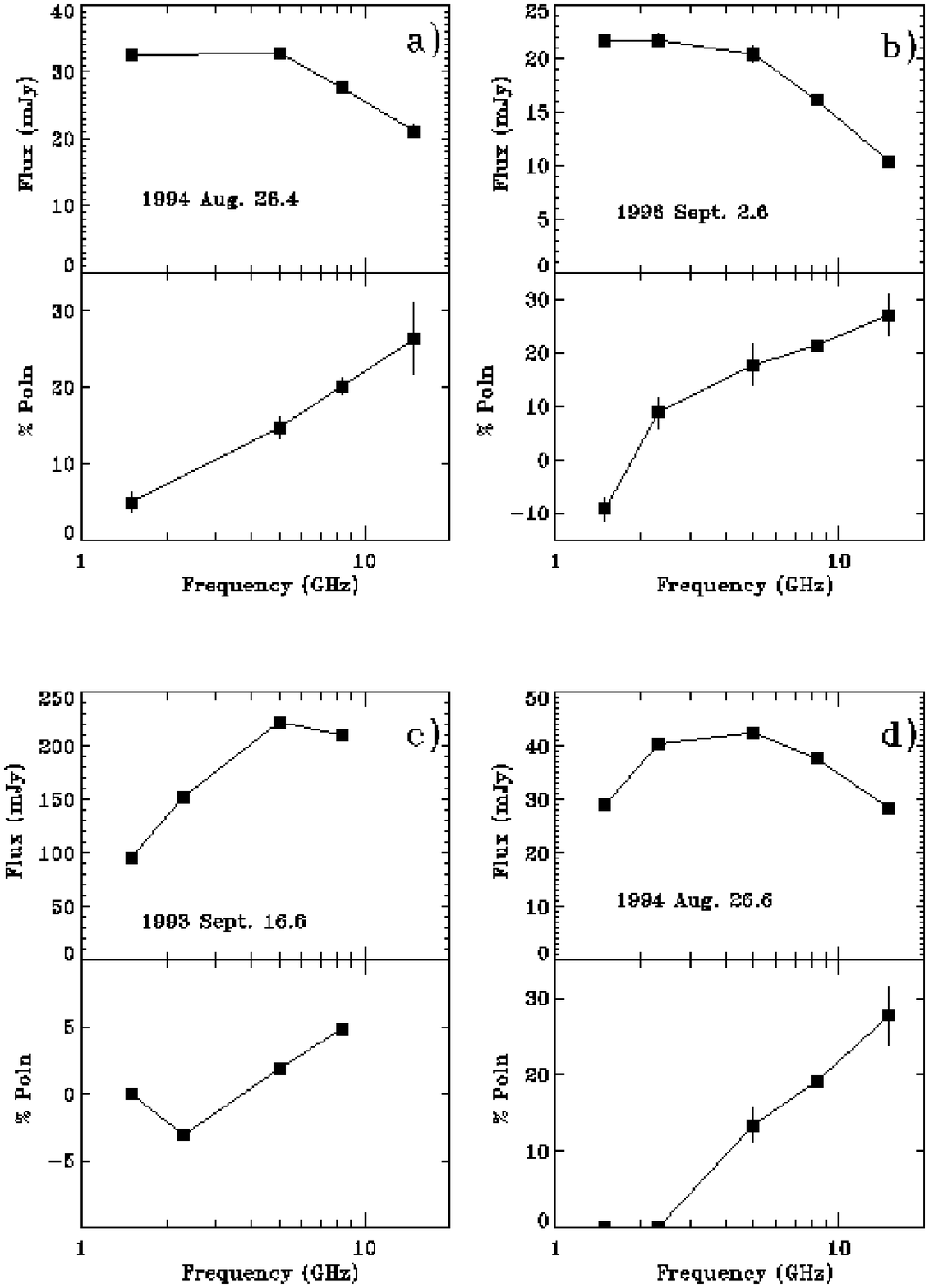}}
\caption[]{Examples of
flux and circular polarization spectra for quiescent emission (panels {\it a} and {\it b})
and during the peak$/$decay of gyrosynchrotron flares (panels {\it c} and {\it d})
from HR~1099.  Spectral indices are listed in Table~\ref{tbl:specindex}.
\label{fig:radiospec} }
\end{center}
\end{figure}

\clearpage
%Figure 1
\begin{figure}[htbp]
\begin{center}
\scalebox{0.7}{
\plotone{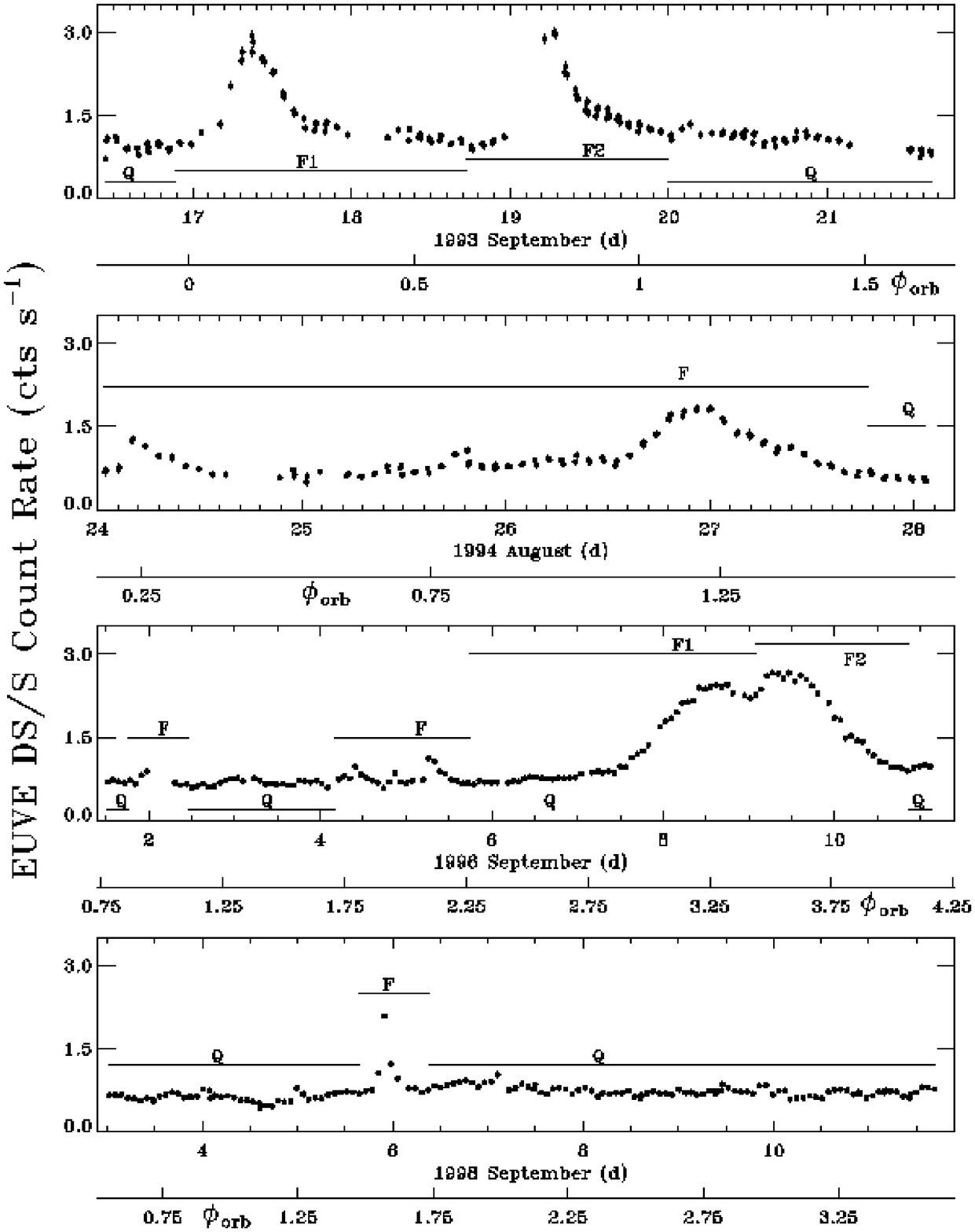}}
\caption[]{Light curves of EUV (80-150 \AA) variability as recorded by {\it EUVE}
during 1993, 1994, 1996 and 1998 observations.  Binary orbital phase is shown in addition to 
UT time.  Time intervals used for spectral extraction are also noted, and are based on
an analysis of light curve variability presented in Section~\ref{moreeuvlc}:  `Q' refers to
quiescent times, `F' to flare times (often several small-amplitude flares), and `F1', `F2', etc.
to discrete, large-amplitude flares.
Each point represents 
$\approx$ 3000 seconds of data, or one point per satellite orbit.  Error bars are 1 $\sigma$.
\label{euvelc}}
\end{center}
\end{figure}
%Figure 10
%%%%%%%%%%%%%%%%%%%%%%%%%%%%%
%Figure cumlum_euve
\begin{figure}[htbp]
\begin{center}
\scalebox{0.8}{
\plotone{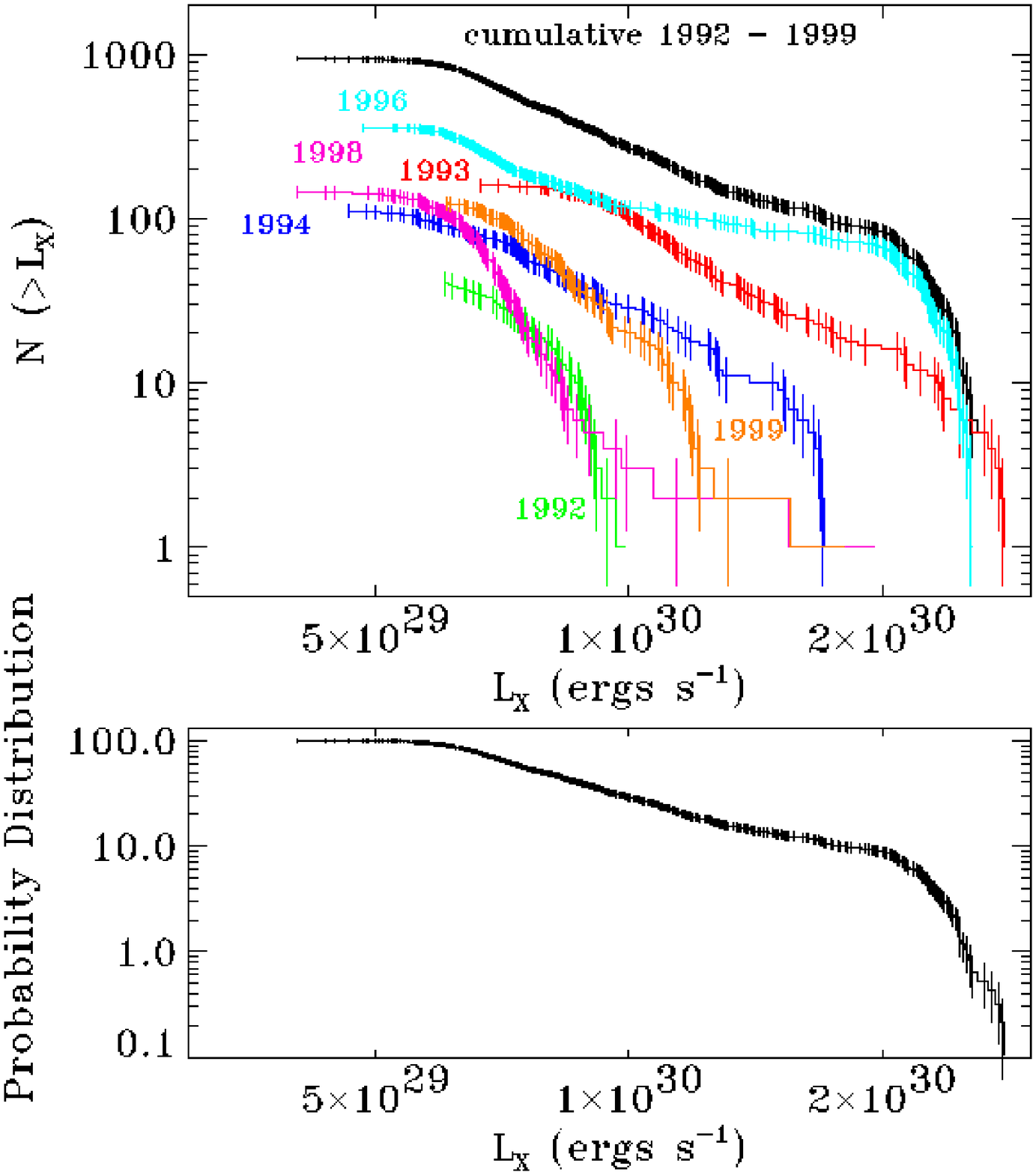}}
\caption[]{{\bf (top)} Cumulative luminosity distributions for the six
DS$/$S observations of HR~1099 from 1992 through 1999.  The number of bins with luminosity
greater than a given luminosity is plotted, along with the same distribution using all the
observations together.  {\bf (bottom)}  Distribution of light curve events from all
DS$/$S observations of HR~1099, expressed as a percent probability of being larger than a given 
L$_{\rm X}$. There is a flattening at low luminosities and a turnover at high luminosities.  The flattening
at
low luminosities represents the non-flaring luminosities which may be considered ``quiescence'',
while the high luminosity turnover (above 2 10$^{30}$ erg s$^{-1}$) represents the peaks of the largest
flares observed.
\label{cumlum_euve}}
\end{center}
\end{figure}
%%%%%%%%%%%%%%%%%%%%%%%%%%%%%

\clearpage
%Figure 11
%%%%%%%%%%%%%%%%%%%%%%%%%%%%%%%%
%Figure euve_flares1
\begin{figure}[htbp]
\begin{center}
\scalebox{0.7}{
\plotone{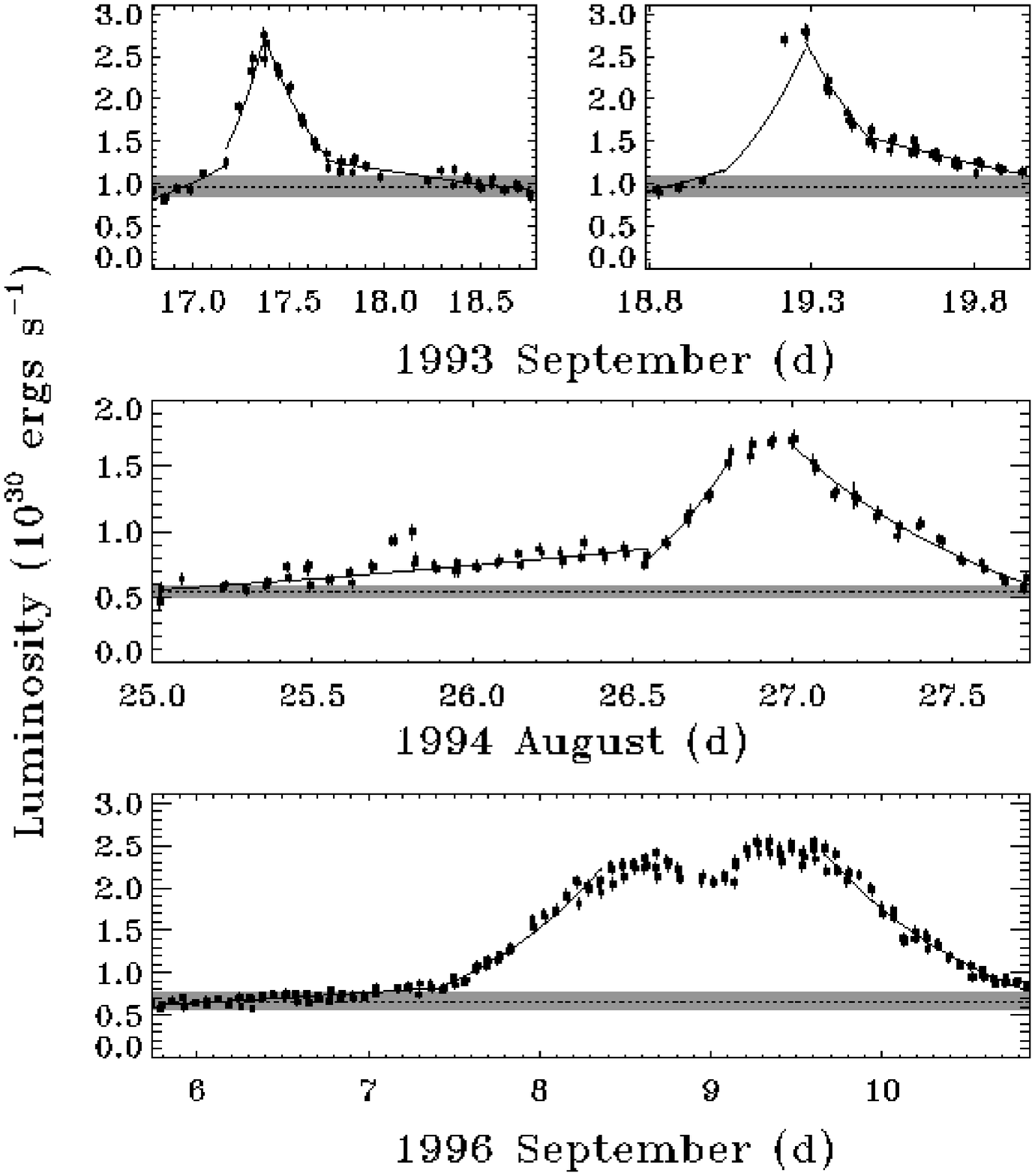}}
\caption[]{Close-up view of four large flares observed with {\it EUVE},
where the count rate has been converted to luminosity.  All four flares seem to show 
a two-stage rise, an initial slow rise followed by a faster phase.  The two flares observed
in 1993 September also show evidence for a two-stage decay, with an initial fast period followed by
a slower decay.  Two large flares occurred in 1996 September too close in time to
separate the decay of the first flare from the rise of the second flare.
\label{euve_flares1}}
\end{center}
\end{figure}
%%%%%%%%%%%%%%%%%%%%%%%%%%%%%%%

\clearpage
%Figure 12
%%%%%%%%%%%%%%%%%%
%Figure composite_flare
\begin{figure}[htbp]
\begin{center}
\scalebox{0.7}{
\plotone{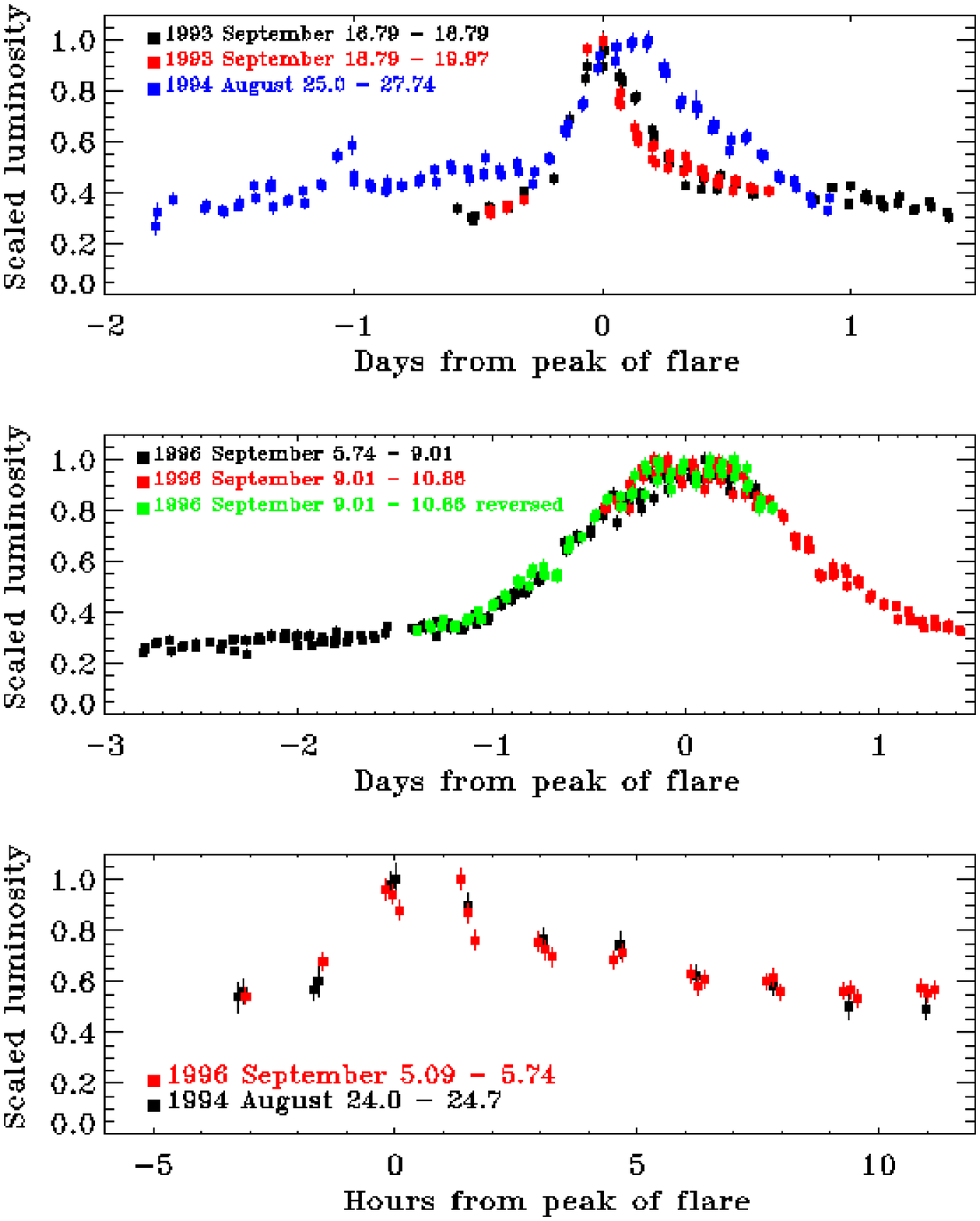}}
\caption[]{{\bf (top)} Flares from 1993 September 16.79 - 18.79 (black), 1993
September 18.79 - 19.97 (red), and 1994 August 25.0 - 27.74 (blue)
shifted in time and flux. 
The two flares in 1993 September are
remarkably similar in shape.  All three flares seem to show an initial slow rise 
before the peak of the flare followed by a fast rise.  The flare in 1994 August 
attained a plateau for $\approx$ 5 hours before decaying again.  While the decays of the two
flares in 1993 September are very similar, the decay from the 1994 August flare is quite
different.  {\bf (middle)}  Flares from 1996 September 5.74 - 9.01 (black) and 
1996 September 9.01 - 10.86 (red),
shifted in time and scaled vertically so the peaks overlap.  
The 1996 September 9.01 - 10.86 flare has also been reflected about the peak to compare the decay
of this flare with the rise of the 1996 September 5.74 - 9.01 flare.  There is a remarkable
amount of symmetry in these two flares.
{\bf (bottom)} Flares from 1994 August 24.0--24.7 (black), and
1996 September 5.09--5.74 (red),
shifted in time and scaled vertically so the peaks overlap.
The flares show a distinct asymmetry, having a fast rise and slow decay that is
typical of impulsive solar flares.
\label{composite_flare}}
\end{center}
\end{figure}
%%%%%%%%%%%%%%%%%%%%%%

\clearpage

%\clearpage
%Figure14 
\begin{figure}[htbp]
\begin{center}
\scalebox{0.8}{
\plotone{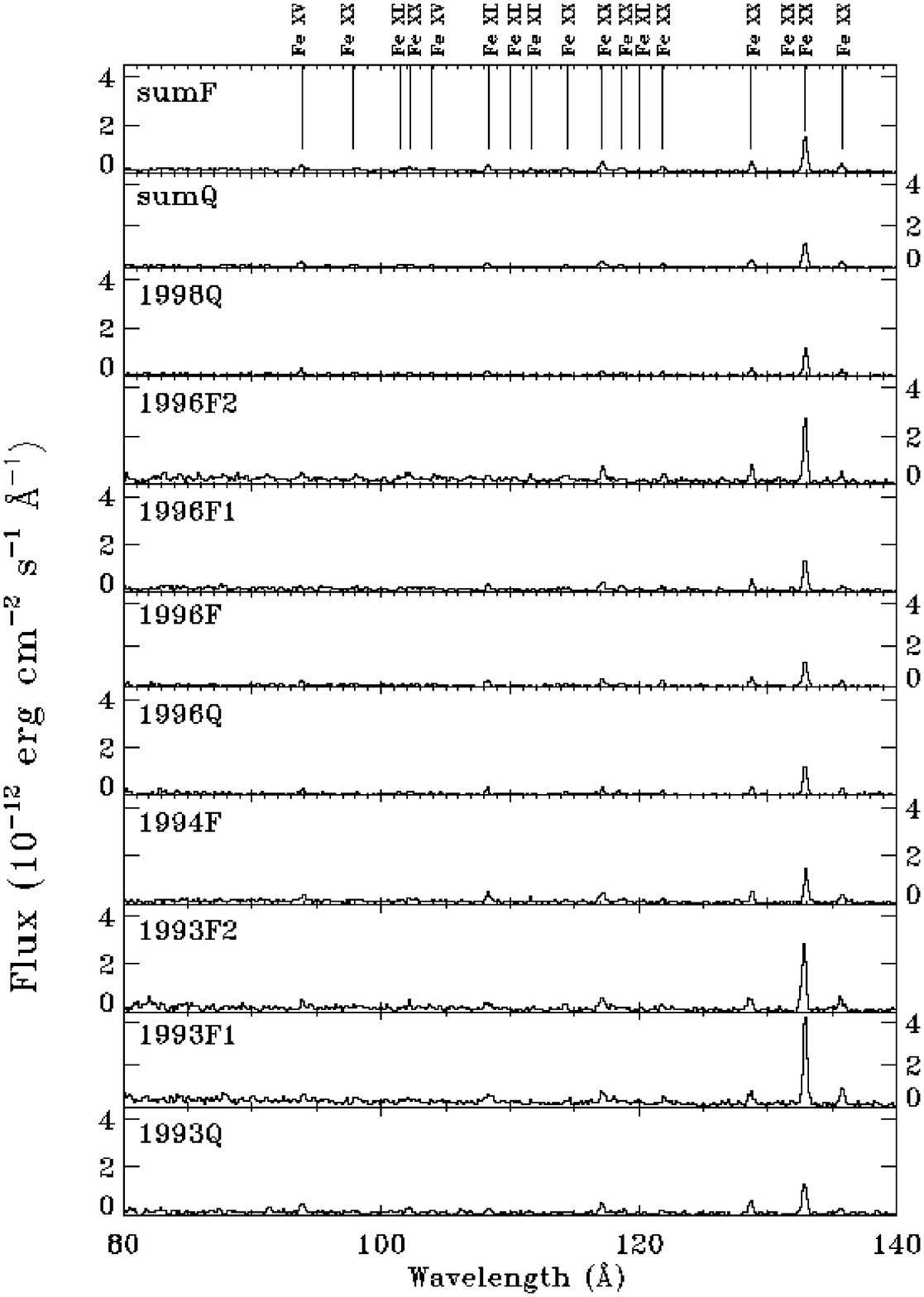}}
\caption[ ]{{\it EUVE} SW spectra from all observations.  Time intervals 
are indicated in Figure~\ref{euvelc}; sumQ and sumF represent the sum of all quiescent
and flaring time intervals, respectively.  
\label{euveswspec}}
\end{center}
\end{figure}

\clearpage
%Figure 15
\begin{figure}[htbp]
\begin{center}
\scalebox{0.8}{
\plotone{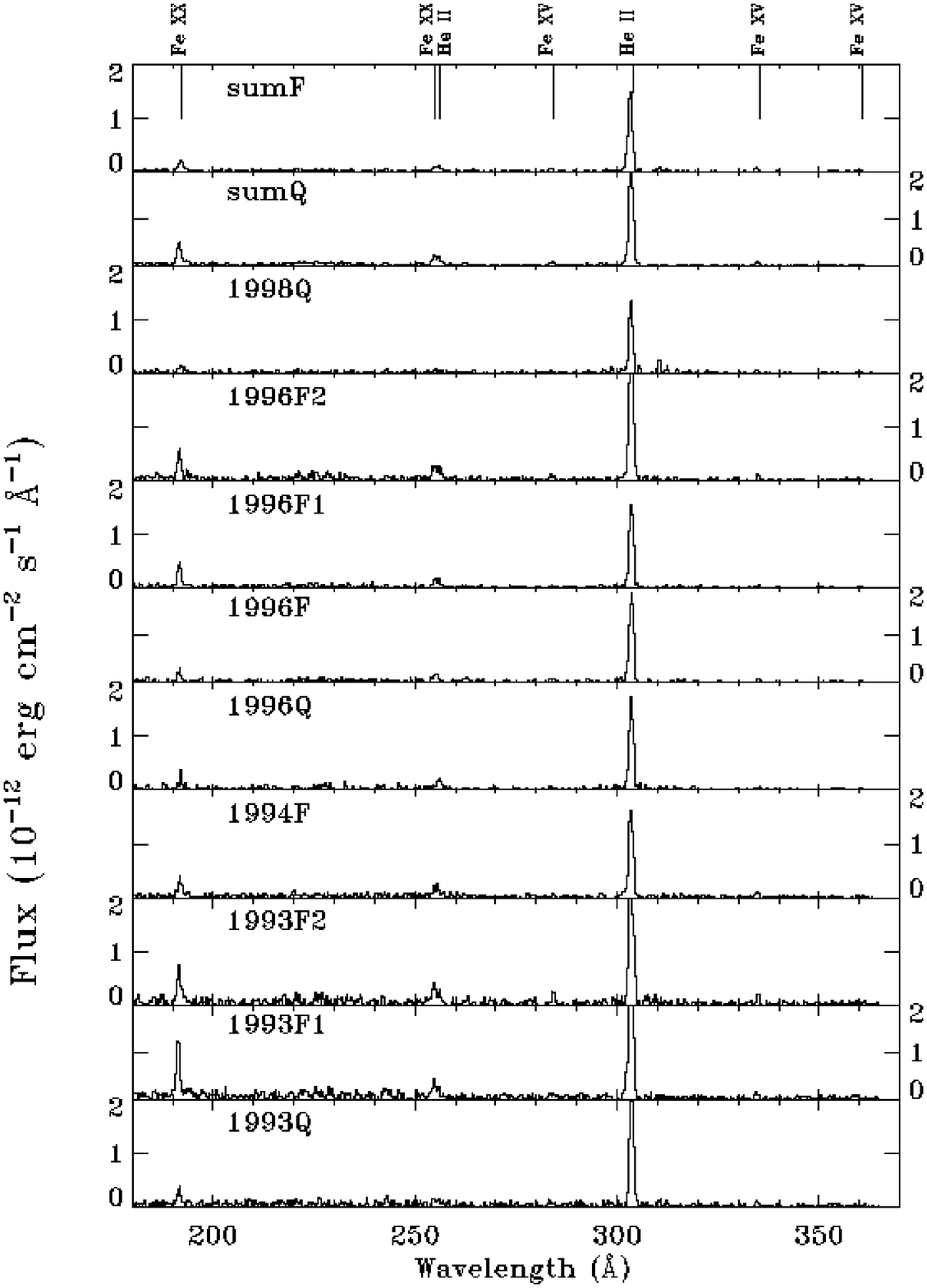}}
\caption[]{{\it EUVE} MW spectra from all observations.  Time intervals 
are indicated in Figure~\ref{euvelc}; sumQ and sumF represent the sum of all quiescent
and flaring time intervals, respectively.
\label{euvemwspec}}
\end{center}
\end{figure}

%Figure 16
%\begin{figure}[htbp]
%\begin{center}
%\rotatebox{90}{\scalebox{0.6}{
%\plotone{f15.eps}}}
%\caption[]{Composite {\it EUVE} spectra from quiescent and flaring time 
%segments {\bf (sumQ)} and {\bf(sumF)}, respectively, including the 1992 and 1999 {\it EUVE} 
%observations.    Detected lines are identified with solid ticks. 
%The quiescent SW spectrum represents a total exposure time of 435 ks; the 
%flare SW spectrum is the total of 234 ks of data.  
%\label{compositespec}}
%\end{center}
%\end{figure}
%%%%%%%%%%%%%%%%%%%%%%%%%%%%%%%%%%%%%%%%%%%%%%%5

\clearpage
%Figure 17
\begin{figure}[htbp]
\begin{center}
\rotatebox{90}{\scalebox{0.6}{
\plotone{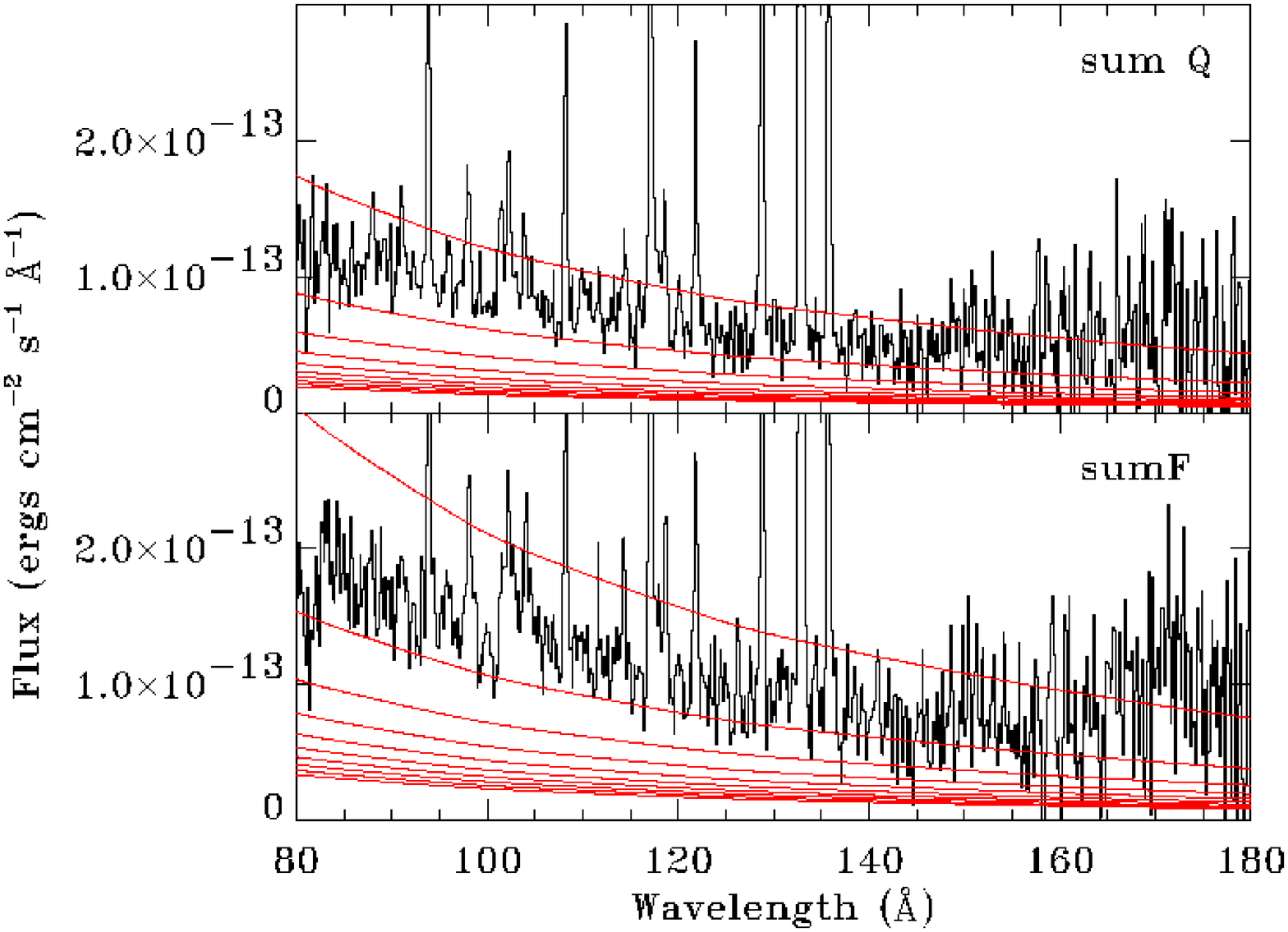}}}
\caption[]{
EUV composite quiescent and flare SW spectrum of HR~1099.  Overlaid in red are continuum spectra calculated
assuming different values of the iron to hydrogen ratio, using the differential
emission measure distribution determined from each spectrum.  From bottom to top, curves decrease
from the solar photospheric abundance to one tenth the solar photospheric value in steps of 0.1.
\label{contfig}}
\end{center}
\end{figure}

\clearpage
%Figure18a
%%%%%%%%%%%%%%%%%%%%%%%%%%%%%%%%%
\begin{figure}[htbp]
\begin{center}
\figurenum{16a}
\scalebox{0.8}{
\plotone{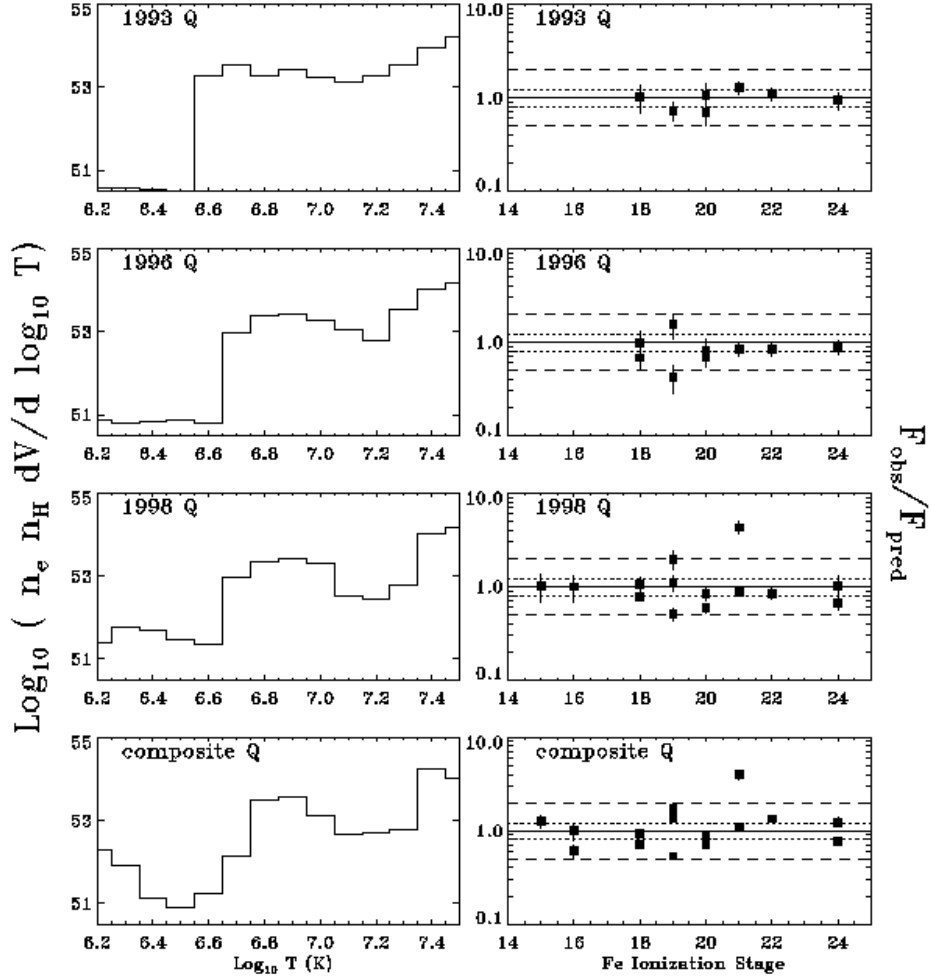}}
\caption[]{(a)
Left panels show DEMs from quiescent {\it EUVE} data; right panels indicate agreement between
observed flux levels and fluxes predicted by the DEM at left.  The solid line indicates 
agreement between observed and predicted fluxes; dashed lines show factor of two agreement
and dotted lines, 20\%.
Time intervals are
shown in the upper left corners and correspond to those depicted in Figure~\ref{euvelc}. 
(b,c) Same as Figure~\ref{hr1099qudem}, for flaring intervals from {\it EUVE} spectra.
\label{hr1099qudem}}
\end{center}
\end{figure}

\clearpage
%Figure 18b
\begin{figure}[htbp]
\begin{center}
\figurenum{16b}
\scalebox{0.8}{
\plotone{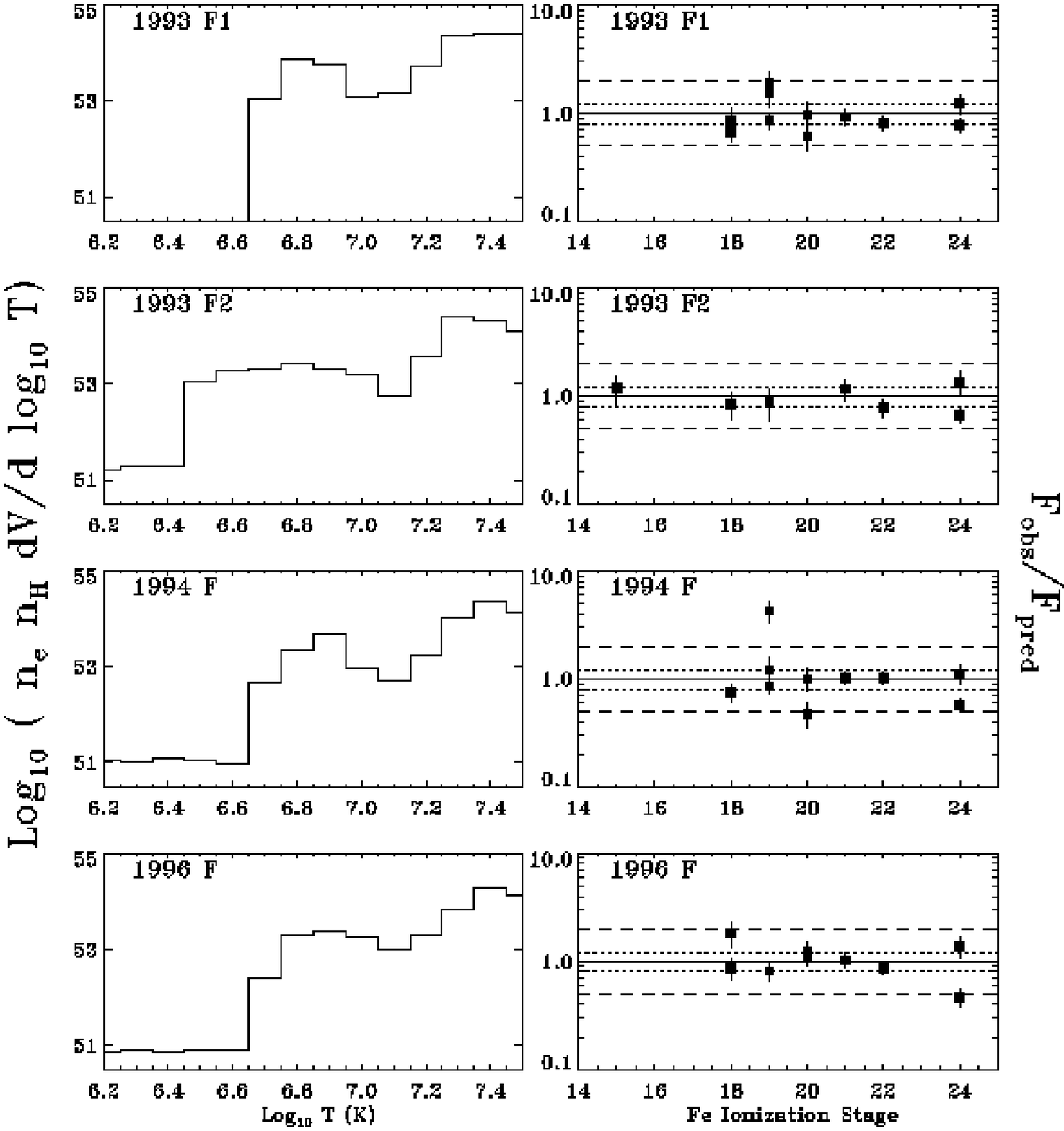}}
\caption[]
{\label{hr1099fldem1}}
\end{center}
\end{figure}

\clearpage
%Figure 18c
\begin{figure}[htbp]
\begin{center}
\figurenum{16c}
\scalebox{0.8}{
\plotone{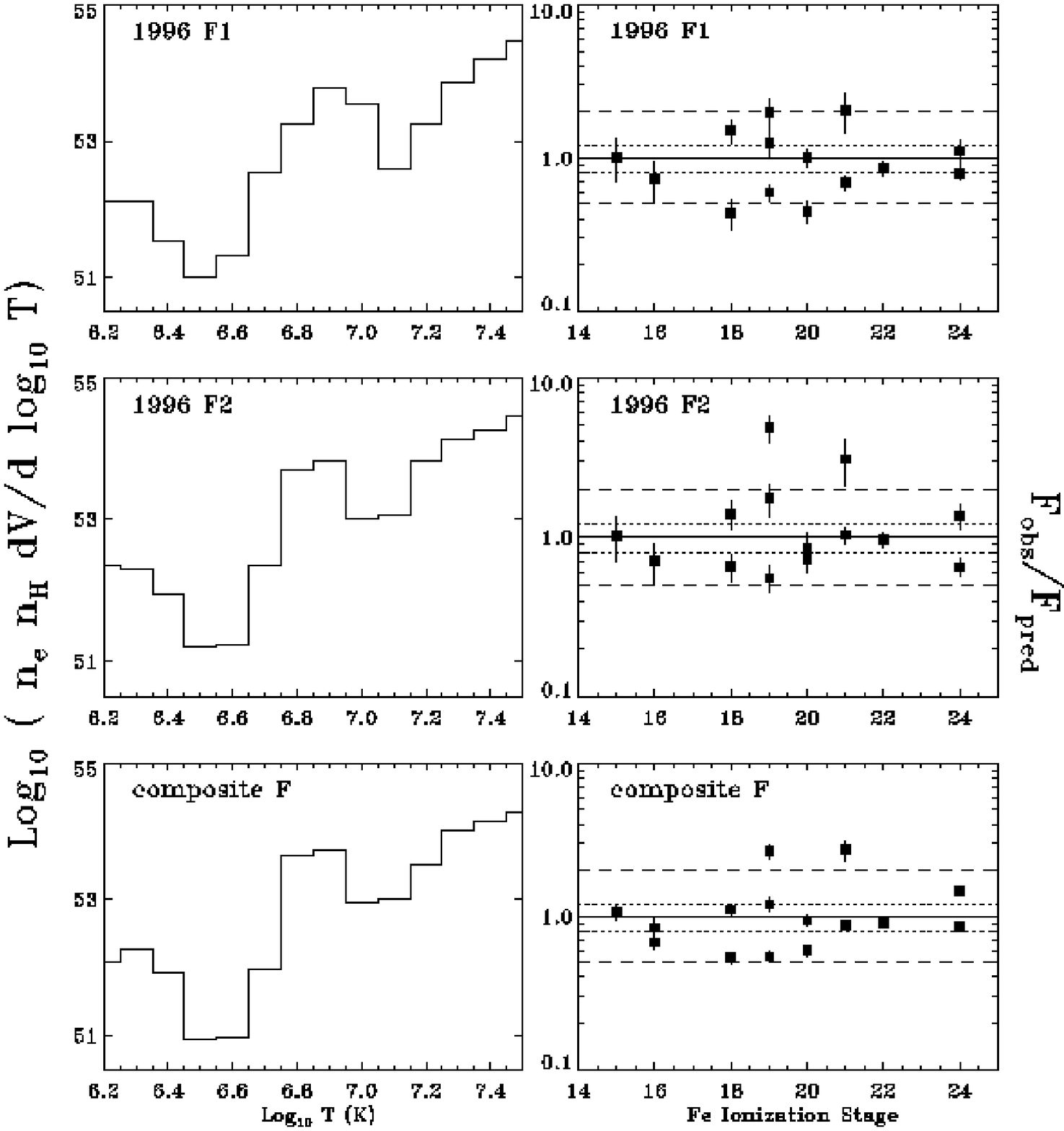}}
\caption[]
{\label{hr1099fldem2}}
\end{center}
\end{figure}
%%%%%%%%%%%%%%%%%%%%%%%%%%%%%%%%%%%%%%%%%%%%%

\addtocounter{figure}{1}
\clearpage
%Figure 19 
\begin{figure}[htbp]
\begin{center}
\scalebox{0.6}{
\plotone{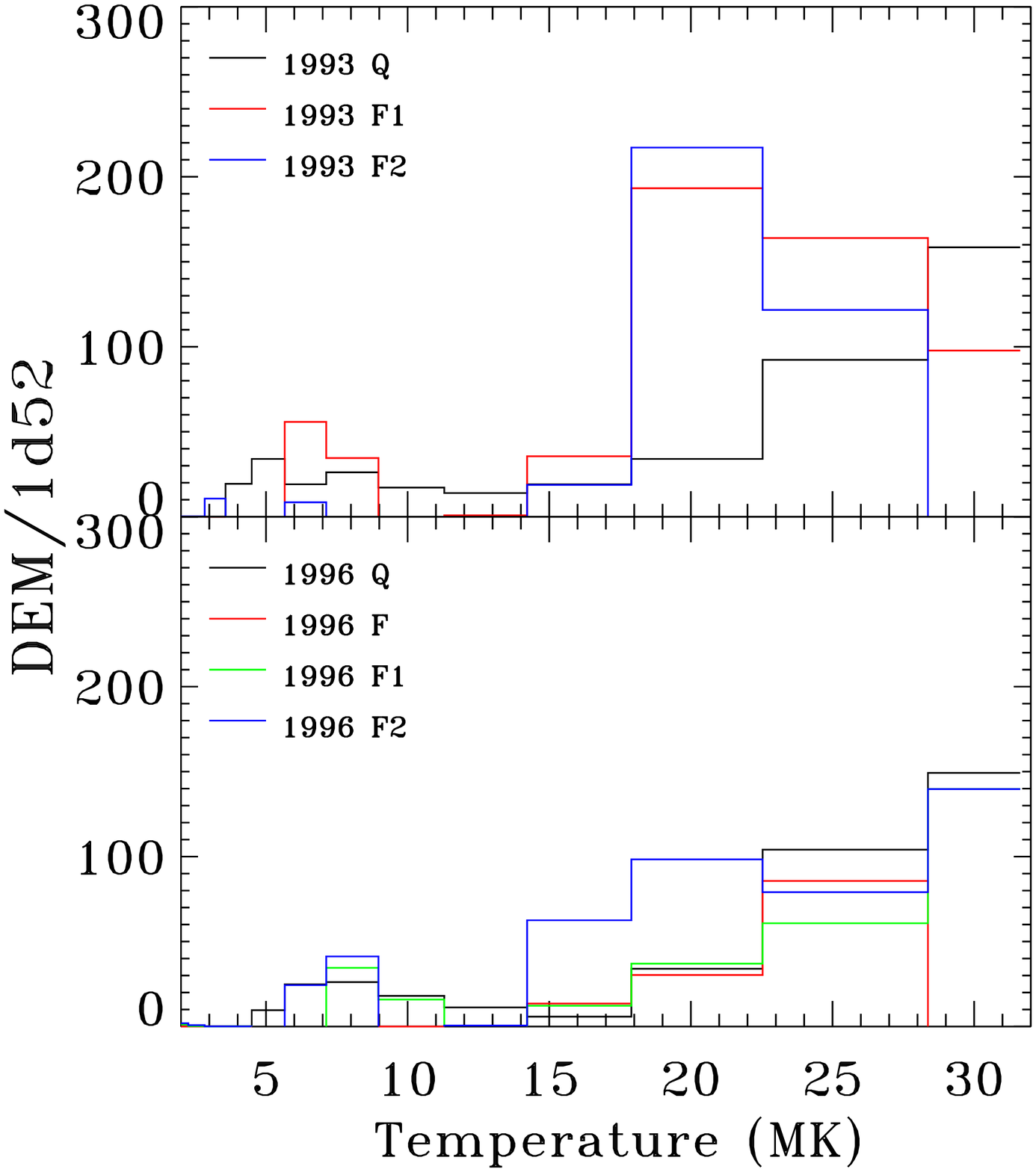}}
\caption[]
{Comparison of quiescent and flare DEMs from 1993 and 1996 {\it EUVE} observations.  The flare DEMs have had
the quiescent DEM subtracted, to examine the flare contribution to the DEM.
This shows up primarily as more plasma at temperatures $>$ 15 MK, and
an enhancement at 6--9 MK. \label{fig:comparedems}}
\end{center}
\end{figure}

%%%%%%%%%%%%%%%%%%%%%%%%%%%%%%%%%%%%%%

\clearpage
%Figure 20
\begin{figure}[htbp]
\begin{center}
\scalebox{0.7}{
\plotone{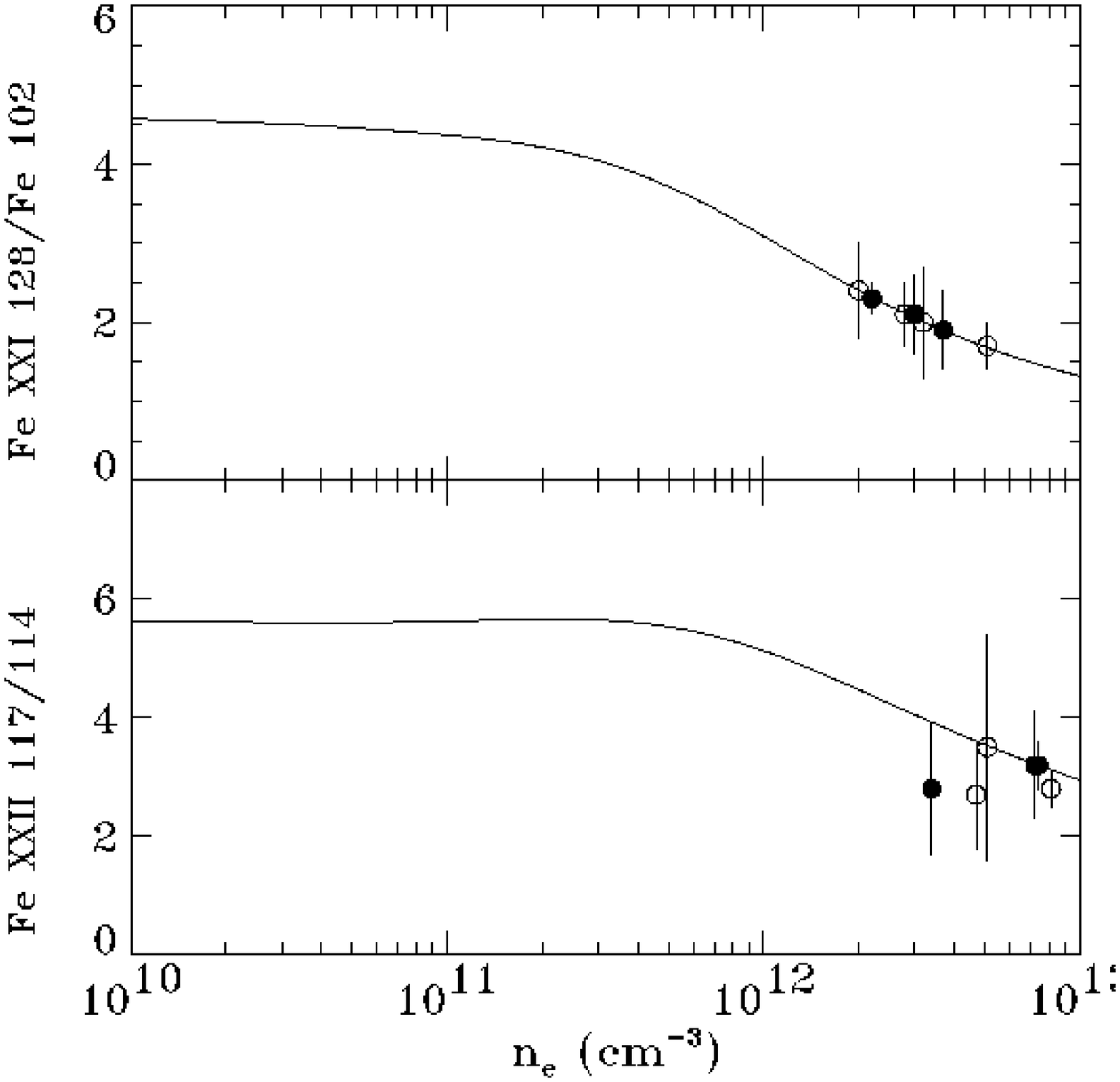}}
\caption[]
{Electron densities derived from Fe~XXI and Fe~XXII energy flux line ratios
in {\it EUVE} spectra of HR~1099.  Circles indicate values of the line ratios; error bars 
correspond to 1 $\sigma$ uncertainties.  Theoretical curves are from \citet{brs}.
Open circles indicate flare segments; filled circles quiescent intervals.
All measurements indicate high electron densities; there is no systematic enhancement of
electron densities during flare segments compared to quiescent segments.
\label{hr1099density}}
\end{center}
\end{figure}

\clearpage
%Figure 21
%%%%%%%%%%%%%%%%%%%%%
%Figure 1994ascalc
\begin{figure}[htbp]
\begin{center}
\scalebox{0.7}{
\plotone{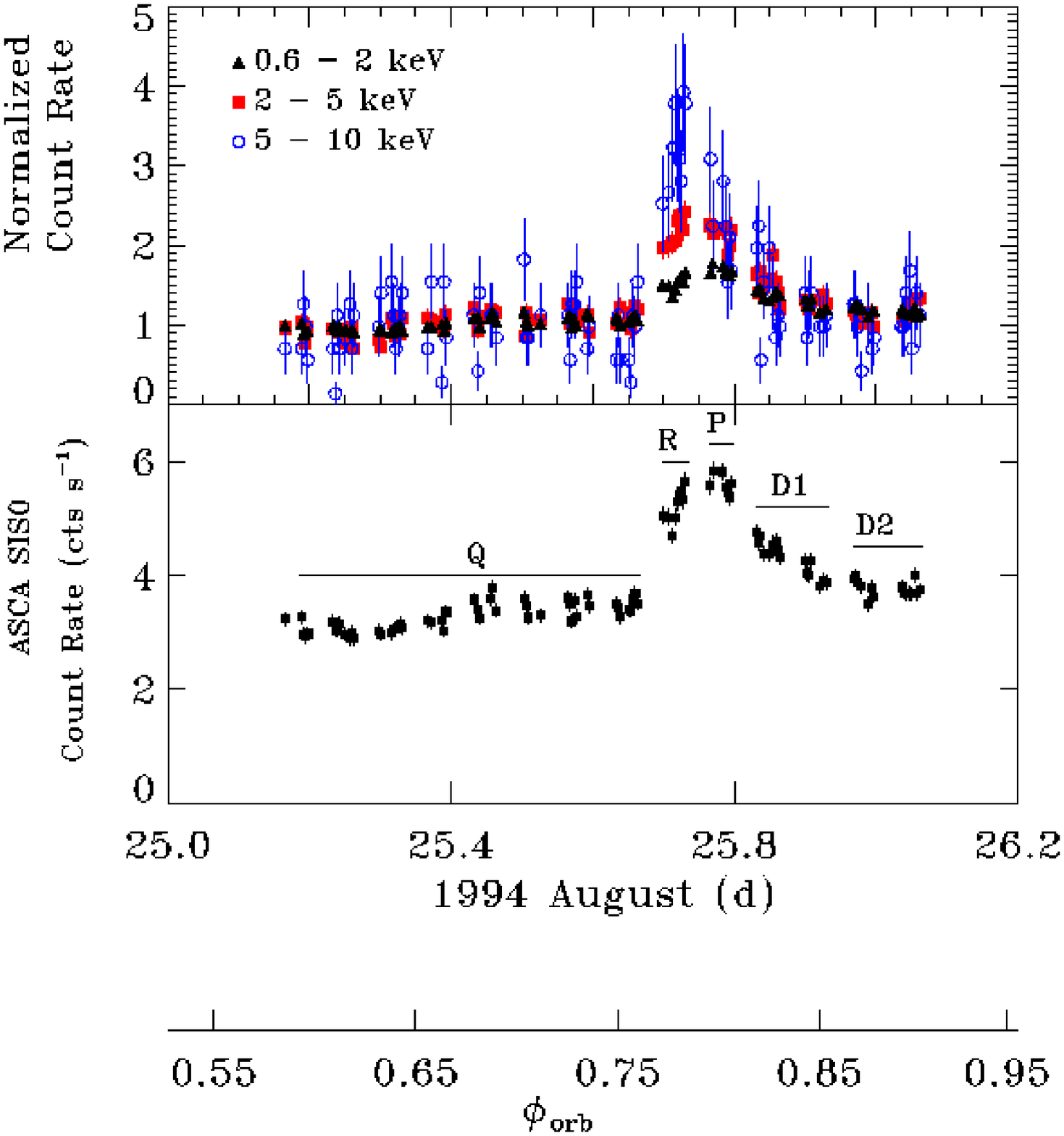}}
\caption[]{
{\bf (top)} Time evolution of {\it ASCA} SIS0 count rates in three different bandpasses, 0.6--2 keV,
2--5 keV, and 5--10 keV, during the observation, normalized to the respective average quiescent
count rates in each bandpass.  Error bars are 1 $\sigma$.  There is a trend of increasing enhancement in count rate
during the flare with increasing bandpass energy.
{\bf (bottom)} {\it ASCA} SIS0 0.6-10 keV light curve during the 1994 observation.
Time intervals used for spectral extraction are indicated; binary orbital phase during the observation
is indicated at the bottom.  Each point is an average over 256 s; error bars are 1 $\sigma$.
\label{1994ascalc}}
\end{center}
\end{figure}
%%%%%%%%%%%%%%%%%%%%%

\clearpage
%Figure 22
%Figure 1994ascaspec
\begin{figure}[htbp]
\begin{center}
\scalebox{0.7}{
\plotone{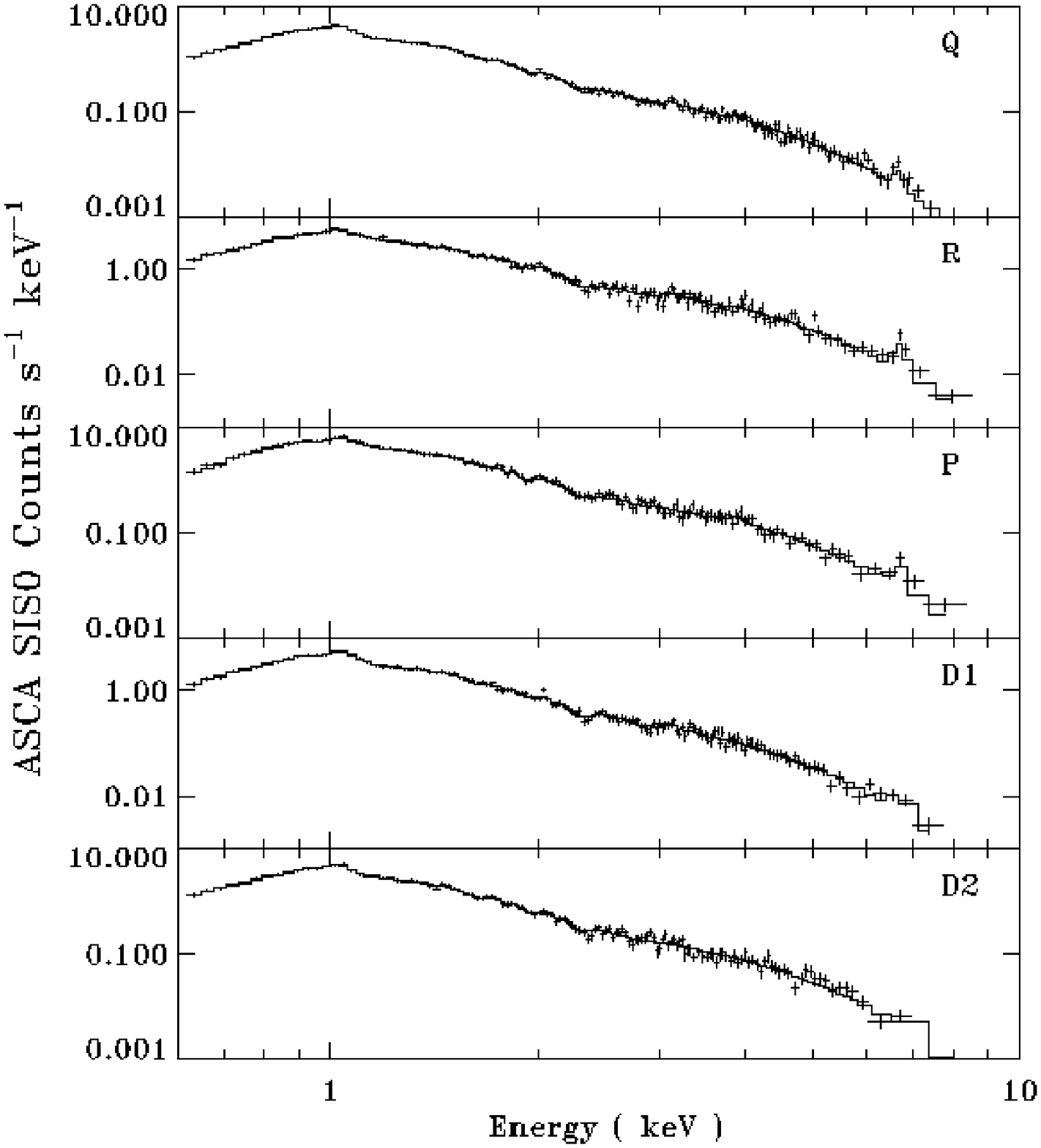}}
\caption[]{Time-resolved {\it ASCA} SIS0 spectra, along with 2T VMEKAL fits to
data.  Time intervals {\bf Q}, {\bf R}, {\bf P}, {\bf D1}, and {\bf D2} are indicated in
Figure~\ref{1994ascalc}.  Column density N$_{H}$ is held fixed at 1.35 10$^{18}$ cm$^{-2}$; see text for details.
\label{1994ascaspec}}
\end{center}
\end{figure}

\clearpage
%Figure 23
%%%%%%%%%%%%%%%%%%%%%%
%Figure asca_demabundchange
\begin{figure}[htbp]
\begin{center}
\rotatebox{90}{\scalebox{0.6}{
\plotone{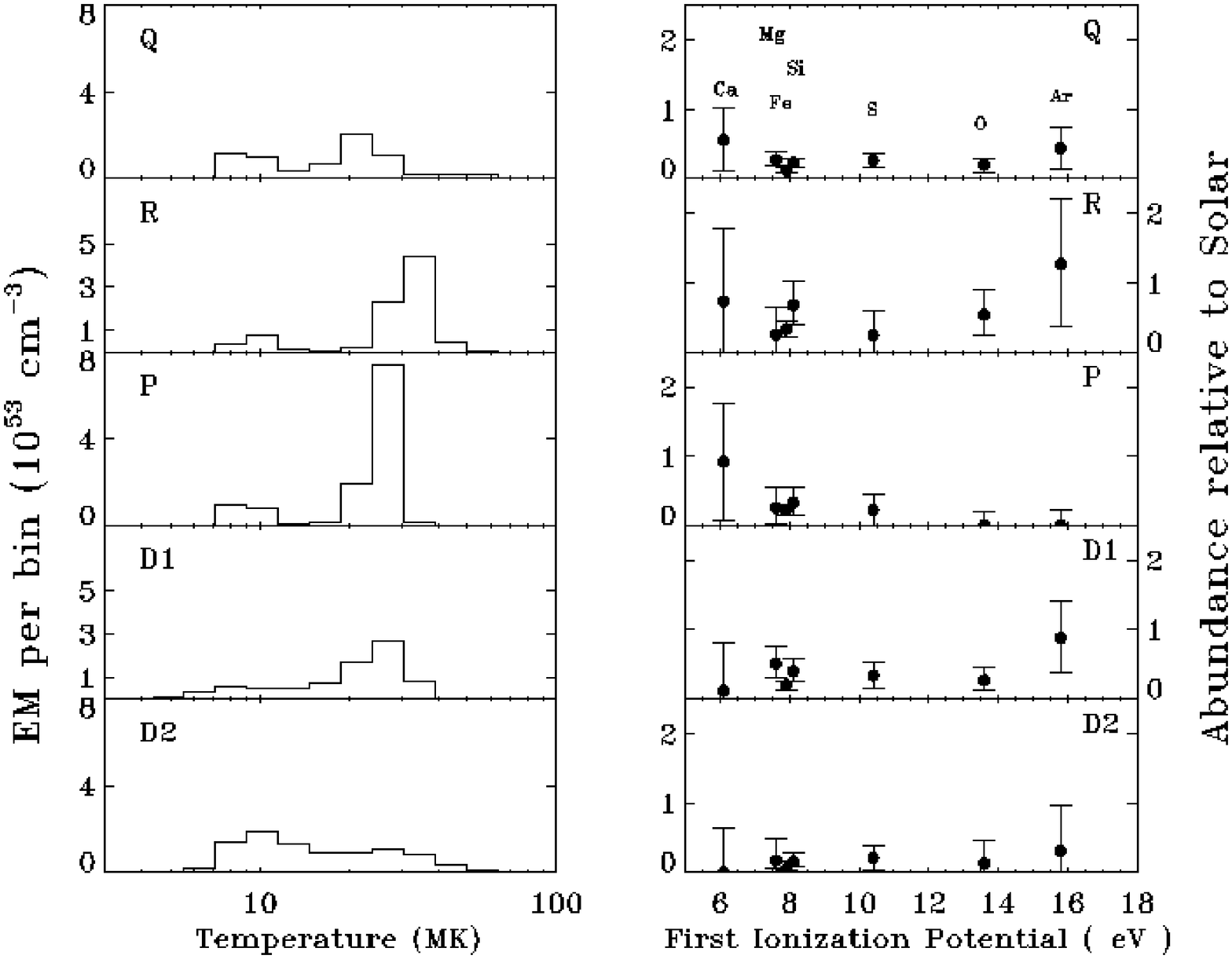}}}
\caption{
{\bf (left)} Variation of the emission measure distribution
determined from {\it ASCA} SIS0+GIS2 spectra over the course of the observation.  
Chebyshev polynomials of order 8--10 have been used.  Abundances from 2T SIS0+GIS2 fits have been
used in these fits.
Time intervals
are indicated in Figure~\ref{1994ascalc}.  {\bf (right)} Variation in the derived abundances
from 2T VMEKAL fits over the course of the observation, plotted against the first ionization potential
(FIP).  Element identifications are shown in the uppermost panel.  Error bars are 90\% confidence
intervals.
\label{asca_dem_abundchange} }
\end{center}
\end{figure}
%%%%%%%%%%%%%%%%%%%%%%%%%%%%%%5

\clearpage
%Figure 24
%%%%%%%%%%%%%%%%
%Figure rxtelc
\begin{figure}[htbp]
\begin{center}
\rotatebox{90}{\scalebox{0.6}{
\plotone{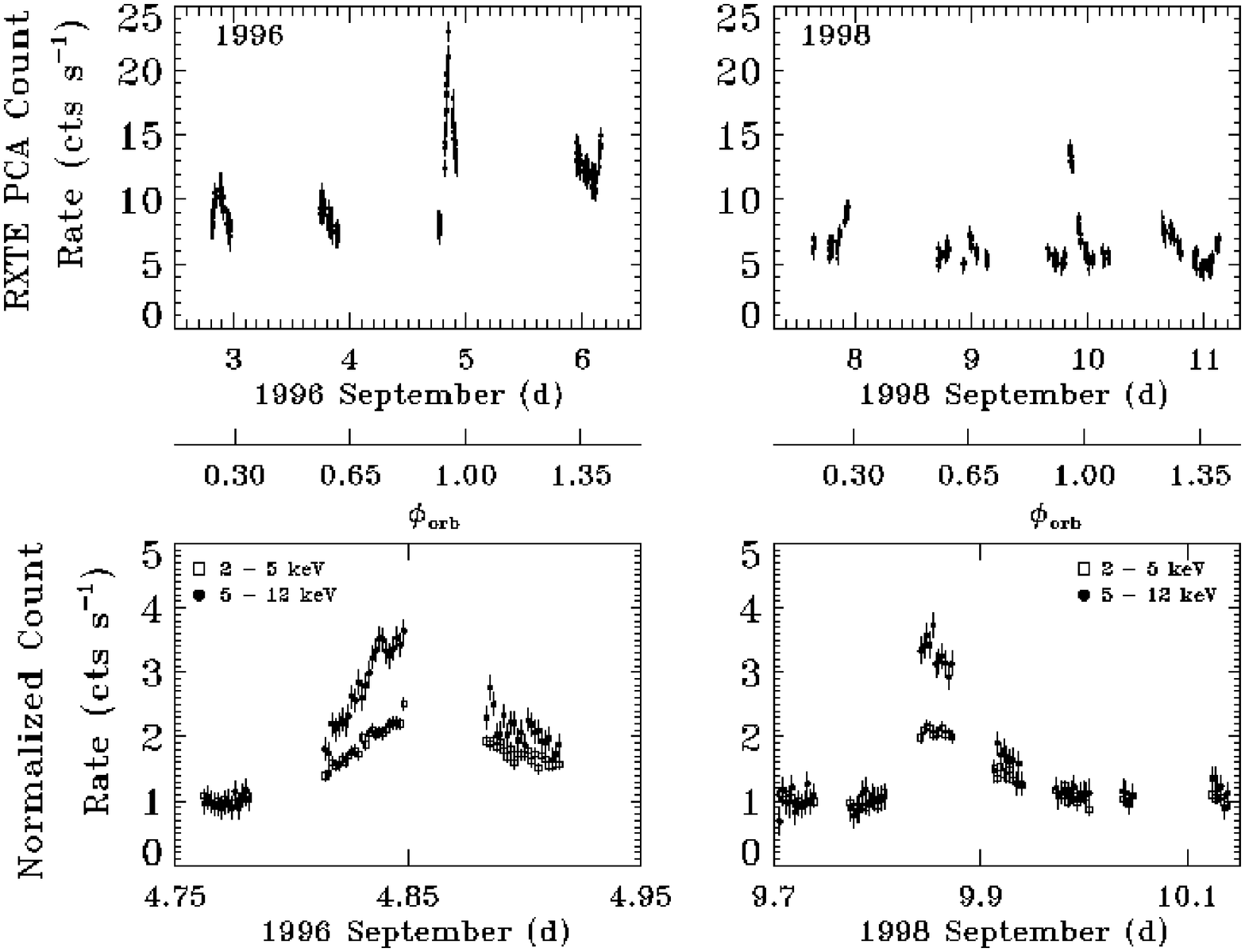}}}
\caption[]{{\bf (top)} {\it RXTE} PCA (2-12 keV) light curves from the 
1996 and 1998 observations.  The 1996 observation has 128 second bins, while the 1998 observation
had 256 second bins.  One sigma error bars are also plotted.
Binary orbital phase during the observation is also indicated.
{\bf (bottom)} Close-up view of two flares in 2-5 keV and 5-12 keV bands.  The
light curves have been divided by the average count rates in each bandpass outside of the flare, to
compare the relative output during the flare in the two bandpasses.
\label{rxtelc}}
\end{center}
\end{figure}
%%%%%%%%%%%%%%%%

\clearpage
%Figure 25
%%%%%%%%%%%%%%%%%%%%%%%%%%%%%
\begin{figure}[htbp]
\begin{center}
\scalebox{0.7}{\plotone{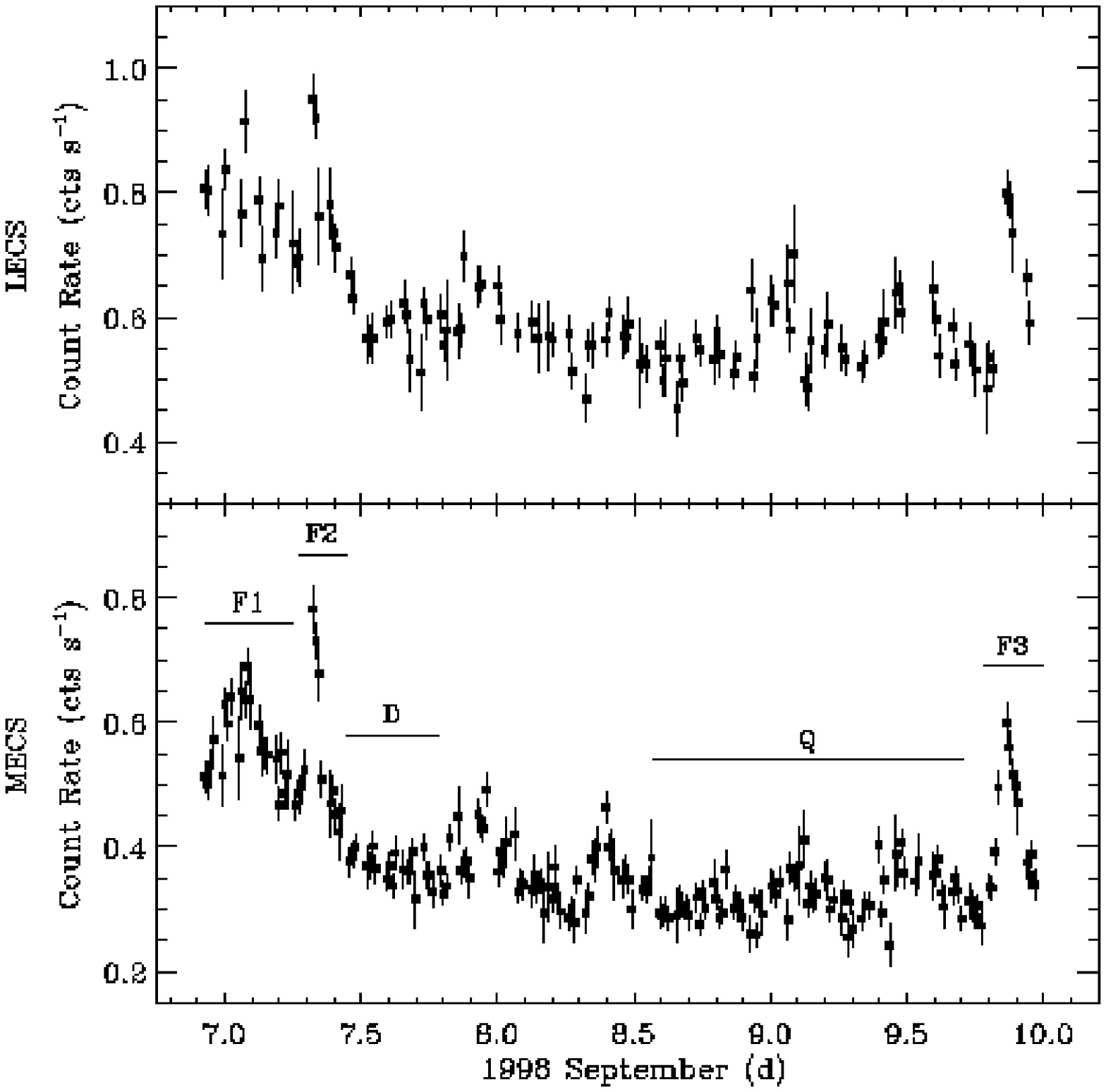}}
\caption[]{{\it BeppoSAX} LECS ({\bf top}) and MECS ({\bf bottom}) light curves 
during the 1998 observation. In the bottom panel the time intervals  used
for spectral extraction are indicated. Data are binned over 900 s; error
bars are 1 $\sigma$.
\label{1998saxlc}}
\end{center}
\end{figure}

\clearpage
%Figure 26
%%%%%%%%%%%%%%%%%%%%%%%%%%%%%
\begin{figure}[htbp]
\begin{center}
\scalebox{0.7}{
\plotone{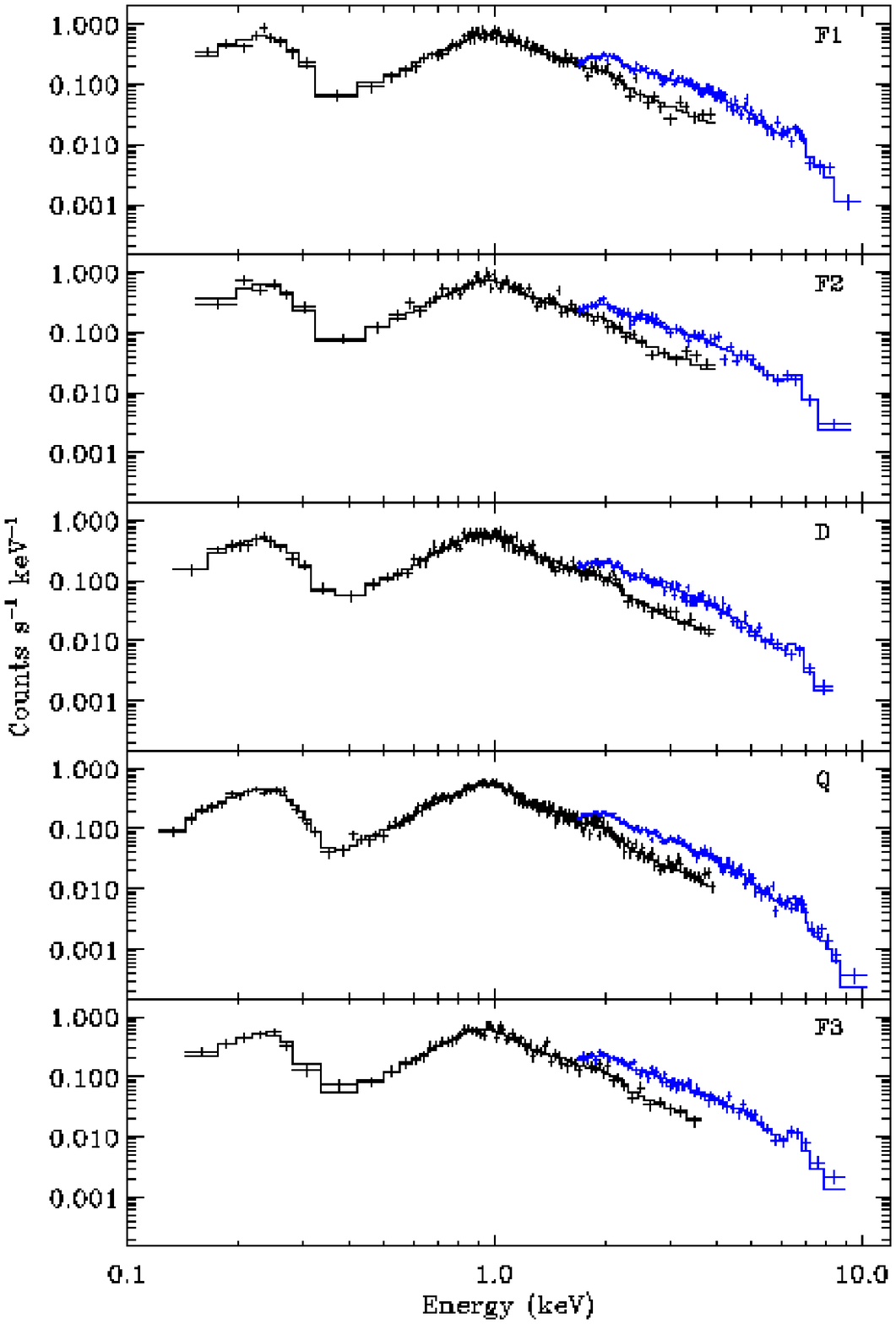}}
\caption[]{Time-resolved {\it BeppoSAX} spectra, along with 2T MEKAL fits to
data. Time intervals {\bf F1}, {\bf F2}, {\bf D}, {\bf Q}, and {\bf F3} are
indicated in
Figure~\ref{1998saxlc}.
\label{1998saxspec}}
\end{center}
\end{figure}

\clearpage
%Figure 27
%%%%%%%%%%%%%%%%%%%%%%%%%%%%%
%Figure 1993multiw
\begin{figure}[htbp]
\begin{center}
\figurenum{25}
\rotatebox{90}{\scalebox{0.7}{
\plotone{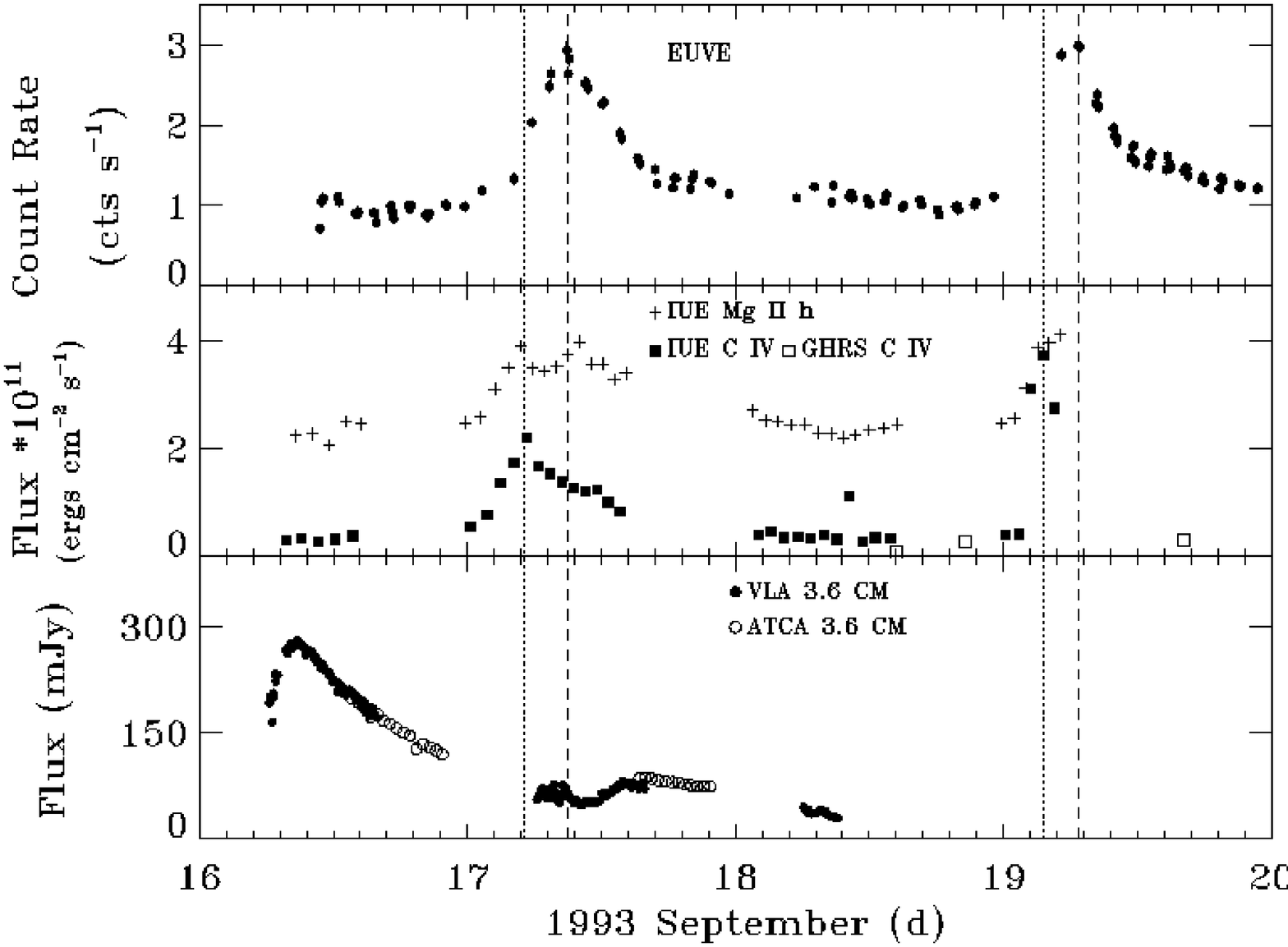}}}
\caption[]{Plot of EUV, UV, and radio variations 
in 1993.  Dashed lines indicate approximate times of peaks in the
EUV radiation; dotted lines indicate times of maximum C~IV emission.
\label{1993multiw}}
\end{center}
\end{figure}

\clearpage

%Figure 26a
%%%%%%%%%%%%%%%%%%%%%%%%%%%%%%
%Figure 1994multiw2
\begin{figure}[htbp]
\begin{center}
\figurenum{26a}
\rotatebox{90}{\scalebox{0.7}{
\plotone{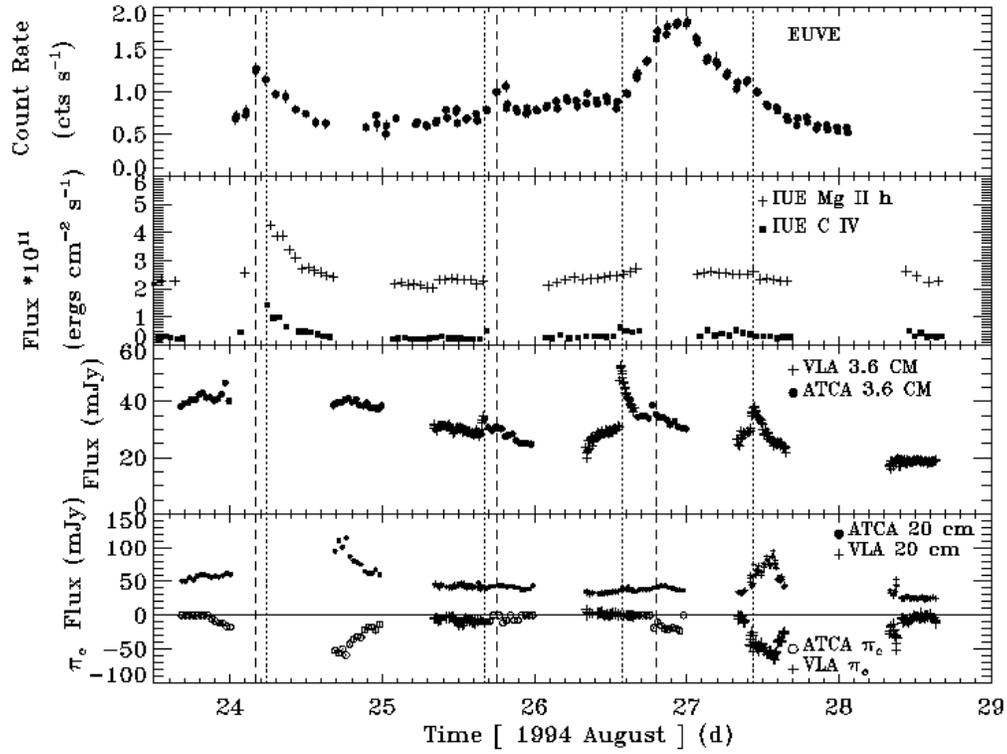}}}
\caption[]{
(a) Plot of EUV, UV, and radio variations in 1994 August.
Dotted lines indicate times of UV and$/$or 3.6 cm peaks; dashed
lines delineate EUV peak times.
(b)Plot of X-ray, EUV, UV, and radio variations over
$\approx$ 20 hours on 1994 August 25.  Dashed$/$dotted lines are
same as for (a).
\label{1994multiw2}}
\end{center}
\end{figure}

%Figure 27a
%Figure 1994multiw1
\begin{figure}[htbp]
\begin{center}
\figurenum{26b}
\scalebox{0.7}{
\plotone{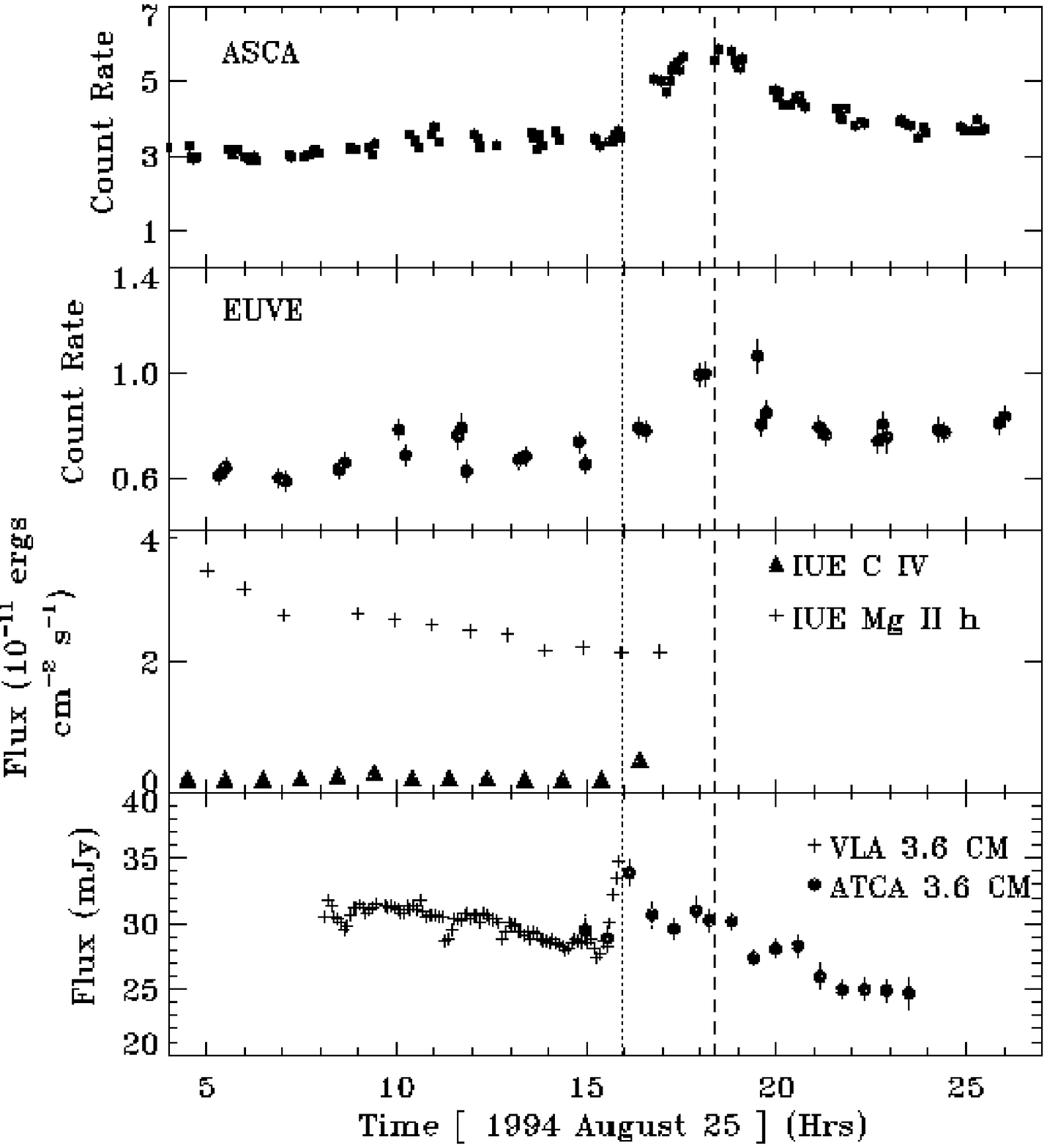}}
\caption[]{
\label{1994multiw1}}
\end{center}
\end{figure}

\clearpage

\addtocounter{figure}{1}
\clearpage
%Figure 27
%%%%%%%%%%%%%%%%%%%%%%%%%%%%%%%
%Figure 1996multiw1
\begin{figure}[htbp]
\begin{center}
\figurenum{27}
\rotatebox{90}{\scalebox{0.7}{
\plotone{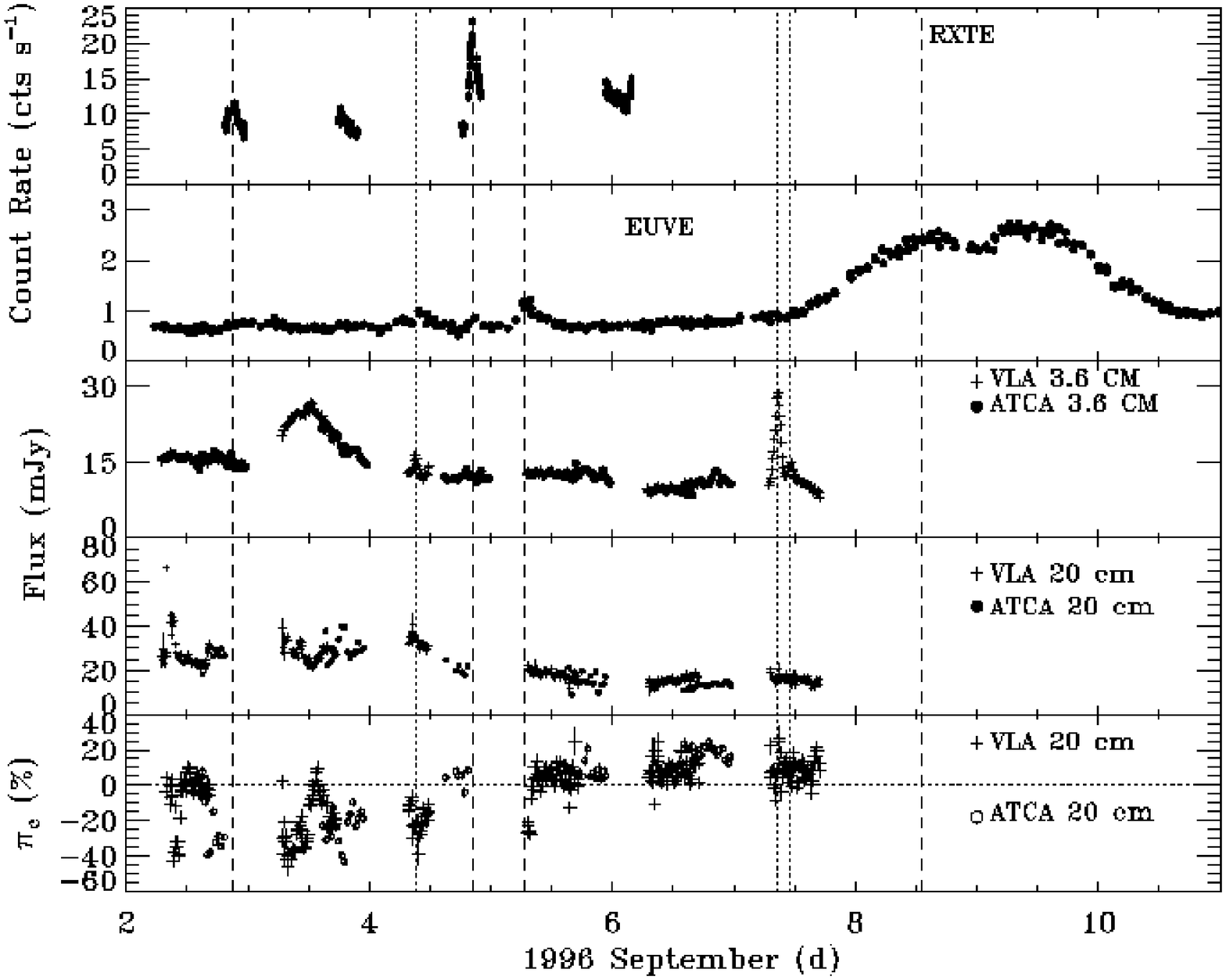}}}
\caption[]{Plot of X-ray, EUV, and radio variations
in 1996.  Dashed lines indicate EUV$/$SXR peaks; dotted lines
indicate radio peaks.
\label{1996multiw1}}
\end{center}
\end{figure}

\clearpage
%Figure 28
%%%%%%%%%%%%%%%%%%%%%%%%%%%%%%
%Figure 1998multiw1
\begin{figure}[htbp]
\begin{center}
\figurenum{28}
\scalebox{0.7}{
\rotatebox{90}{\plotone{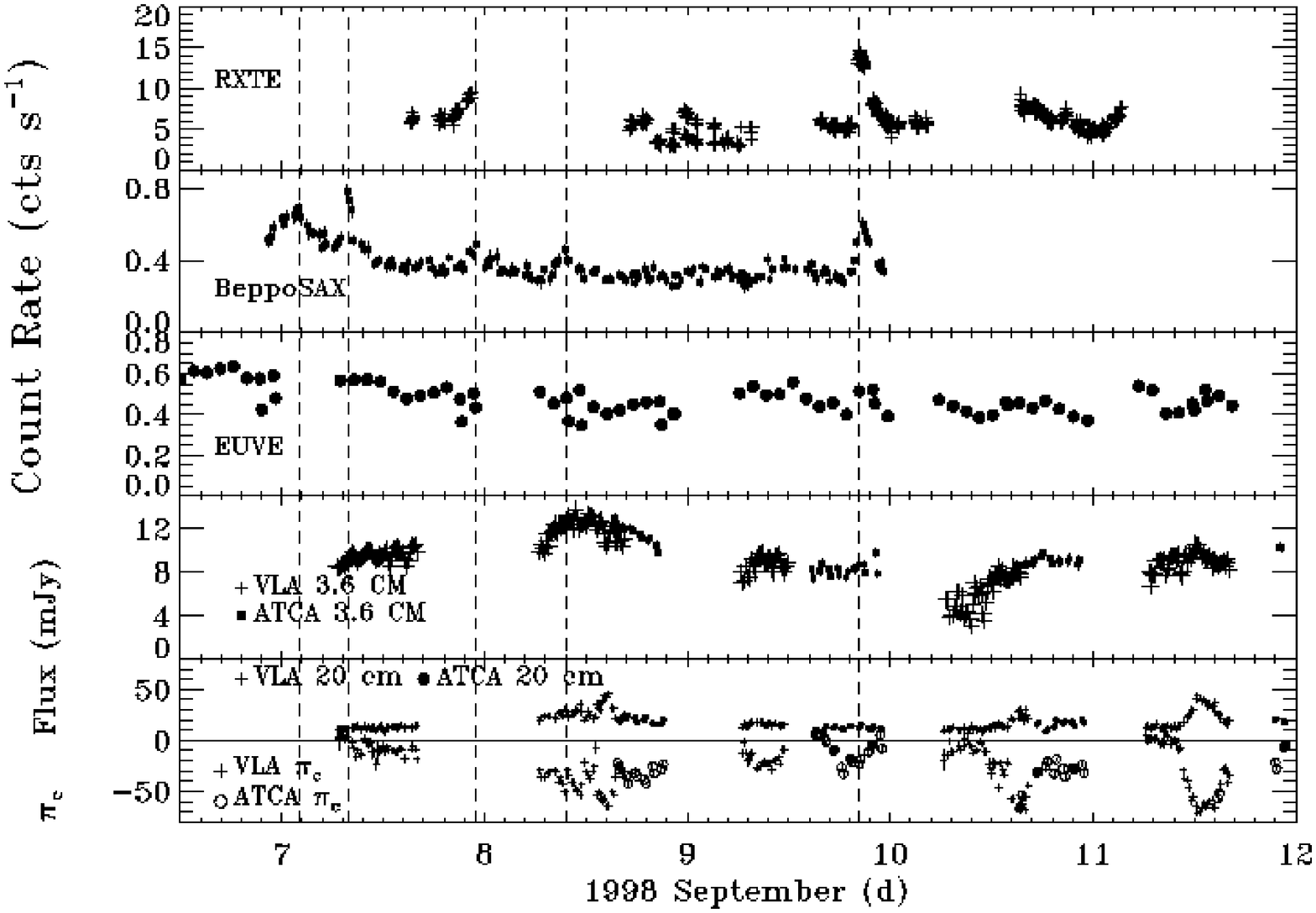}}}
\caption[]{X-ray, EUV, and radio variations
in 1998.  Dashed lines indicate times of EUV/X-ray flare peaks.
 \label{1998multiw1}}
\end{center}
\end{figure}
%%%%%%%%%%%%%%%%%%%%%%%%%%%%%%%%%%%%%%%%

\clearpage
\addtocounter{figure}{1}
%Figure 29
\begin{figure}[htbp]
\begin{center}
\figurenum{29}
\scalebox{0.7}{
\plotone{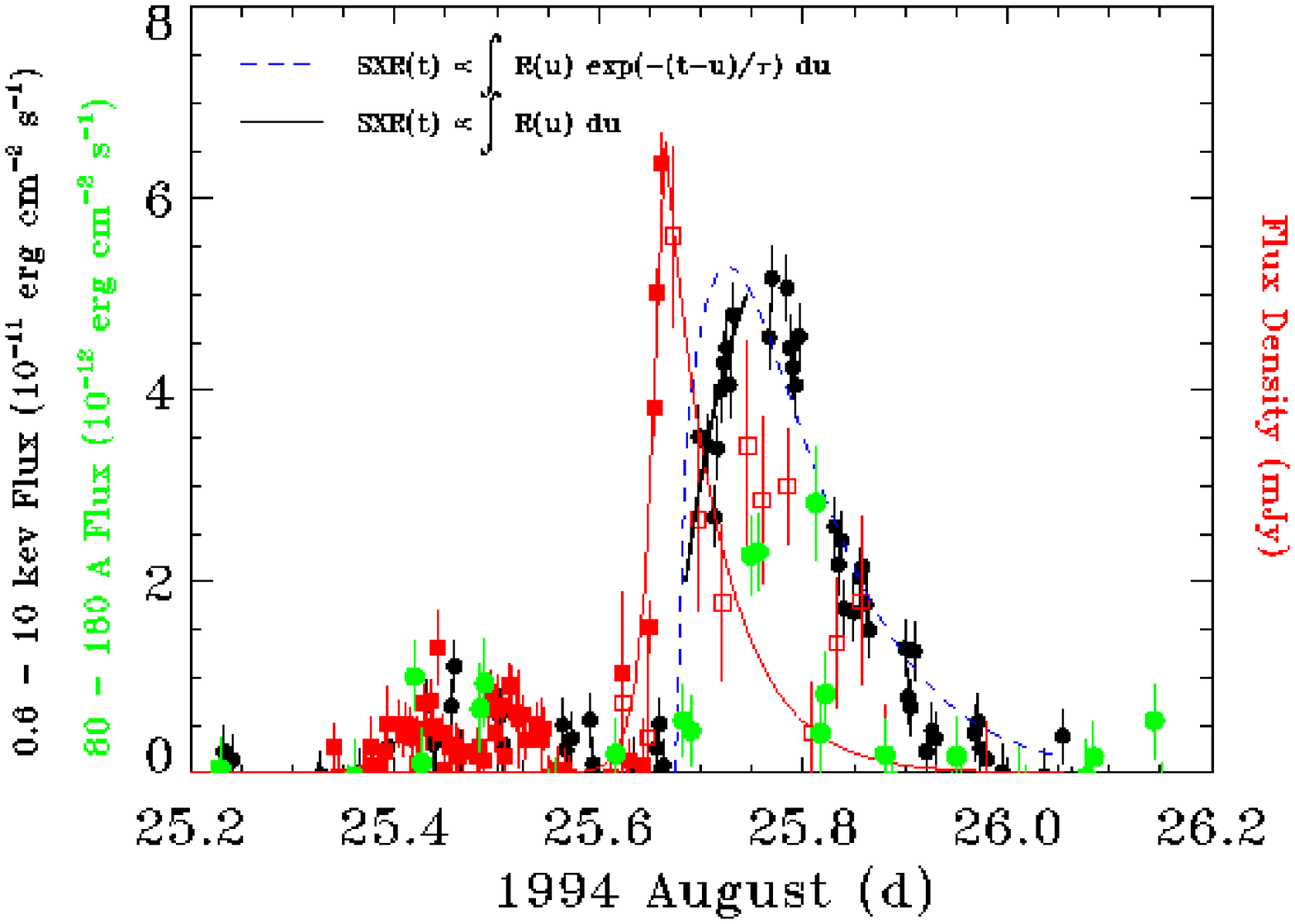}}
\caption[]{Example of the Neupert effect in a flare on HR~1099.  Red squares are 3.6 cm radio flare fluxes
(filled=VLA data, open=ATCA data), with an
estimate of the quiescent emission subtracted; red curve is a fit to the data, using an exponentially
decaying time profile.  Black circles are {\it ASCA} 0.6--10 keV flare fluxes, 
green circles {\it EUVE} 80--180 \AA\ DS flare fluxes; an estimate of the X-ray/EUV quiescent
emission has been subtracted from these.  Blue dotted line illustrates the convolution of the radio profile with
an exponential function; $\tau$ is the radiative decay timescale, set to 6000 seconds, and appropriate for
plasma at T$\sim$ 10$^{7}$K, n$_{e}$ $\sim$ 10$^{11}$ cm$^{-3}$.  This is the expected temporal trend of the luminosity of 
the thermal plasma responding to energy deposition as represented by the radio profile.  The black curve
shows the expected increase in luminosity if the radiative timescale becomes infinite; this approximation
is only applicable to the rise phase of the X-ray flare.
\label{fig:neupert}}
\end{center}
\end{figure}

\end{document}